\documentclass[fleqn,usenatbib]{mnras}

\usepackage{amsmath}
\usepackage[T1]{fontenc}
\DeclareRobustCommand{\VAN}[3]{#2}
\let\VANthebibliography\thebibliography
\def\thebibliography{\DeclareRobustCommand{\VAN}[3]{##3}\VANthebibliography}

\usepackage{txfonts}
\usepackage{graphicx}	
\usepackage{amsfonts} 
\usepackage{algorithm}
\usepackage[noend]{algpseudocode}
\usepackage{array}
\usepackage{textcomp}
\usepackage{stfloats}
\usepackage{url}
\usepackage{verbatim}
\usepackage{graphicx}
\usepackage{caption}
\usepackage{subcaption}
\usepackage[dvipsnames]{xcolor}
\usepackage{adjustbox}
\usepackage{hyperref}
\usepackage{booktabs}

\usepackage{placeins}

\definecolor{linkcolor}{RGB}{92,92,192}
\definecolor{blackcolor}{RGB}{0,0,0}
\hypersetup{colorlinks=true,linkcolor=blackcolor,citecolor=linkcolor,urlcolor=linkcolor}

\usepackage{tikz}
\usepackage{pgfplots}

\usepackage{algpseudocode}
\usepackage{algorithmicx}

\usepackage{bm}

\usetikzlibrary{positioning}

\usepackage{natbib}
\usepackage{multirow}

\usepackage{mathrsfs}

\usepackage{subcaption} 

\title[S-R2D2]{S-R2D2: a spherical extension of the R2D2 deep neural network series paradigm for wide-field radio-interferometric imaging}

\author[A. Tajja et al.]{
A. Tajja$^{1,2}$,
A. Aghabiglou$^{2}$,
E. Tolley$^{1}$,
J-P. Kneib$^{1}$,
J-P. Thiran$^{1}$,
and Y. Wiaux$^{2}$\thanks{E-mail: y.wiaux@hw.ac.uk} \\
$^{1}$ EPFL, École Polytechnique Fédérale de Lausanne, Lausanne, Switzerland.\\
$^{2}$ Institute of Sensors, Signal, and Systems, Heriot-Watt University, Currie, Edinburgh EH14 4AS, UK.}

\date{Accepted XXX. Received YYY; in original form ZZZ}
\pubyear{2025}

\pgfplotsset{compat=1.18}

\begin{document}
\label{firstpage}

\maketitle

\pagerange{\pageref{firstpage}--\pageref{lastpage}}
\begin{abstract}
\noindent
Recently, the R2D2 paradigm, standing for ``Residual-to-Residual DNN series for high-Dynamic-range imaging'', was introduced for image formation in Radio Interferometry (RI) as a learned version of the traditional algorithm CLEAN. The first incarnations of R2D2 are limited to planar imaging on small fields of view, failing to meet the spherical-imaging requirement of modern telescopes observing wide fields. To address this limitation, we propose the spherical-imaging extension S-R2D2. Firstly, as R2D2, S-R2D2 encapsulates its minor cycles in existing 2D-Euclidean deep neural network (DNN) architectures, but adapts its iterative scheme to incorporate the wide-field measurement model mapping a spherical image to visibility data. We implemented this model as the composition of an efficient Fourier-based interpolator mapping the spherical image onto the equatorial plane, with the standard RI operator mapping the equatorial-plane image to visibility data. Importantly, the interpolation step must inevitably be performed at a lower-than-optimal resolution on the plane, to meet the high-resolution requirement on the sphere of wide-field imaging while preserving scalability. Therefore, secondly, we design S-R2D2's DNN training loss to jointly learn to correct the interpolation approximations and identify residual image structures on the sphere, ensuring consistency with the spherical ground truth using the adjoint plane-to-sphere interpolator. Finally, we demonstrate through simulations S-R2D2's capability to perform fast and accurate reconstructions of spherical monochromatic intensity images, across high-resolution, high-dynamic-range settings.
\end{abstract}


\section{Introduction}
\label{sec:introduction}
\noindent
Modern radio interferometers (RI), such as the Murchison Widefield Array (MWA; \citealp{MWA1,MWA2,MWA3}), the Low-Frequency Array (LOFAR; \citealp{LOFAR1,LOFAR2}), or the future Square Kilometre Low-Frequency Array (SKA-Low; \citealp{SKA_Low_1,SKA_Low_2}) telescopes, aim to probe the sky on wide fields of view (FOV). Unlike in the small-field regime, where signals can be approximated as planar images, the spherical curvature can no longer be neglected in such wide-field settings. This places additional challenges on the design of image formation algorithms, now having to map observed visibility data to a spherical image \citep{Thompson2017}.\\

\noindent
On the one hand, traditionally, wide-field signals have been reconstructed with the CLEAN algorithm \citep{hogbom1974aperture,Clark1980}, using a faceting strategy \citep{Cornwell_Perley_1992,Tasse_2018}, where the spherical FOV is modelled as a collection of small-field planar patches. Faceting strategies image each sky patch separately, while ensuring fidelity to the observed visibility data before assembling them into a final image. This requires correcting both in-facet distortions and between-facet inconsistencies. Although parallelisable and scalable \citep{Monnier}, faceting introduces additional challenges in properly accounting for curvature effects, particularly towards the extents of the FOV, where these effects become more pronounced.

\noindent
On the other hand, triggered by early work of \citealp{wiaux2009compressed} in the context of compressive sensing and optimisation theory, a realm of alternative algorithms to CLEAN have been proposed in recent years to improve the precision of the RI image formation process \citep{Gardsen,Junklewitz,Terris2022}, with most recent efforts unlocking the potential of deep learning \citep{Connor_2022,Dabbech_2022,terris2024}. Most recently, a learned version of CLEAN, the ``Residual-to-Residual Deep neural network series for large scale high-Dynamic-range imaging'' (R2D2) paradigm, has been demonstrated to deliver a unique regime of joint precision and speed \citep{Dabbech_2024,aghabiglou2024r2d2,aghabiglou2025}. However, being in their infancy, these evolutions are not endowed with as comprehensive a set of functionalities as CLEAN. R2D2 in particular is limited to forming 2D-Euclidean images on small FOVs.\\

\noindent
The purpose of this article is to enhance the R2D2 DNN series paradigm with a capability to solve the wide-field inverse problem on the sphere. We propose an alternative reconstruction approach to faceting that preserves the spherical topology of the target image throughout the reconstruction process, in order to address curvature effects without the need for additional corrections.\\

\noindent
With regards to modelling the target spherical image and the associated wide-field RI measurement operator, various methods have been proposed in the literature. A first approach represents the sky intensity target in the spherical harmonic basis and maps the visibilities to its harmonic coefficients \citep{Shaw_2014,Eastwood_2018,Kriele2022}. A second approach considers the full 3D geometry of the problem and leverages the 3D nonuniform fast Fourier transform (FFT) \citep{kashani2023,TOLLEY2025100920}. Leveraging such models to lift R2D2 onto the sphere would require designing or adapting DNNs processing spherical signals \citep{Perraudin_2019,SPHARM_Net,ocampo2023}, \emph{de facto} severely limiting the diversity of available architectures, and their speed and accuracy, when compared to the versatility and flexibility of deep learning machinery available to process 2D-Euclidean images. A third formulation, developed by \citet{McEwen_Wiaux_WFOV}, models the discrete wide-field inverse problem as an augmentation of the standard small-field one, which can be represented via a 2D nonuniform FFT, with two key additions: (i) the non-zero baseline component in the telescope pointing direction, referred to as $w$-component, that imprints additional Fourier convolution kernels on top of those inherent to the nonuniform FFT \citep{dabbech_2017}; (ii) an interpolation operation that maps the spherical signal onto the equatorial plane before 2D nonuniform FFT. Our work aims to demonstrate that leveraging the unique factorised form of such a model offers a unique opportunity to lift R2D2 onto the sphere while taking full advantage of efficient 2D-Euclidean DNN architectures.\\

\noindent
The proposed spherical extension of R2D2, denoted S-R2D2, firstly integrates within its scheme, a fast implementation of this measurement model. We note that our proposed proof of concept does not depend on the explicit structure of the 2D Fourier kernels defining the planar part of the measurement operator. Therefore, without loss of generality, we omit the effect of the $w$-component in this first proof of concept. The measurement operator thus simply reads as the composition of the 2D nonuniform FFT with an efficient sphere-to-plane Fourier-based interpolator. Secondly, S-R2D2 refines R2D2's training loss by integrating the adjoint plane-to-sphere interpolator, enabling it to accurately learn residual image structures on the sphere. However, to jointly target a sufficiently large resolution on the sphere and preserve scalability, the forward interpolator and its adjoint must operate at a lower-than-optimal resolution on the plane, which results in a critical precision-efficiency trade-off. To address this suboptimal configuration, we rely on S-R2D2's iterative structure and the DNNs' training stage to learn to correct inevitable interpolation approximations. Finally, while analysing the precision-efficiency trade-off, we compare S-R2D2 to R2D2 and demonstrate its capability to achieve fast and precise reconstruction of spherical monochromatic intensity images in simulation, for a fixed high-resolution on the sphere and various high-dynamic-range configurations.\\

\noindent
The remainder of the article is structured as follows. In Section~\ref{sec:section_2}, we review the wide-field measurement model and the R2D2 imaging algorithm. Section~\ref{sec:section_3} presents the developed Fourier-based interpolator and introduces the S-R2D2 method. Section~\ref{sec:section_4} describes the procedure for generating a wide-field RI image dataset for DNN training. Section~\ref{sec:section_5} provides a comparative evaluation of R2D2 and S-R2D2 in a wide-field setting, including an analysis of the precision-efficiency trade-off. Finally, concluding remarks are made in Section~\ref{sec:conclusion}.\\

\noindent
In the remainder of the article, the subscript $\cdot_{\mathrm{p}}$ denotes a planar quantity, either a planar image, a planar DNN or a planar operator. The subscript $\cdot_{\mathrm{s}}$ denotes a spherical quantity, either a spherical signal, a spherical DNN, or a spherical operator.

\section{RI Imaging}\label{sec:section_2}
\noindent
In this section, derived from the general continuous measurement model, we formulate the wide-field inverse problem. We also review the deep learning series approach R2D2. 

\subsection{Wide-Field Measurement Model}
\label{subsec:sec_2_subsec_1}
\begin{figure}
\centering
\adjustbox{valign=t}{\begin{subfigure}[b]{0.1665\textwidth}
\includegraphics[width=\textwidth]{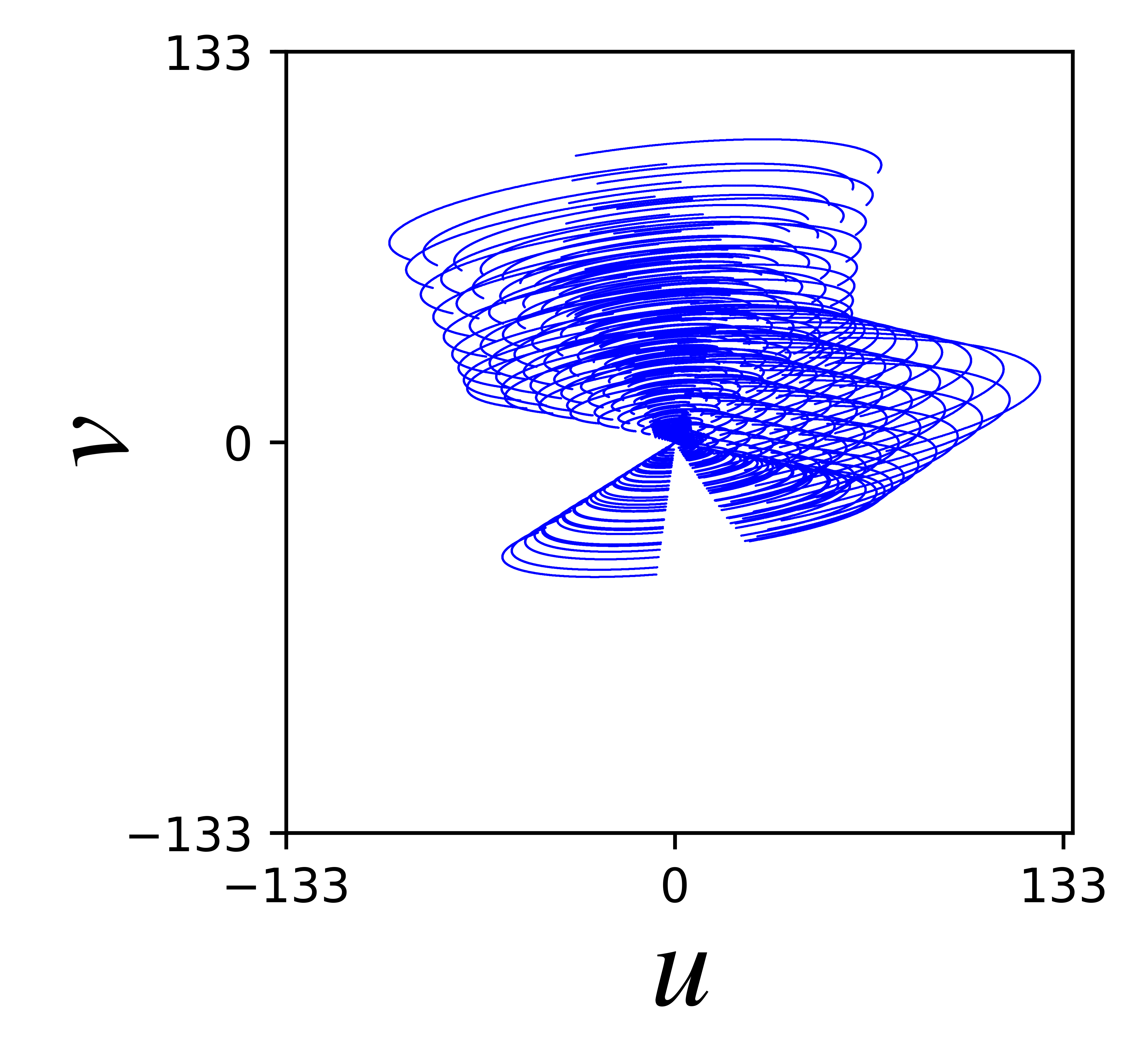}
\vspace{-2.25em} 
\caption{}
\end{subfigure}}
\hfill
\adjustbox{valign=t}{\begin{subfigure}[b]{0.152\textwidth}
\includegraphics[width=\textwidth]{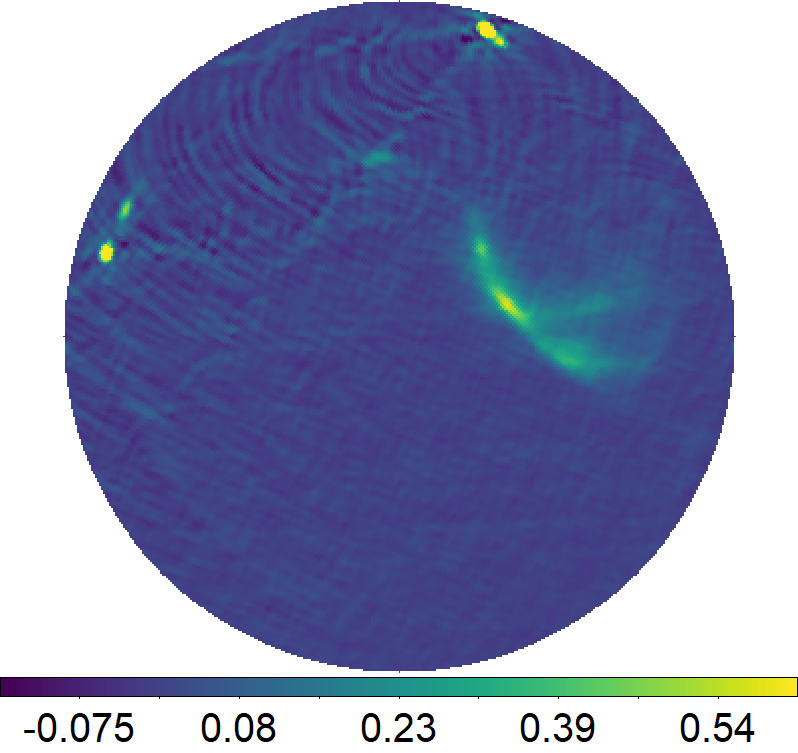}
\vspace{-1.525em}
\caption{}
\end{subfigure}}
\hfill
\adjustbox{valign=t}{\begin{subfigure}[b]{0.152\textwidth}
\includegraphics[width=\textwidth]{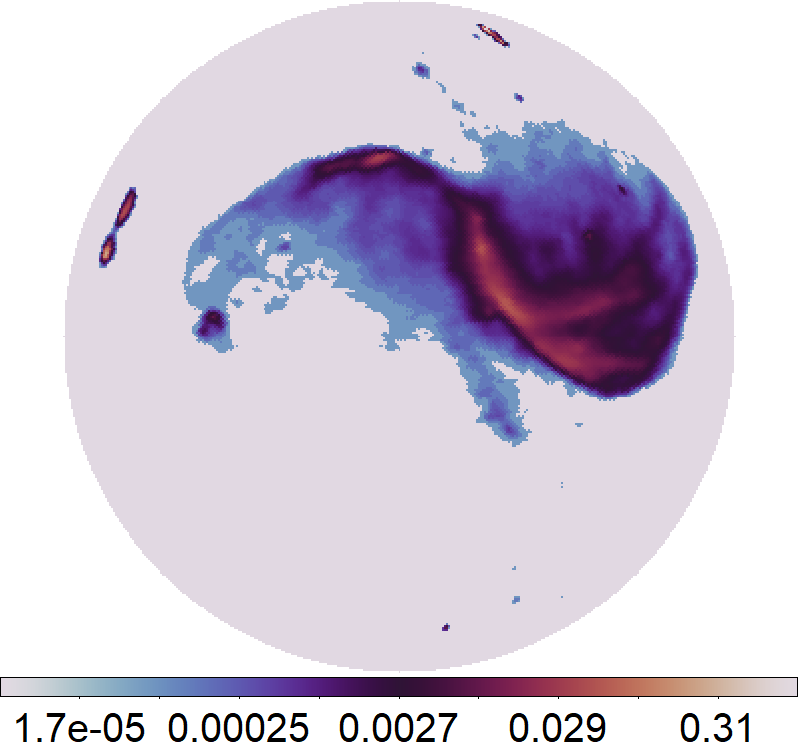}
\vspace{-1.525em}
\caption{}
\end{subfigure}}
\caption{Illustration of a wide-field RI inverse problem, imaged on the sphere. Panel (a) shows a VLA-type sampling pattern. Panel (b) shows, in a linear scale, the back-projection of RI measurements onto the sphere, generated from the measurement equation \eqref{eqn:vis_discrete_2D_sphere}, using the sampling pattern (a) and the ground truth (c). This spherical ground truth (c) is generated from the Messier~106 radio galaxy \citep{shimwell2022} with the procedure depicted in Section~\ref{subsec:sec_4_subsec_1}, and displayed in logarithmic scale with the logarithmic exponent equals to its dynamic range ($\textrm{DR}=1.4\!\times\!10^5$). Solving this inverse problem consists of reconstructing the target (c) from the back-projected measurements (b). We visualise the Northern hemisphere in the orthographic projection perspective in panels (b) and (c).}
\label{fig:inverse problem}
\end{figure}
\begin{figure*}
\vspace{-0.55cm}
\centering
\hspace{-0.35cm}
\vspace{-0.15cm}
\begin{subfigure}[t]{0.29\textwidth}
    
    \includegraphics[width=\textwidth]{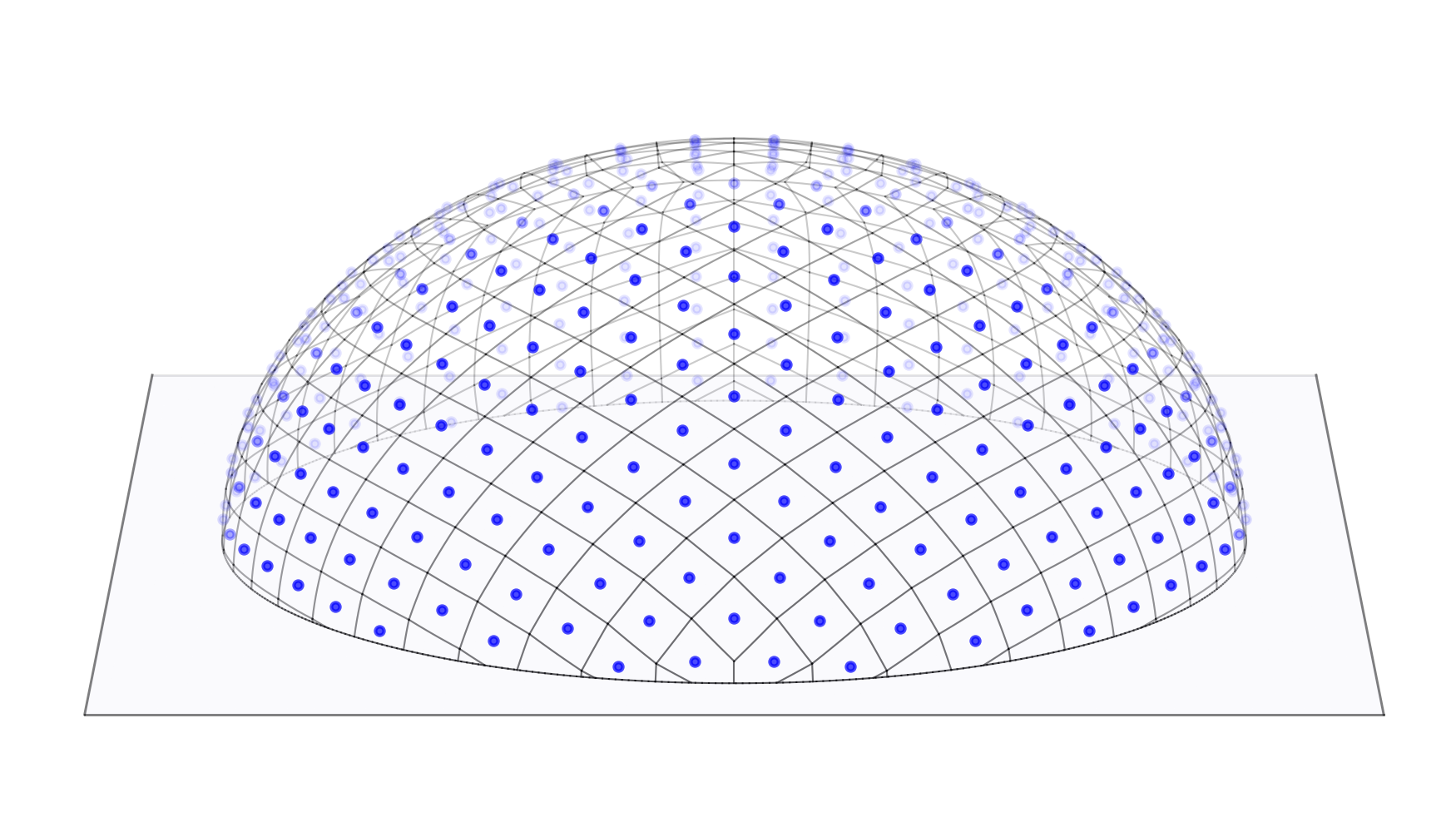}
    \vspace{-0.425cm}
    \caption{}
\end{subfigure}
\hspace{0.15cm}
\begin{subfigure}[t]{0.32\textwidth}
    \includegraphics[width=\textwidth]{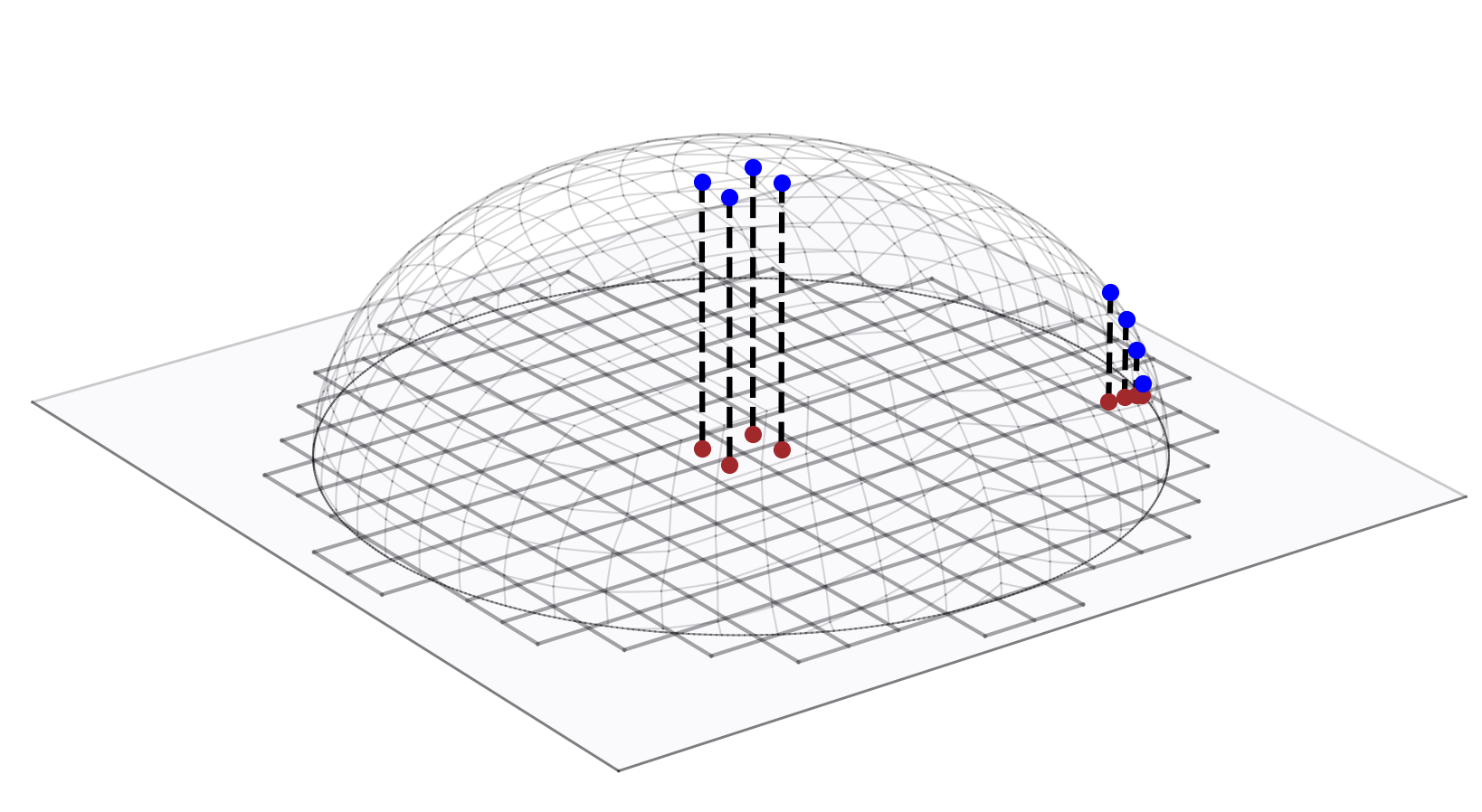}
    \vspace{-0.425cm}
    \caption{}
\end{subfigure}
\hspace{0.245cm}
\begin{subfigure}[t]{0.305\textwidth}
    \includegraphics[width=\textwidth]{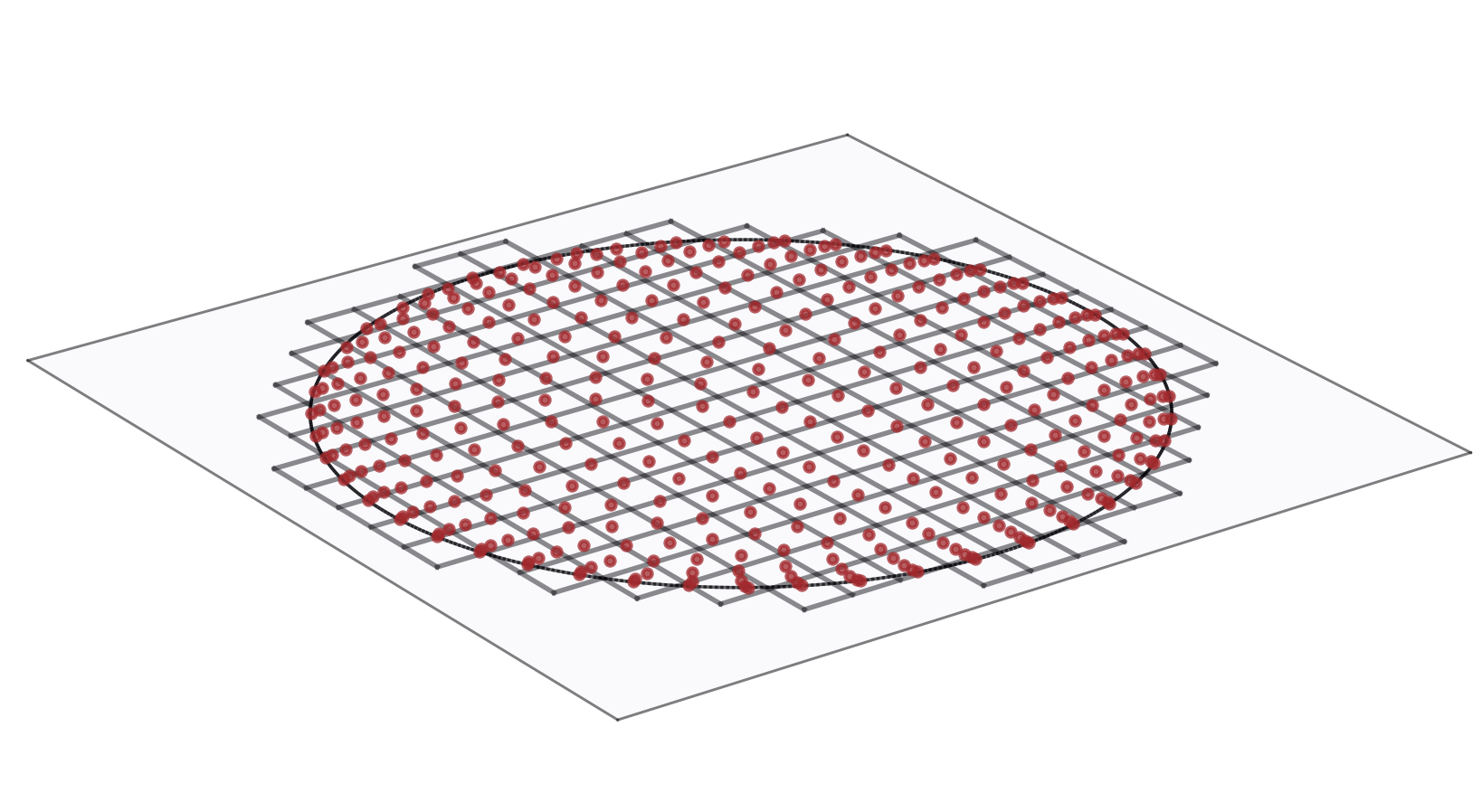}
    \vspace{-0.425cm}
    \caption{}
\end{subfigure}
\caption{Panel (a) illustrates the sample arrangement on the sphere using the uniform equal-area HEALPix scheme. The invertible projection, depicted with dashed lines in panel (b), maps a point on the sphere with coordinates 
$(l,m,n)$ to the equatorial plane by setting $n=0$. Thus, without loss of information, spherical samples, represented on the HEALPix grid, are equivalently nonuniform samples on the equatorial plane, as shown in panel (c). Moreover, panel (b) showcases that at the north pole, the frequency content is preserved, because the distances after projection are preserved. The maximum projected frequency content arises at the extents of the FOV, where the frequency content is projected onto much higher frequency content on the $(l, m)$ plane.}
\label{fig:HealPix_scheme}
\end{figure*}

\noindent
Radio-interferometric (RI) telescopes point their antennas in a common direction $\mathbf{\hat{s}_0}$ belonging to the unit celestial sphere $\mathrm{S}^2$ and image celestial regions with a specific FOV. The observed FOV is constrained by the primary beam pattern $\mathrm{A}(\boldsymbol{\tau})$ of each telescope, defined relative to the pointing direction $\mathbf{\hat{s}_0}$ with $\boldsymbol{\tau}=\mathbf{\hat{s}}-\mathbf{\hat{s}_0}$, for any arbitrary direction $\mathbf{\hat{s}}$. At a given time, each pair of antennas probes a visibility $\mathrm{y}$, which is a noisy complex measurement of the radio-emission target $\mathrm{x}(\boldsymbol{\tau})$, which is assumed to be a monochromatic intensity signal, thus real and non-negative. This measurement model can be expressed as:
\begin{equation}
\label{eqn:vis_int_sphere}
\mathrm{y}(\mathbf{b}) = \int_{\mathrm{S}^2} \mathrm{A}(\boldsymbol{\tau})\mathrm{x}(\boldsymbol{\tau})\mathrm{e}^{-i2\pi \mathbf{b}.\boldsymbol{\tau}}\mathrm{d}\Omega(\boldsymbol{\tau}),
\end{equation}
with the baseline $\mathbf{b} = (u,v,w)$ representing the separation between two antennas in units of the observing wavelength, where $w$ is measured in the pointing direction and $u$ and $v$ are measured in the perpendicular plane \citep{Thompson2017}. The rotation-invariant measure on the sphere, $\mathrm{d}\Omega(\boldsymbol{\tau})$, can be expressed in the spherical coordinates as $\mathrm{d}\Omega(\boldsymbol{\tau})=\sin(\theta)\mathrm{d}\theta\mathrm{d}\varphi$, with colatitude $\theta \in [0,\pi]$ and longitude $\varphi \in [0,2\pi)$. Without loss of generality and any particular FOV assumption, equation \eqref{eqn:vis_int_sphere} can be reformulated by doing a change of coordinate from spherical to Cartesian, as:
\begin{equation}
\label{eqn:vis_int}
\mathrm{y}(u,v,w) = \int_{\mathrm{D}^2} \mathrm{A}(l,m)\mathrm{x}(l,m)\mathrm{C}(w,l,m)\mathrm{e}^{-i2\pi[ul+vm]}\frac{\mathrm{d}l\mathrm{d}m}{n(l,m)},
\end{equation}
where we expressed the unit disk $\mathrm{D}^2$, $\boldsymbol{\tau}=(l,m,n(l,m)-1)$, the $w$-component $\mathrm{C}(w,l,m)=\mathrm{e}^{-i2\pi w(n(l,m)-1)}$, $n(l,m)=\sqrt{1-(l^2+m^2)}$ and thus $\mathrm{d}\Omega(\boldsymbol{\tau})=\mathrm{d}l\mathrm{d}m/n(l,m)$. Then, by gathering all visibility measurements $\mathrm{y}$, the wide-field inverse problem \eqref{eqn:vis_int}, illustrated in Figure~\ref{fig:inverse problem}, can be discretised as \citep{McEwen_Wiaux_WFOV}:
\begin{equation}
\label{eqn:vis_discrete_2D_sphere}
    \mathbf{y} = \mathbf{\Phi}_{\mathrm{p}}\mathbf{\Gamma}\mathbf{x}_{\mathrm{s}}^{\star} + \mathbf{n} = \mathbf{\Phi}_{\mathrm{s}}\mathbf{x}_{\mathrm{s}}^{\star} + \mathbf{n},
\end{equation}
where $\mathbf{y} \in\mathbb{C}^{\mathrm{M}}$ is the measurement vector, and $\mathbf{x}_{\mathrm{s}}^{\star} \in \mathbb{R}^{\mathrm{N}_{\mathrm{s}}}_{+}$ is the spherical intensity target, that we choose to represent in the uniform equal-area HEALPix scheme on the sphere \citep{Gorski_2005}. The Gaussian noise $\mathbf{n}\in\mathbb{C}^{\mathrm{M}}$ is characterized by a mean of 0 and a given standard deviation. Under a small-field assumption, the RI operator $\mathbf{\Phi}_{\mathrm{p}}$ can be modelled as a standard 2D nonuniform FFT, where the underlying sampling pattern is measurement-specific. In a wide-field regime, $\mathbf{\Phi}_{\mathrm{p}}$ is augmented to account for additional effects. Firstly, it incorporates the $w$-component and the primary beam term, which are neglected in the remainder of this article without loss of generality, as our study does not depend on the explicit structure of $\mathbf{\Phi}_{\mathrm{p}}$'s Fourier kernels. Secondly, before applying $\mathbf{\Phi}_{\mathrm{p}}$, an interpolator $\mathbf{\Gamma}$ is required to grid the spherical samples $\mathbf{x}_{\mathrm{s}}^{\star}$ onto a uniform planar grid, which is the inevitable input format of $\mathbf{\Phi}_{\mathrm{p}}$. As illustrated in Figure~\ref{fig:HealPix_scheme}, the spherical samples $\mathbf{x}_{\mathrm{s}}^{\star}$ are equivalently nonuniform samples on the equatorial plane. Then, the sphere-to-plane interpolator $\mathbf{\Gamma}$ is equivalently a nonuniform-to-uniform interpolator on the equatorial plane. The design of $\mathbf{\Gamma}$ is discussed in detail in Section~\ref{sec:section_3}. Furthermore, we note that in a small-field framework, $\mathbf{x}_{\mathrm{s}}^{\star}$ can be approximated as a uniformly sampled planar image $\mathbf{x}_{\mathrm{p}}^{\star}$, and $\mathbf{\Gamma}$ simplifies to $\mathbf{I}$.
\subsection{R2D2 Algorithm}
\label{subsec:sec_2_subsec_2}

\noindent
Introduced in \citet{aghabiglou2024r2d2,aghabiglou2025}, the imaging algorithm R2D2 was developed to achieve fast and accurate reconstructions of RI signals in a high-dynamic-range and small-field context. Depicted in Algorithm~\ref{algo:R2D2_training}, its scheme relies on an iteration-dependent supervised training of $\mathrm{I}$ 2D-Euclidean convolutional DNNs $(\mathcal{N}_{\mathrm{p}}^{\boldsymbol{\theta}^i})_{i=1}^{\mathrm{I}}$. Similarly to CLEAN's minor cycles, that identify remaining components through projections onto a sparse dictionary, R2D2's DNNs progressively identify residual image content, providing additional learned regularisation and improving the
estimation of the target. The DNNs share the same U-Net architecture and the same hyperparameters (learning rate, batch size; \citealp{aghabiglou2024r2d2,aghabiglou2025}) but are iteration-specific, each with its own learned parameters $\boldsymbol{\theta}^{i}$. At iteration $i\in\{1,2,...,\mathrm{I}\}$, $\mathcal{N}_{\mathrm{p}}^{\boldsymbol{\theta}^i}$ takes as input both the current image estimate $\mathbf{x}_{\mathrm{p}}^{i-1}$ and its associated visibility-data residual back-projected onto the image domain $\mathbf{r}_{\mathrm{p}}^{i-1}$:
\begin{align}
\left\{
\begin{aligned}
    \label{eqn:r2d2_iteration_i}
    \mathbf{x}_{\mathrm{p}}^{i} &= \left[\mathbf{x}_{\mathrm{p}}^{i-1} +\mathcal{N}_{\mathrm{p}}^{\boldsymbol{\theta}^i}(\mathbf{r}_{\mathrm{p}}^{i-1}, \mathbf{x}_{\mathrm{p}}^{i-1})\right]_+\\
    \mathbf{r}_{\mathrm{p}}^{i} &= \mathbf{x}_{\mathrm{p}}^{\mathrm{d}} - \mathrm{\kappa} \textrm{Re}(\mathbf{\Phi}_{\mathrm{p}}^{\dagger}\mathbf{\Phi}_{\mathrm{p}}\mathbf{x}_{\mathrm{p}}^{i}),
\end{aligned}
\right.
\end{align}
where $\left[\cdot\right]_+$ denotes the projection onto the non-negative orthant. This projection is essential to ensure the non-negativity of the reconstructed image, as the target signal is assumed to be a monochromatic intensity image, thus real and non-negative. The initial conditions are $\mathbf{x}_{\mathrm{p}}^{0}=0$ and $\mathbf{r}_{\mathrm{p}}^{0}=\mathbf{x}_{\mathrm{p}}^{\mathrm{d}}= \mathrm{\kappa} \textrm{Re}(\mathbf{\Phi}_{\mathrm{p}}^{\dagger}\mathbf{y})$. The back-projection of $\mathbf{y}$ onto the image domain, denoted $\mathbf{x}_{\mathrm{p}}^{\mathrm{d}}$, is called the dirty image and the normalisation factor $\mathrm{\kappa} = 1/\max \left[\textrm{Re}(\mathbf{\Phi}_{\mathrm{p}}^{\dagger}\mathbf{\Phi}_{\mathrm{p}}\boldsymbol{\delta})\right]> 0$, where $\boldsymbol{\delta}$ is the central discrete dirac image taking the value 1 at the central pixel and 0 elsewhere. Moreover, while all DNNs are image-domain residual DNNs, the first network $\mathcal{N}_{\mathrm{p}}^{\boldsymbol{\theta}^1}$ is, in particular, also an end-to-end DNN, producing the first estimation of the target image from $(\mathbf{r}_{\mathrm{p}}^{0},\mathbf{x}_{\mathrm{p}}^{0})=(\mathbf{x}_{\mathrm{p}}^{\mathrm{d}},0)$. At any iteration $i\in\{1,2,...,\mathrm{I}\}$, $\mathcal{N}_{\mathrm{p}}^{\boldsymbol{\theta}^i}$'s output is a learned residual image, that captures remaining high-frequency (resolution) and faint-sources (dynamic-range) information. Thus, added to current estimate $\mathbf{x}^{i-1}_{\mathrm{p}}$, it refines our estimation of the target $\mathbf{x}_{\mathrm{p}}^{\star}$. More precisely, while ensuring the non-negativity of the next estimate $\mathbf{x}^{i}_{\mathrm{p}}$, $\mathcal{N}_{\mathrm{p}}^{\boldsymbol{\theta}^i}$ is trained on $\mathrm{L}^{i}$ data to minimise the difference between the current estimate $\mathbf{x}^{i-1}_{\mathrm{p}}$ and the target $\mathbf{x}_{\mathrm{p}}^{\star}$ using the residual information $\mathbf{r}_{\mathrm{p}}^{i-1}$  as: 
\begin{equation}
    \label{eqn:loss_function}
    \boldsymbol{\theta}^i= \underset{\boldsymbol{\theta}^i}{\arg\min}\left( \frac{1}{\mathrm{L}^{i}} \sum_{l=1}^{\mathrm{L}^{i}} \left\lVert \mathbf{x}_{\mathrm{p},l}^{\star} - \left[ \mathbf{x}_{\mathrm{p},l}^{i-1} +\mathcal{N}_{\mathrm{p}}^{\boldsymbol{\theta}^i} \left( \mathbf{r}_{\mathrm{p},l}^{i-1}, \mathbf{x}_{\mathrm{p},l}^{i-1} \right) \right]_+ \right\rVert_1 \right),
\end{equation}
where $l \in\{1,2,...,\mathrm{L}^{i}\}$ indexes individual images in the training dataset. Importantly, the Fourier nature of the measurement operator $\mathbf{\Phi}_{\mathrm{p}}$ enables a fast FFT-based implementation, and thus efficient generation of the datasets $(\mathbf{x}_{\mathrm{p},l}^{i-1},\mathbf{r}_{\mathrm{p},l}^{i-1})_{l=1}^{\mathrm{L}^{i}}$ at any iteration $i\in\{1,2,...,\mathrm{I}\}$.\\

\noindent
Furthermore, at each iteration $i\in\{1,2,...,\mathrm{I}\}$, a pruning procedure reduces the current dataset size $\mathrm{L}^{i}$ such that $\mathrm{L}^{i}=\alpha^{i-1} \mathrm{L}^{i-1}$ where $\alpha^{i-1} \in [0,1]$ represents the fraction of images removed. This pruning criterion removes the pair of data $(\mathbf{x}_{\mathrm{p},l}^{\star},(\mathbf{x}_{\mathrm{p},l}^{i},\mathbf{r}_{\mathrm{p},l}^{i}))$ from the training dataset if $\mathbf{r}_{\mathrm{p},l}^{i}$ reaches the image-domain noise level $\mathbf{\Phi}_{\mathrm{p}}^{\dagger}\mathbf{n}_{l}$ and satisfies:
\begin{equation}
    \label{eqn:pruning criterion}
    \|\mathbf{r}_{\mathrm{p},l}^{i}\|_{2} \leq \|\mathbf{\Phi}_{\mathrm{p}}^{\dagger}\mathbf{n}_{l}\|_2.  
\end{equation}
The removal of the pair $(\mathbf{x}_{\mathrm{p},l}^{\star},(\mathbf{x}_{\mathrm{p},l}^{i},\mathbf{r}_{\mathrm{p},l}^{i}))$ based on the pruning criterion \eqref{eqn:pruning criterion} is better understood in a simulation framework, where the visibilities $\mathbf{y}_{l}$ are generated from the ground truth $\mathbf{x}_{\mathrm{p},l}^{\star}$ using the equation \eqref{eqn:vis_discrete_2D_sphere} (with $\mathbf{\Gamma}=\mathbf{I}$), which reduces $\mathbf{r}_{\mathrm{p},l}^{i}$ to:
\begin{equation}
\label{eqn:simulation_residual}
    \mathbf{r}_{\mathrm{p},l}^{i} = \mathrm{\kappa} \textrm{Re}\left[\mathbf{\Phi}_{\mathrm{p}}^{\dagger}\mathbf{\Phi}_{\mathrm{p}}(\mathbf{x}_{\mathrm{p},l}^{\star}-\mathbf{x}_{\mathrm{p},l}^{i})+\mathbf{\Phi}_{\mathrm{p}}^{\dagger}\mathbf{n}_{l}\right].
\end{equation}
Then, as $\mathbf{\Phi}_{\mathrm{p}}$ is linear, if $\mathbf{r}_{\mathrm{p},l}^{i}$ satisfies the pruning criterion \eqref{eqn:pruning criterion} and is thus similar to noise, the current reconstruction $\mathbf{x}_{\mathrm{p},l}^{i}$ is sufficiently accurate, and the underlying inverse problem is solved for the pair $(\mathbf{x}_{\mathrm{p},l}^{\star},(\mathbf{x}_{\mathrm{p},l}^{i},\mathbf{r}_{\mathrm{p},l}^{i}))$, that is thus removed from the training dataset. While reducing the dataset size and thus accelerating the training stage, the pruning step improves $\mathcal{N}_{\mathrm{p}}^{\boldsymbol{\theta}^i}$'s training by eliminating solved inverse problems, enabling it to focus on the remaining relevant ones. In practice, we terminate the training stage and set the number of iterations $\mathrm{I}$ when the reconstruction quality, for a chosen metric, stabilises on the validation dataset.\\

\noindent
During the reconstruction stage, we aim to recover the unknown target signal $\mathbf{x}_{\mathrm{p}}^{\star}$ from the probed visibilities $\mathbf{y}$. Given the $\mathrm{I}$ trained DNNs $(\mathcal{N}_{\mathrm{p}}^{\boldsymbol{\theta}^i})_{i=1}^{\mathrm{I}}$, and starting with the initial estimation $\mathbf{x}_{\mathrm{p}}^{0}=0$ and residual $\mathbf{r}_{\mathrm{p}}^{0}=\mathbf{x}_{\mathrm{p}}^{\mathrm{d}}=\mathrm{\kappa} \textrm{Re}(\mathbf{\Phi}_{\mathrm{p}}^{\dagger}\mathbf{y})$, we gradually recover the dynamic range and the frequency content of the unknown $\mathbf{x}_{\mathrm{p}}^{\star}$ and build our final estimation $\mathbf{\hat{x}}_{\mathrm{p}} = \mathbf{x}_{\mathrm{p}}^{\mathrm{I}}$ using Algorithm~\ref{algo:R2D2_reconstruction}. Furthermore, we note that in the absence of a non-negativity assumption on $\mathbf{x}_{\mathrm{p}}^{\star}$, the non-negative constraints in \eqref{eqn:r2d2_iteration_i} and \eqref{eqn:loss_function} can be omitted. Then, the final reconstruction would take the straightforward series expression $\mathbf{\hat{x}}_{\mathrm{p}} = \mathbf{x}_{\mathrm{p}}^{\mathrm{I}}= \sum_{i=1}^{\mathrm{I}} {\mathcal{N}_{\mathrm{p}}^{\boldsymbol{\theta}^i}(\mathbf{r}_{\mathrm{p}}^{i-1},\mathbf{x}_{\mathrm{p}}^{i-1})}$, motivating the denomination of ``DNN series''.

\noindent
\begin{minipage}{0.48\textwidth}
\begin{algorithm}[H]
  \caption{: R2D2 Training} 
  \label{algo:R2D2_training}
  \begin{algorithmic}[1]
  \small
      \State \textbf{Input:} ground-truth images $(\mathbf{x}^{\star}_{l,p})_{l=1}^{\mathrm{L}^{1}}$, initial back-projected visibility-data residuals $(\mathbf{r}_{\mathrm{p},l}^{0}=\mathbf{x}_{\mathrm{p},l}^{\mathrm{d}})_{l=1}^{\mathrm{L}^{1}}$, initial estimations $(\mathbf{x}_{\mathrm{p},l}^{0}=0)_{l=1}^{\mathrm{L}^{1}}$, measurement operator $\mathbf{\Phi}_{\mathrm{p}}$ and $\mathrm{\kappa}$ > 0.
      \State \textbf{While} $i\leq \mathrm{I}$:
        \State \quad \textbf{Repeat}
          \Statex \quad  Take gradient descent step on:
          \Statex \quad \quad $\qquad \nabla_{\theta^{i}} \left( \frac{1}{\mathrm{L}^{i}} \sum\limits_{l=1}^{\mathrm{L}^i}\left\lVert \mathbf{x}_{\mathrm{p},l}^{\star} - \left[ \mathbf{x}_{\mathrm{p},l}^{i-1} +\mathcal{N}_{\mathrm{p}}^{\boldsymbol{\theta}^i} \left( \mathbf{r}_{\mathrm{p},l}^{i-1}, \mathbf{x}_{\mathrm{p},l}^{i-1} \right) \right]_+ \right\rVert_1 \right)
            $
        \Statex \quad \textbf{Until} $\mathcal{N}_{\mathrm{p}}^{\boldsymbol{\theta}^i}$ converged
      \State \quad  $\mathbf{x}_{\mathrm{p},l}^{i} = \mathbf{x}_{\mathrm{p},l}^{i-1} +\mathcal{N}_{\mathrm{p}}^{\boldsymbol{\theta}^i}(\mathbf{r}_{\mathrm{p},l}^{i-1}, \mathbf{x}_{\mathrm{p},l}^{i-1})$
      \State \quad  $\mathbf{x}_{\mathrm{p},l}^{i} =\left[ \mathbf{x}_{\mathrm{p},l}^{i}\right]_+$
      \State \quad $\mathbf{r}_{\mathrm{p},l}^{i} = \mathbf{x}_{\mathrm{p},l}^{\mathrm{d}} - \mathrm{\kappa} \textrm{Re}(\mathbf{\Phi}_{\mathrm{p}}^{\dagger}\mathbf{\Phi}_{\mathrm{p}}\mathbf{x}_{\mathrm{p},l}^{i})$
      \State \quad Pruning Step: $\mathrm{L}^{i+1}=\alpha^i \mathrm{L}^{i}, \ \alpha^i\in[0,1]$
    \State \textbf{End}
  \end{algorithmic}
\end{algorithm}
\end{minipage}

\noindent
\begin{minipage}{0.48\textwidth}
\begin{algorithm}[H]
  \caption{: R2D2 Reconstruction} 
  \label{algo:R2D2_reconstruction}
  \begin{algorithmic}[1]
  \small
      \State \textbf{Input:} trained DNNs $(\mathcal{N}_{\mathrm{p}}^{\boldsymbol{\theta}^i})_{i=1}^{\mathrm{I}}$, probed visibilities $\mathbf{y}$, initial back-projected visibility-data residual $\mathbf{r}_{\mathrm{p}}^{0}=\mathbf{x}_{\mathrm{p}}^{\mathrm{d}}=\mathrm{\kappa}\textrm{Re}(\mathbf{\Phi}_{\mathrm{p}}^{\dagger}\mathbf{y})$, initial estimation $\mathbf{x}_{\mathrm{p}}^{0} = 0$, measurement operator $\mathbf{\Phi}_{\mathrm{p}}$ and $\mathrm{\kappa}$ > 0.
      \State \textbf{While} $i\leq \mathrm{I}$:
      \State \quad  $\mathbf{x}_{\mathrm{p}}^{i} = \mathbf{x}_{\mathrm{p}}^{i-1} +\mathcal{N}_{\mathrm{p}}^{\boldsymbol{\theta}^i}(\mathbf{r}_{\mathrm{p}}^{i-1}, \mathbf{x}_{\mathrm{p}}^{i-1})$
      \State \quad  $\mathbf{x}_{\mathrm{p}}^{i} =\left[ \mathbf{x}_{\mathrm{p}}^{i}\right]_+$
      \State \quad $\mathbf{r}_{\mathrm{p}}^{i} = \mathbf{x}_{\mathrm{p}}^{\mathrm{d}} - \mathrm{\kappa} \textrm{Re}(\mathbf{\Phi}_{\mathrm{p}}^{\dagger}\mathbf{\Phi}_{\mathrm{p}}\mathbf{x}_{\mathrm{p}}^{i})$
    \State \textbf{End}
    \State \textbf{Return:} $\mathbf{\hat{x}}_{\mathrm{p}} = \mathbf{x}_{\mathrm{p}}^{\mathrm{I}}.$
  \end{algorithmic}
\end{algorithm}
\end{minipage}

\section{Proposed Methodology}\label{sec:section_3}
\noindent
In this section, we discuss the design of the interpolator $\mathbf{\Gamma}$, part of the wide-field model \eqref{eqn:vis_discrete_2D_sphere}, and present an efficient Fourier-based interpolator. Finally, we introduce S-R2D2, a spherical adaptation of R2D2 that aims to extend its performance to the wide-field framework. 

\subsection{Interpolation Challenges and Limitations}\label{subsec:sec_3_subsec_1}
\noindent
The purpose of the sphere-to-plane interpolator $\mathbf{\Gamma}$ is to grid spherical samples onto a uniform grid on the equatorial plane, while preserving the information of the underlying signal. Designing $\mathbf{\Gamma}$ necessitates to firstly determine the required resolution on the equatorial plane for a given resolution on the sphere, and secondly define the underlying interpolation function. For simplicity, in the remainder of the article, we use the notion of resolution and number of pixels interchangeably, as we fixed the FOV $\mathrm{\theta}_{\textrm{FOV}}$.\\

\noindent
Firstly, to ensure information is preserved, \citet{McEwen_Wiaux_WFOV} established a resolution rule that determines the minimum required number of pixels on the equatorial plane $\mathrm{N}_{\mathrm{p}}$ for a given number of pixels, within the FOV, on the sphere $\mathrm{N}_{\mathrm{s}}$: 
\begin{equation}
\label{eqn:resolution_rule}
\frac{\mathrm{N}_{\mathrm{p}}}{\mathrm{N}_{\mathrm{s}}} = \frac{\tan^2 \left(\frac{\mathrm{\theta}_{\textrm{FOV}}}{2}\right)}{2\left(1-\cos\left(\frac{\mathrm{\theta}_{\textrm{FOV}}}{2}\right)\right)}.
\end{equation}
The resolution rule follows from the requirement that the band-limit\footnotemark\footnotetext{A signal is band-limited on the plane (rsp. on the sphere) if represented by a finite number of coefficients in the Fourier (rsp. spherical harmonic) basis.} of the uniform planar grid, related to $\mathrm{N}_{\mathrm{p}}$ via the Nyquist-Shannon theorem, must be at least equal to the maximum projected frequency of a given band-limited\footnotemark \ spherical signal.\footnotetext{For a HEALPix pixelisation and a harmonic band-limit on the sphere $l_{\max}$, we can choose $\mathrm{N}_{\mathrm{s}}=(\sqrt{8}l_{\max}/\pi)^2\!\times\!(1-\cos(\mathrm{\theta}_{\textrm{FOV}}/2))$ \citep{hivon_2010}.} Moreover, this maximum is reached at the extents of the FOV, as illustrated in Figure~\ref{fig:HealPix_scheme}. The ratio \eqref{eqn:resolution_rule} increases rapidly with $\mathrm{\theta}_{\textrm{FOV}}$ as:
\begin{equation}
\label{eqn:resolution_rule_equivalent_at_infinity}
 \frac{\mathrm{N}_{\mathrm{p}}}{\mathrm{N}_{\mathrm{s}}}   
 \underset{\scriptstyle  \mathrm{\theta}_{\textrm{FOV}} \to \pi^{-}}{\sim} 
 \frac{1}{2}\tan^2\left(\frac{\mathrm{\theta}_{\textrm{FOV}}}{2}\right) 
 \underset{\scriptstyle  \mathrm{\theta}_{\textrm{FOV}} \to \pi^{-}}{\sim} 
 \frac{2}{(\pi-\mathrm{\theta}_{\textrm{FOV}})^2} 
 \underset{\scriptstyle  \mathrm{\theta}_{\textrm{FOV}} \to \pi^{-}}{\longrightarrow} +\infty,
\end{equation}
where we used that $\tan(\mathrm{x})=\cot(\pi/2-\mathrm{x})$ and $\cot(\mathrm{y}) \underset{\scriptstyle  \mathrm{y} \to 0}{\sim} 1/\mathrm{y}$. As a numerical example, for $\mathrm{\theta}_{\textrm{FOV}}=170^{\circ}, \ \mathrm{N}_{\mathrm{p}} \approx 70\!\times\!\mathrm{N}_{\mathrm{s}}$. Therefore, it implies that for very wide FOV $\mathrm{\theta}_{\textrm{FOV}}$ and a fine resolution on the sphere $\mathrm{N}_{\mathrm{s}}$, the necessary resolution on the plane $\mathrm{N}_{\mathrm{p}}$ becomes computationally impractical. Importantly, in practice, using the Nyquist-Shannon theorem, the number of pixels $\mathrm{N}_\mathrm{p}$ is limited by the bandwidth of the probed measurements $\mathbf{y}$. Then, as reciprocally indicated by equation \eqref{eqn:resolution_rule_equivalent_at_infinity}, for a large $\mathrm{\theta}_{\textrm{FOV}}$ and our limited value $\mathrm{N}_{\mathrm{p}}$ fixed by $\mathbf{y}$, following the resolution rule would lead to a poor target resolution on the sphere $\mathrm{N}_{\mathrm{s}}$, failing to meet the high-resolution requirement of wide-field imaging. Thus, we must bypass the resolution rule to target a sufficiently large resolution on the sphere, albeit at the cost of inevitable loss of information.\\ 

\noindent
Secondly, as discussed in Section~\ref{subsec:sec_2_subsec_1}, $\mathbf{\Gamma}$ is equivalently a nonuniform-to-uniform interpolator on the equatorial plane as the spherical samples are nonuniform samples on the equatorial plane. Importantly, preserving information while interpolating nonuniform samples onto a uniform grid on the plane requires solving a Fourier weight inverse problem \citep{STROHMER2000297}. Indeed, the underlying interpolation function is obtained as a weighted sum of shifted sinc functions, with the weights being the solutions of this weight inverse problem. However, solving it in our large-scale wide-field context is impractical. Consequently, we must use a simplified interpolation function, inevitably leading to loss of information.

\subsection{Proposed Interpolation Procedure} \label{subsec:sec_3_subsec_2}
\noindent
Firstly, despite the inevitable approximations in the design of the underlying interpolation functions of $\mathbf{\Gamma}$, and thus of $\mathbf{\Gamma}^{\dagger}$, we can implement an efficient approximation of the uniform-to-nonuniform interpolator $\mathbf{\Gamma}^{\dagger}$. Indeed, we can interpolate on the plane, $\mathrm{N}_{\mathrm{p}}$ uniform samples onto $\mathrm{N}_{\mathrm{s}}$ nonuniform samples, with a weighted sum of shifted sinc functions where the weights are the sample values. Importantly, this can be simply implemented as a FFT followed by the type-2 nonuniform FFT. Moreover, this approximation of $\mathbf{\Gamma}^{\dagger}$ ensures information is preserved while interpolating for signals assumed to be simultaneously band-limited on the sphere and on the plane, with resolutions $\mathrm{N}_{\mathrm{p}}$ and $\mathrm{N}_{\mathrm{s}}$ that satisfy the resolution rule \eqref{eqn:resolution_rule}, motivating our chosen approximation. Then, we approximate $\mathbf{\Gamma}$ as the corresponding adjoint, which can be simply implemented as the type-1 nonuniform FFT (which is the adjoint of the type-2 nonuniform FFT) followed by an inverse FFT. For simplicity, in the remainder of the article, $\mathbf{\Gamma}$ and $\mathbf{\Gamma}^{\dagger}$ refer to their Fourier-based approximations. Furthermore, such Fourier-based implementations of $\mathbf{\Gamma}$ and $\mathbf{\Gamma}^{\dagger}$ are computationally efficient and support automatic differentiation, enabling their integration into dataset generation and DNN training loss computation.\\

\begin{figure}
\centering
\adjustbox{valign=t}{\begin{subfigure}[b]{0.157\textwidth}
\includegraphics[width=\textwidth]{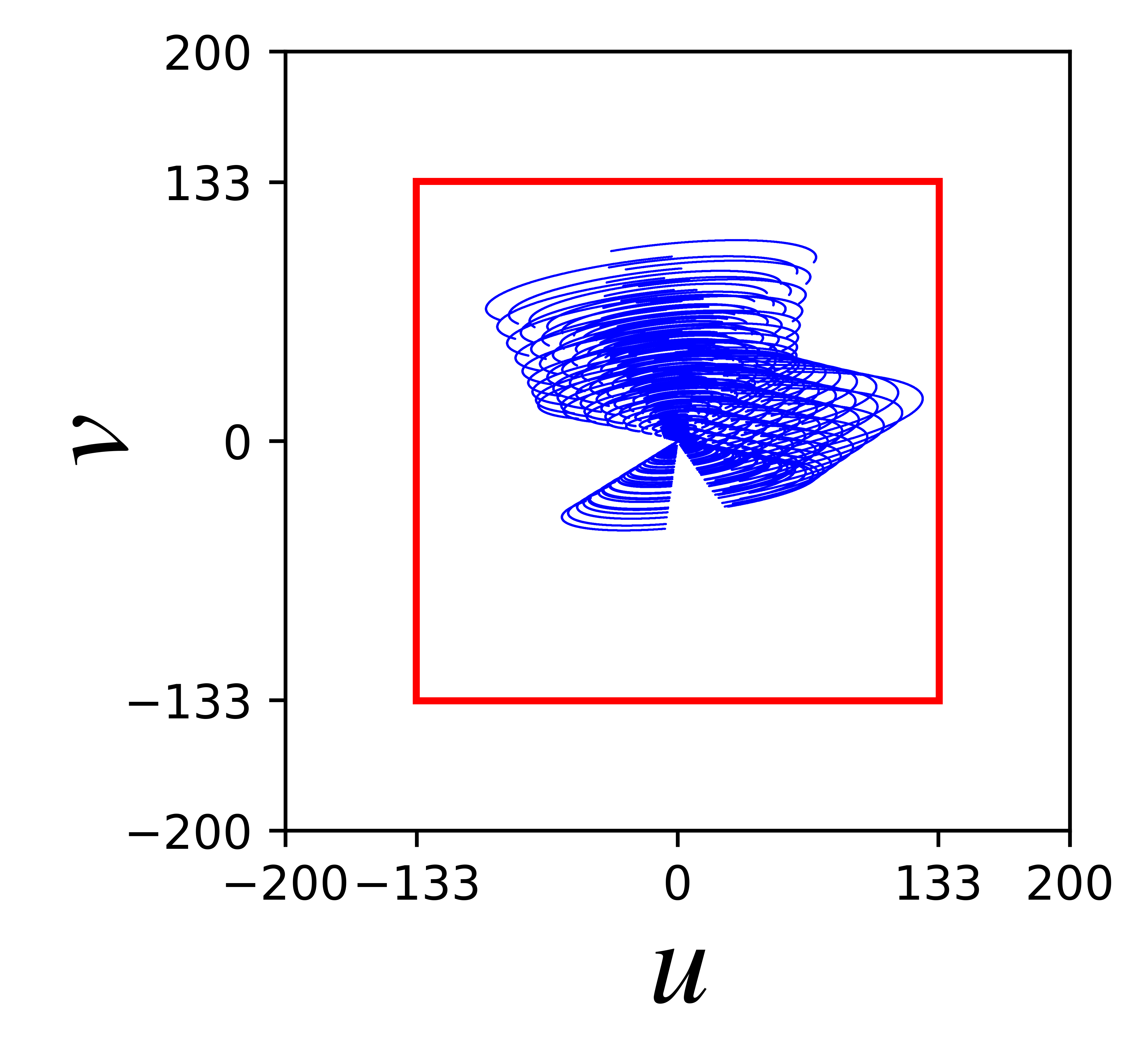}
\vspace{-2.25em} 
\caption{}
\end{subfigure}}
\adjustbox{valign=t}{\begin{subfigure}[b]{0.157\textwidth}
\includegraphics[width=\textwidth]{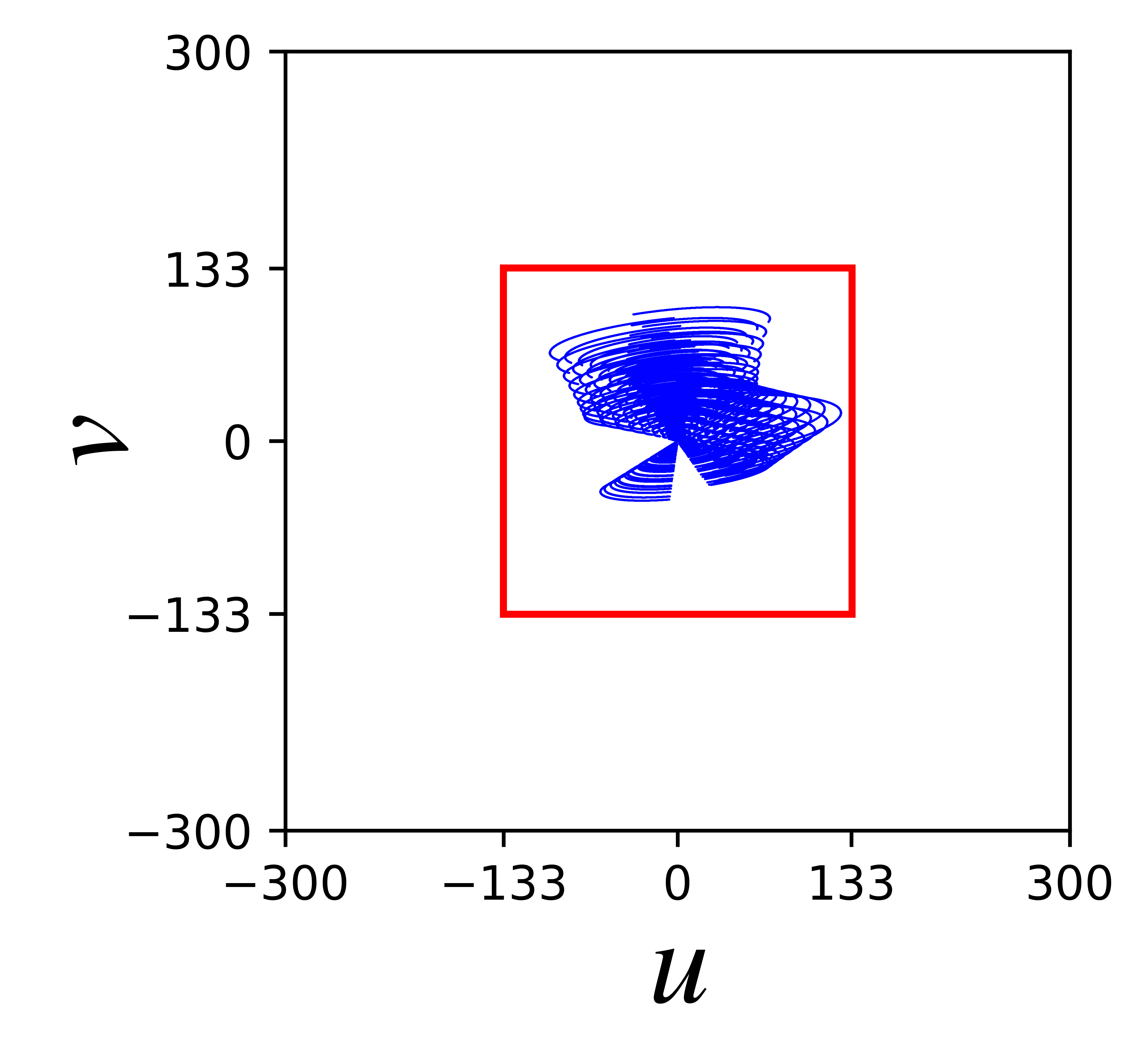}
\vspace{-2.25em}
\caption{}
\end{subfigure}}
\adjustbox{valign=t}{\begin{subfigure}[b]{0.157\textwidth}
\includegraphics[width=\textwidth]{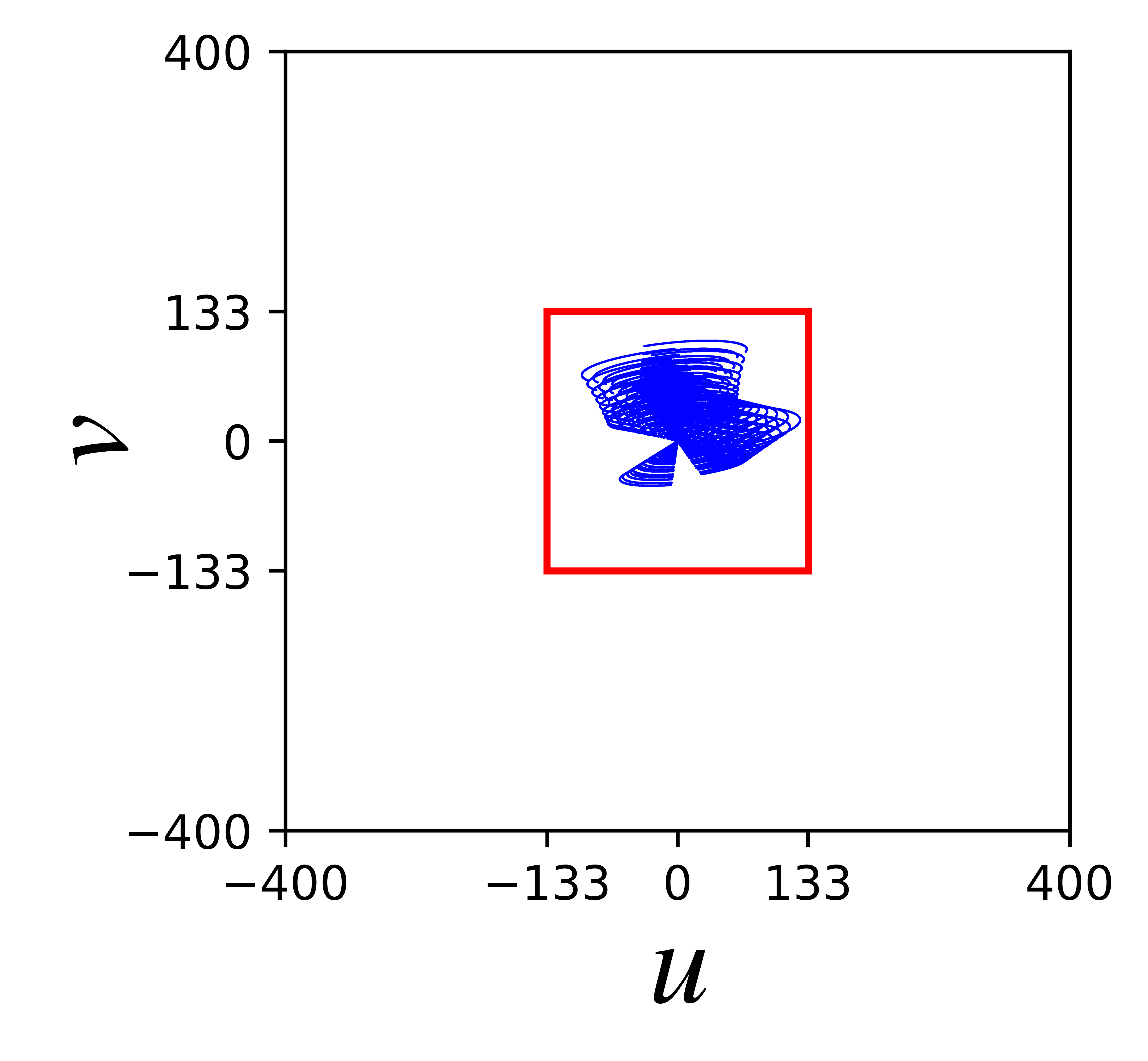}
\vspace{-2.25em}
\caption{}
\end{subfigure}}
\caption{Illustration of the effect of the application of the super-resolution equation \eqref{eqn:SR_rule} on the visibility domain using three different $\textrm{SR}$ levels: $\textrm{SR}=1.5$ (panel (a)), $\textrm{SR}=2.25$ (panel (b)) and $\textrm{SR}=3$ (panel (c)), which respectively represent visibility domains with $\mathrm{N}_{\mathrm{p}}=400^2$, $\mathrm{N}_{\mathrm{p}}=600^2$ and $\mathrm{N}_{\mathrm{p}}=800^2$ pixels. The sampling pattern is framed and preserved across all $\textrm{SR}$ levels.}
\label{fig:effect_of_SR}
\end{figure}

\noindent
Secondly, to mitigate the suboptimal configuration arising from the necessity to bypass the resolution rule \eqref{eqn:resolution_rule}, we can boost the resolution $\mathrm{N}_\mathrm{p}$ fixed by $\mathbf{y}$ up to a chosen super-resolution ($\textrm{SR}$) factor, such as:
\begin{equation}
    \label{eqn:SR_rule}
    \mathrm{N}_{\mathrm{p}}=\mathrm{N}_{\mathrm{p}}^{\min}\!\times\! \left( \frac{\textrm{SR}}{\textrm{SR}_{\mathrm{0}}} \right)^{2},
\end{equation}
where $\mathrm{N}_{\mathrm{p}}^{\min}/\textrm{SR}_{\mathrm{0}}^2$ corresponds to the value $\mathrm{N}_{\mathrm{p}}$ fixed by $\mathbf{y}$, with $\textrm{SR}_{\mathrm{0}}$ being the baseline factor for $\textrm{SR}$ enhancement and $\mathrm{N}_{\mathrm{p}}^{\min}$ being the lowest desirable resolution on the plane which corresponds to maintaining the same pixel size on the sphere and the plane. Then, the spatial bandwidth of the probed sampling pattern is equal to $\sqrt{\mathrm{N}_{\mathrm{p}}^{\min}}/\textrm{SR}_{\mathrm{0}}$. As illustrated in Figure~\ref{fig:effect_of_SR}, boosting $\mathrm{N}_\mathrm{p}$ while following \eqref{eqn:SR_rule} ensures consistency as it enables us to map different interpolated signals $\mathbf{\Gamma}\mathbf{x}_{\mathrm{s}}^{\star}$, generated from the same spherical signal $\mathbf{x}_{\mathrm{s}}^{\star}$ using different resolutions $\mathrm{N}_\mathrm{p}$, onto the same visibilities $\mathbf{y}$ by applying the measurement model \eqref{eqn:vis_discrete_2D_sphere}.\\ 

\begin{figure*}
 \centering
 \setlength\tabcolsep{4pt}
 \begin{tabular}{cccc}

\includegraphics[width=0.265\linewidth, trim=0 45 0 0]{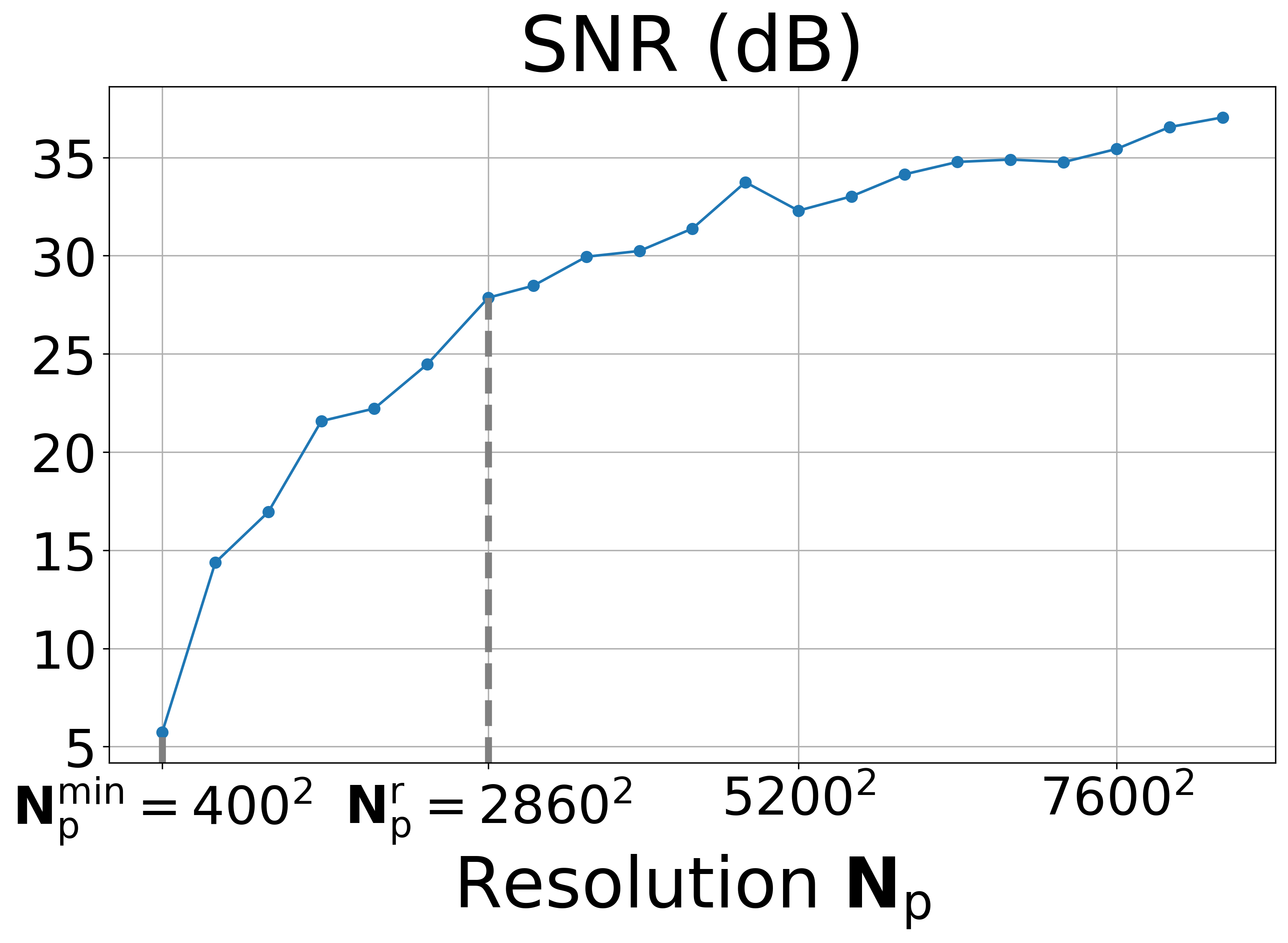} &
\hspace{1.1cm}\includegraphics[width=0.185\linewidth]{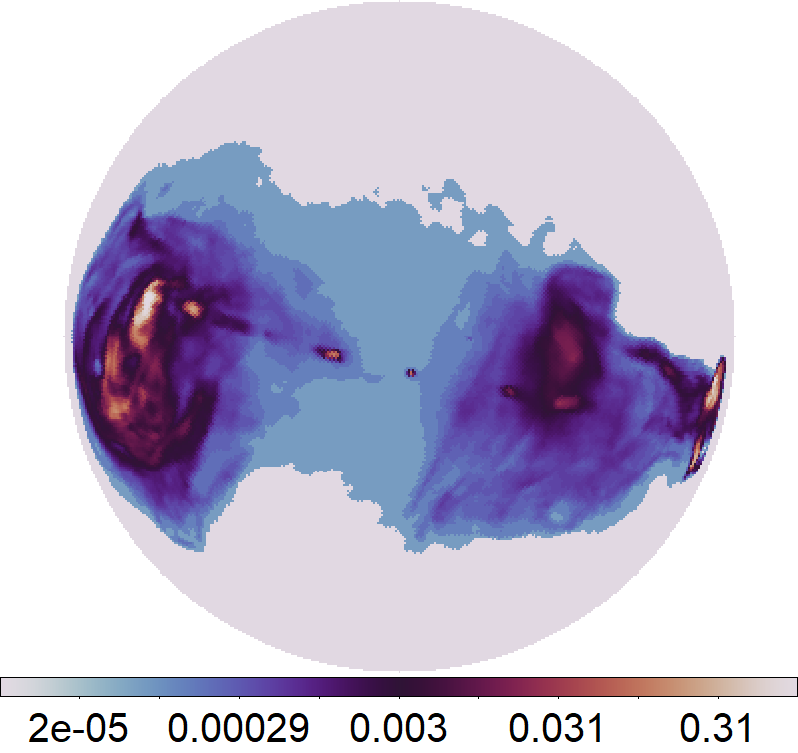} & \hspace{0.3cm}
\includegraphics[width=0.185\linewidth]{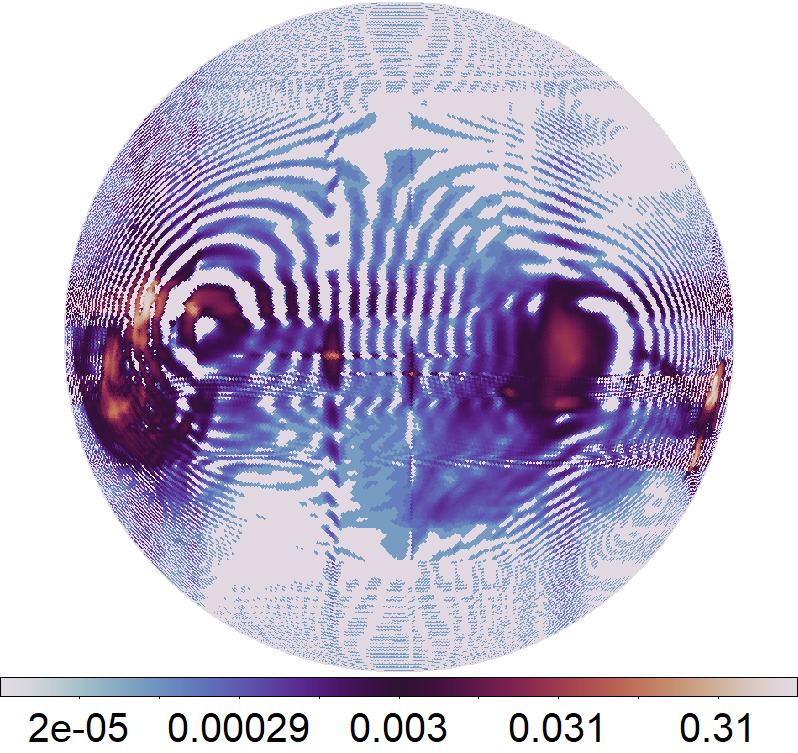} & \hspace{0.3cm}
\includegraphics[width=0.185\linewidth]{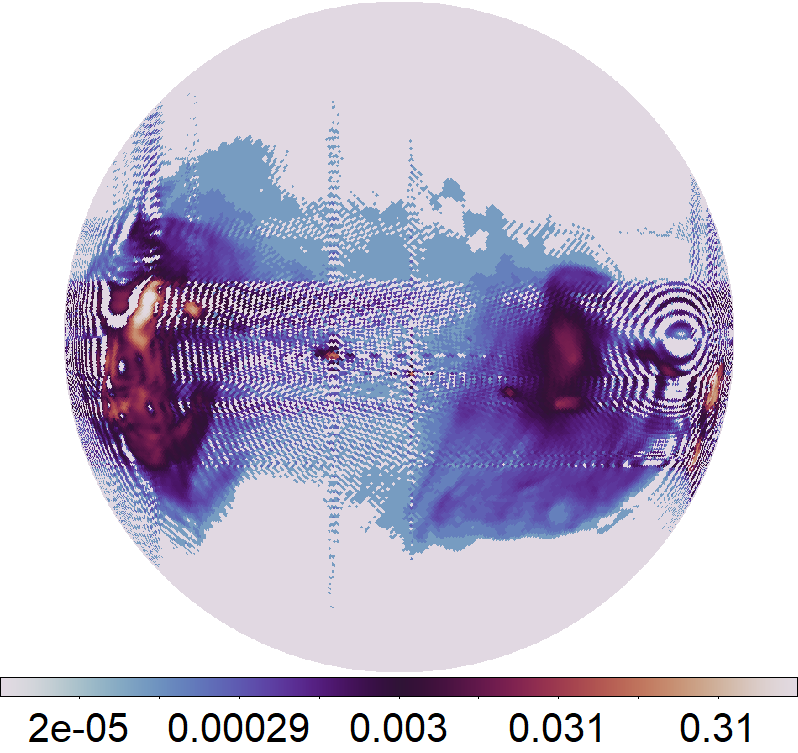} \\

&\hspace{1.1cm} \scriptsize Ground Truth $\mathbf{x}_{\mathrm{s}}^{\star}$ & \hspace{0.3cm} \scriptsize $\mathbf{\Gamma}^{\dagger}\mathbf{\Gamma}\mathbf{x}_{\mathrm{s}}^{\star}$; $\mathrm{N}_{\mathrm{p}}=400^2$; (8.6, 3.5)~dB & \hspace{0.3cm} \scriptsize $\mathbf{\Gamma}^{\dagger}\mathbf{\Gamma}\mathbf{x}_{\mathrm{s}}^{\star}$; $\mathrm{N}_{\mathrm{p}}=600^2$; (8.7, 5.9)~dB \\
\addlinespace[3pt]

\includegraphics[width=0.265\linewidth, trim=0 45 0 0]{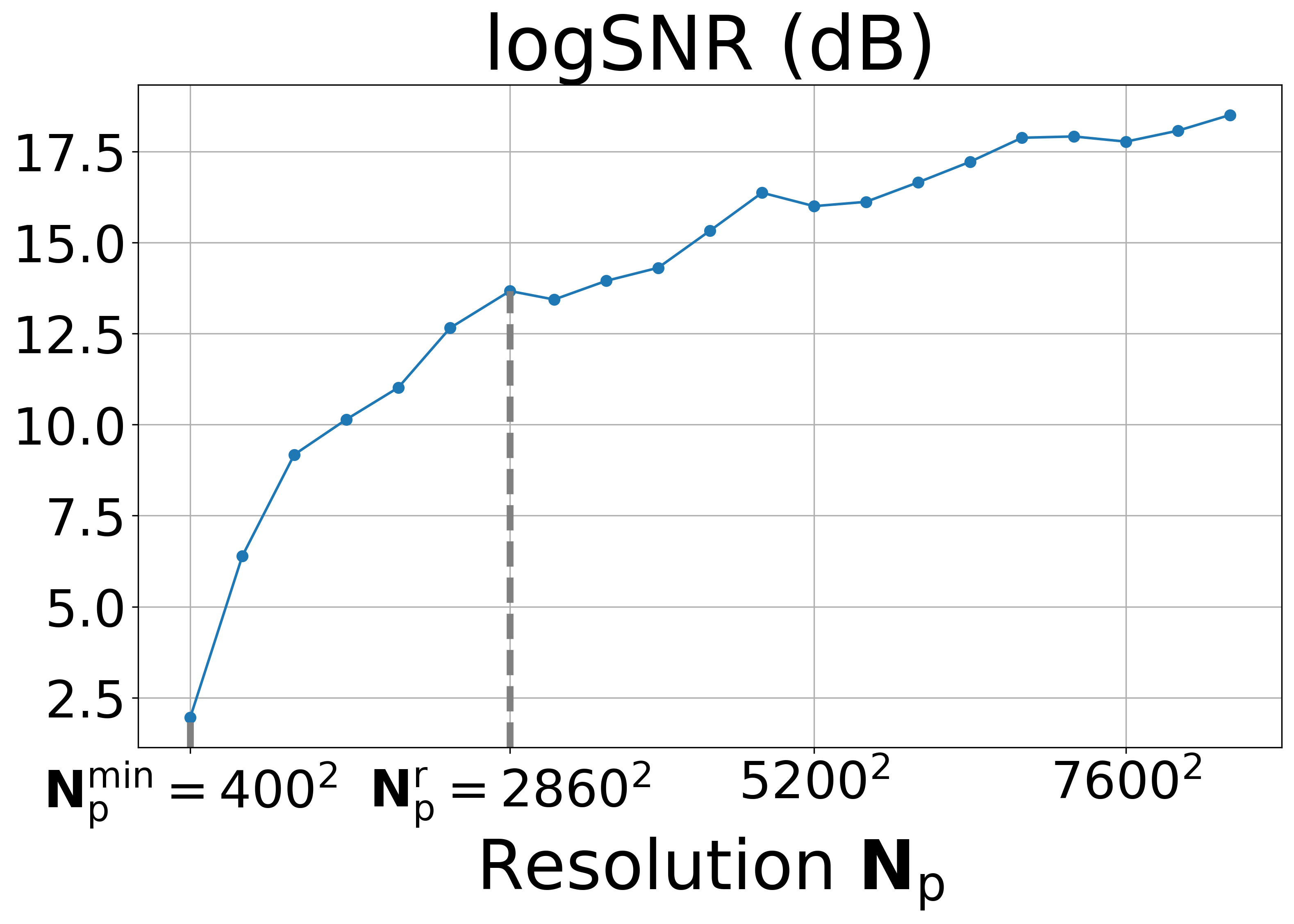} &
 \hspace{1.1cm} \includegraphics[width=0.185\linewidth]{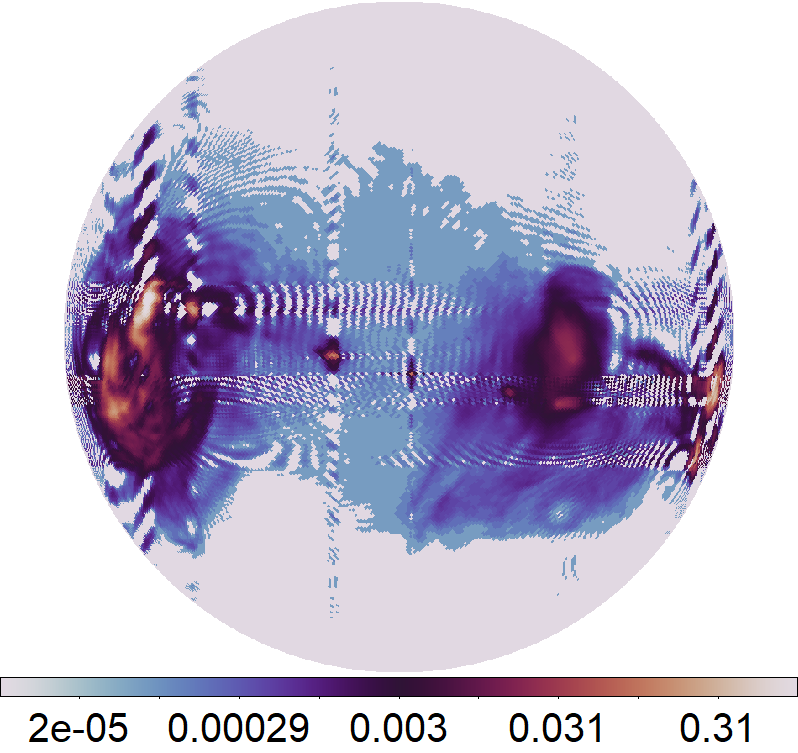} & \hspace{0.2cm}
\includegraphics[width=0.185\linewidth]{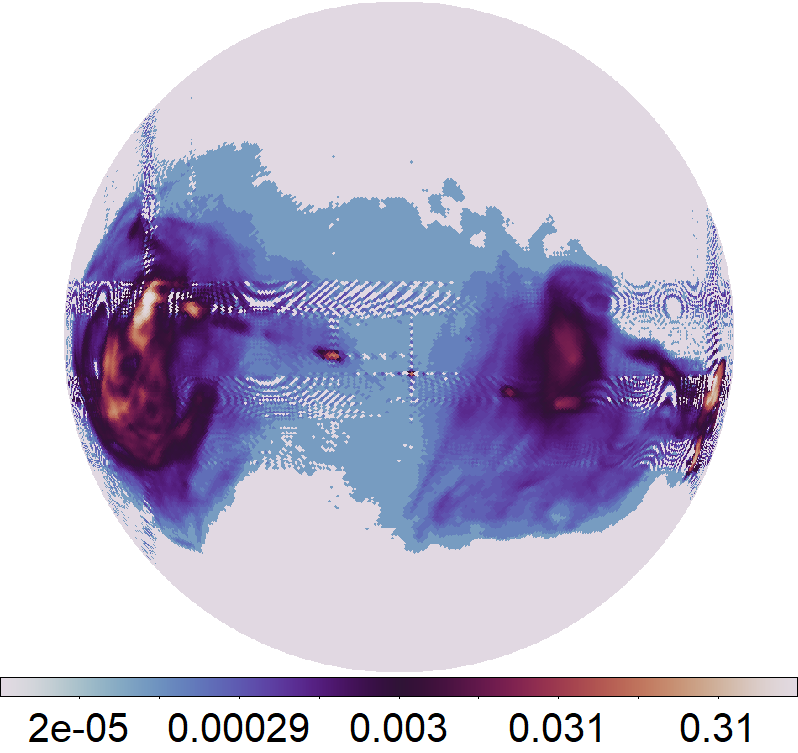} & \hspace{0.2cm}
\includegraphics[width=0.185\linewidth]{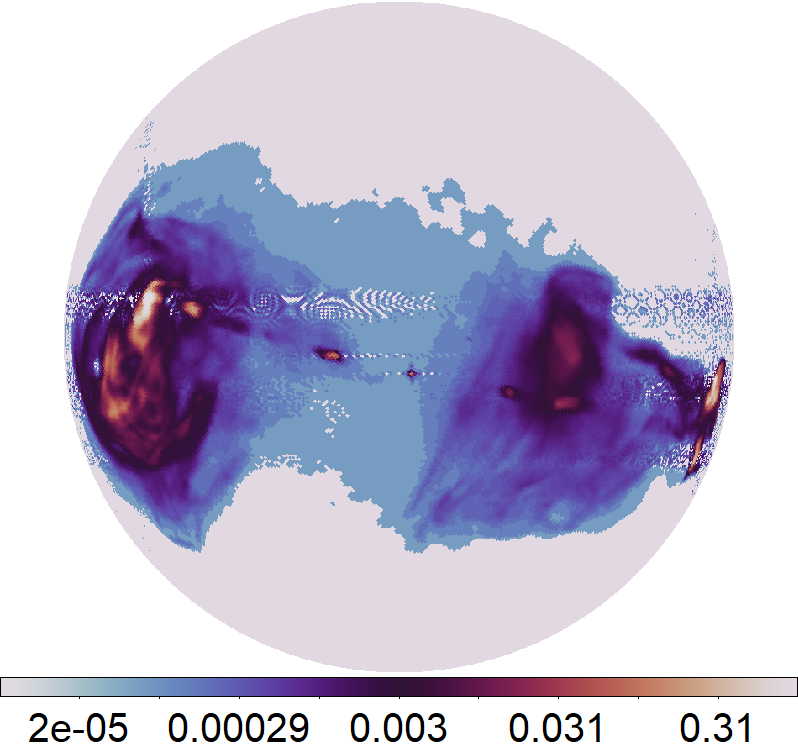} \\

&\hspace{1.1cm} \scriptsize $\mathbf{\Gamma}^{\dagger}\mathbf{\Gamma}\mathbf{x}_{\mathrm{s}}^{\star}$; $\mathrm{N}_{\mathrm{p}}=800^2$; (17.5, 6.9)~dB & \hspace{0.2cm} \scriptsize $\mathbf{\Gamma}^{\dagger}\mathbf{\Gamma}\mathbf{x}_{\mathrm{s}}^{\star}$; $\mathrm{N}_{\mathrm{p}}=2860^2$; (25.2, 11.1)~dB & \hspace{0.2cm} \scriptsize $\mathbf{\Gamma}^{\dagger}\mathbf{\Gamma}\mathbf{x}_{\mathrm{s}}^{\star}$; $\mathrm{N}_{\mathrm{p}}=8400^2$; (37.6, 15.1)~dB \\
 \end{tabular}
 \caption{Illustration of the effect of the operator $\mathbf{\Gamma}^{\dagger}\mathbf{\Gamma}$ on ground-truth signals $\mathbf{x}_{\mathrm{s}}^{\star}$. The two metric plots, on the left, show respectively the evolution of $\textrm{SNR}$ (top) and $\textrm{logSNR}$ (bottom) metrics (defined in Section~\ref{subsec:sec_5_subsec_2}) between $\mathbf{x}_{\mathrm{s}}^{\star}$ and $\mathbf{\Gamma}^{\dagger}\mathbf{\Gamma}\mathbf{x}_{\mathrm{s}}^{\star}$, as a function of the resolution on the plane $\mathrm{N}_{\mathrm{p}}$. The values represent the averages computed over the 60 signals part of the test dataset, defined in Section~\ref{subsec:sec_4_subsec_3}. $\mathrm{N}_{\mathrm{s}}$ is fixed for all images and its choice is detailed in Section~\ref{subsec:sec_5_subsec_1}. Lower values of $\mathrm{N}_{\mathrm{p}}$ result in reduced interpolation accuracy with critical aliasing artifacts. The signals $\mathbf{x}_{\mathrm{s}}^{\star}$ do not satisfy any band-limit assumption. Therefore, we do not observe a saturation in metric values beyond the resolution-rule value, denoted $\mathrm{N}_{\mathrm{p}}^{\mathrm{r}}$. On the right, we visualise a specific ground truth $\mathbf{x}_{\mathrm{s}}^{\star}$ alongside the corresponding $\mathbf{\Gamma}^{\dagger}\mathbf{\Gamma}\mathbf{x}_{\mathrm{s}}^{\star}$ for different $\mathrm{N}_{\mathrm{p}}$. The ground truth was generated with a dynamic range $\textrm{DR} = 1.1\!\times\!10^5$ from the radio galaxy 3c354 (NRAO Archives), following the procedure described in Section~\ref{subsec:sec_4_subsec_1}. Values of ($\textrm{SNR}$, $\textrm{logSNR}$) are reported below each panel. We visualise the Northern hemisphere in the orthographic projection perspective and in logarithmic scale (with the logarithmic exponent being equal to $\textrm{DR}$), clipping the negative values for visualisation purposes.}  
 \label{fig:gamma_gamma_diaga}
\end{figure*}

\noindent
Figure~\ref{fig:gamma_gamma_diaga} illustrates the accuracy of $\mathbf{\Gamma}$ and $\mathbf{\Gamma}^{\dagger}$, by examining the effect of $\mathbf{\Gamma}^{\dagger}\mathbf{\Gamma}$ on spherical signals as a function of $\mathrm{N}_{\mathrm{p}}$ and for a fixed value $\mathrm{N}_{\mathrm{s}}$. The optimal scenario corresponds to $\mathbf{\Gamma}^{\dagger}\mathbf{\Gamma}=\mathrm{I}$. Indeed, regularisation-based reconstruction methods, such as R2D2, integrate the operator $\mathbf{\Phi}_{\mathrm{s}}^{\dagger}\mathbf{\Phi}_{\mathrm{s}}=\mathbf{\Gamma}^{\dagger}\mathbf{\Phi}_{\mathrm{p}}^{\dagger}\mathbf{\Phi}_{\mathrm{p}}\mathbf{\Gamma}$ in their scheme, and thus $\mathbf{\Gamma}^{\dagger}\mathbf{\Gamma}=\mathrm{I}$ ensures that transitions between the equatorial plane and the sphere occur without loss of information during the reconstruction process. However, insufficient values of $\mathrm{N}_{\mathrm{p}}$ result in suboptimal interpolators, which can significantly degrade the reconstruction performance. Therefore, achieving this optimal scenario requires large values of $\mathrm{N}_{\mathrm{p}}$, which necessitates, in our RI context, to super-resolve the baseline resolution $\mathrm{N}_{\mathrm{p}}=\mathrm{N}_{\mathrm{p}}^{\min}$ with a significantly large $\textrm{SR}$ factor. However, any reconstruction algorithm has a limited $\textrm{SR}$ capability. Indeed, as $\textrm{SR}$ increases, the reconstruction task becomes more challenging, since it consists of recovering the frequency content of the target signal at a higher resolution $\mathrm{N}_{\mathrm{p}}$. Therefore, this introduces a critical precision-efficiency trade-off, where increasing $\mathrm{N}_{\mathrm{p}}$ enhances accuracy but also increases reconstruction complexity, requiring careful balancing within the algorithm. Additionally, these inherent interpolation limitations suggests incorporating specific corrective mechanisms into the scheme of the employed algorithm to mitigate them.\\

\noindent
Furthermore, we note that in a real data context, any inexactitude in the measurement operator, such as interpolation approximations, creates additional limitations arising from modelling discrepancy between the real probed visibilities and the unknown target. In contrast, in simulation, the modelling imperfections are not seen by the simulated RI visibilities, generated with the approximate measurement operator. Consequently, the challenges associated with real data introduce additional complexities that are not present in simulations and remain beyond the scope of this article.

\subsection{S-R2D2 Algorithm}\label{subsec:sec_3_subsec_3}
\noindent
Our goal is to leverage the efficient implementation of the wide-field measurement operator $\mathbf{\Phi}_{\mathrm{s}}=\mathbf{\Phi}_{\mathrm{p}}\mathbf{\Gamma}$ and adapt R2D2 to integrate it in order to jointly account for interpolation approximations and solve the wide-field inverse problem \eqref{eqn:vis_discrete_2D_sphere} on the sphere.\\

\noindent
A first natural approach consists of using R2D2's scheme \eqref{eqn:r2d2_iteration_i} and plugging in $\mathbf{\Phi}_{\mathrm{s}}$ as follows:
\begin{align}
\left\{
\begin{aligned}
    \label{eqn:spherical_r2d2}
    \mathbf{x}_{\mathrm{s}}^{i} &=\left[\mathbf{x}_{\mathrm{s}}^{i-1} +\mathcal{N}_{\mathrm{s}}^{\boldsymbol{\theta}^i}(\mathbf{r}_{\mathrm{s}}^{i-1},\mathbf{x}_{\mathrm{s}}^{i-1})\right]_+ \\
    \mathbf{r}_{\mathrm{s}}^{i} &= \mathbf{x}_s^{\mathrm{d}} - \mathrm{\kappa} \textrm{Re}(\mathbf{\Phi}_{\mathrm{s}}^{\dagger}\mathbf{\Phi}_{\mathrm{s}}\mathbf{x}_{\mathrm{s}}^{i}).
\end{aligned}
\right.
\end{align}
However, here the DNNs $\mathcal{N}_{\mathrm{s}}^{\boldsymbol{\theta}^i}$ necessarily process spherical signals as input $(\mathbf{r}_{\mathrm{s}}^{i-1},\mathbf{x}_{\mathrm{s}}^{i-1})$ and output a spherical signal $\mathbf{x}_{\mathrm{s}}^{i}$. Thus, while preserving R2D2's scheme, $\mathcal{N}_{\mathrm{s}}^{\boldsymbol{\theta}^i}$'s architecture must be suited to the spherical setting and thus, cannot be a 2D-Euclidean U-Net network as in R2D2. Such spherical convolutional networks have been developed to process deep learning operations directly on the sphere and on a HEALPix grid. However, in the context of our inverse problem, preliminary experiments with the architecture developed by \citet{Perraudin_2019} revealed significant challenges in speed, accuracy, and hyperparameter fine-tuning.\\

\noindent
Therefore, we considered a different approach and instead leveraged the 2D-Euclidean U-Net architecture $\mathcal{N}_{\mathrm{p}}^{\boldsymbol{\theta}^i}$ used in R2D2, which has demonstrated its efficiency in a RI context. Then, we propose the S-R2D2 extension, detailed in Algorithm~\ref{algo:S_R2D2_training}, to adapt R2D2's scheme \eqref{eqn:r2d2_iteration_i} to jointly use 2D-Euclidean DNNs and ensure accurate reconstructions on the sphere. More precisely, S-R2D2 constructs the residual $\mathbf{r}_{\mathrm{p}}^{i}$ and reconstruction $\mathbf{x}_{\mathrm{p}}^{i}$ updates (inputs of $\mathcal{N}_{\mathrm{p}}^{\boldsymbol{\theta}^i}$) on the plane, using the truncated part $\mathbf{\Phi}_{\mathrm{p}}^{\dagger}\mathbf{\Phi}_{\mathrm{p}}\mathbf{\Gamma}$ of the operator $\mathbf{\Phi}_{\mathrm{s}}^{\dagger}\mathbf{\Phi}_{\mathrm{s}}=\mathbf{\Gamma}^{\dagger}\mathbf{\Phi}_{\mathrm{p}}^{\dagger}\mathbf{\Phi}_{\mathrm{p}}\mathbf{\Gamma}$, while incorporating the remaining plane-to-sphere interpolator $\mathbf{\Gamma}^{\dagger}$ in an additional step to ensure consistency with the spherical target as follows:
\begin{align}
\left\{
\begin{aligned}
    \label{eqn:s_r2d2}
    \mathbf{x}_{\mathrm{p}}^{i} &= \mathbf{x}_{\mathrm{p}}^{i-1} +\mathcal{N}_{\mathrm{p}}^{\boldsymbol{\theta}^i}(\mathbf{r}_{\mathrm{p}}^{i-1}, \mathbf{x}_{\mathrm{p}}^{i-1})\\
    \mathbf{x}_{\mathrm{s}}^{i} &=\left[ \mathbf{\Gamma}^{\dagger}\mathbf{x}_{\mathrm{p}}^{i}\right]_+ =\left[ \mathbf{x}_{\mathrm{s}}^{i-1} +\mathbf{\Gamma}^{\dagger}\mathcal{N}_{\mathrm{p}}^{\boldsymbol{\theta}^i}(\mathbf{r}_{\mathrm{p}}^{i-1}, \mathbf{x}_{\mathrm{p}}^{i-1})\right]_+ \\
    \mathbf{r}_{\mathrm{p}}^{i} &= \mathbf{x}_{\mathrm{p}}^{\mathrm{d}} - \mathrm{\kappa} \textrm{Re}(\mathbf{\Phi}_{\mathrm{p}}^{\dagger}\mathbf{\Phi}_{\mathrm{p}}\mathbf{\Gamma}\mathbf{x}_{\mathrm{s}}^{i}),
\end{aligned}
\right.
\end{align}
where $\mathbf{r}_{\mathrm{p}}^{0}=\mathbf{x}_{\mathrm{p}}^{\mathrm{d}}=\mathrm{\kappa} \mathbf{\Phi}_{\mathrm{p}}^{\dagger}\mathbf{y}$ with $\mathbf{y}=\mathbf{\Phi}_{\mathrm{p}} \mathbf{\Gamma}\mathbf{x}_{\mathrm{s}}^{\star}+\mathbf{n}$, and $\mathbf{x}_{\mathrm{p}}^{0}=0$. The constant $\mathrm{\kappa}$ was previously defined in Section~\ref{subsec:sec_2_subsec_2}. Moreover, we define the visibility-data residual back-projected onto the sphere as $\mathbf{r}_{\mathrm{s}}^{i} =\mathbf{\Gamma}^{\dagger}\mathbf{r}_{\mathrm{p}}^{i}$.
At any iteration $i\in\{1,2,...,\mathrm{I}\}$, $\mathcal{N}_{\mathrm{p}}^{\boldsymbol{\theta}^i}$ is trained on $\mathrm{L}^{i}$ data to minimise the difference between the current spherical estimate $\mathbf{x}^{i-1}_{\mathrm{s}}$ and the real, non-negative target $\mathbf{x}_{\mathrm{s}} ^{\star}$ (assumed to be a monochromatic intensity image) using the residual information $\mathbf{r}_{\mathrm{p}}^{i-1}$, while ensuring the non-negativity of the next spherical estimate $\mathbf{x}^{i}_{\mathrm{s}}$, as follows:
\begin{equation}
\label{eqn:training_procedure_S_R2D2}
    \boldsymbol{\theta}^i= \underset{\boldsymbol{\theta}^i}{\arg\min}  \left( \frac{1}{\mathrm{L}^{i}} \sum_{l=1}^{\mathrm{L}^{i}} \left\lVert \mathbf{x}_{\mathrm{s},l}^{\star} - \left[\mathbf{\Gamma}^{\dagger} \left( \mathbf{x}_{\mathrm{p},l}^{i-1} +\mathcal{N}_{\mathrm{p}}^{\boldsymbol{\theta}^i} \left( \mathbf{r}_{\mathrm{p},l}^{i-1}, \mathbf{x}_{\mathrm{p},l}^{i-1} \right) \right) \right]_+ \right\rVert_1 \right),
\end{equation}
\noindent
where $l \in\{1,2,...,\mathrm{L}^{i}\}$ indexes individual images in the training dataset. Since $\mathbf{\Gamma}^{\dagger}$ is integrated as an additional step, alternative plane-to-sphere interpolators could have been considered. However, incorporating the Fourier-based interpolator $\mathbf{\Gamma}^{\dagger}$ into the training loss enables the DNNs to jointly learn to iteratively reconstruct $\mathbf{x}_{\mathrm{s}}^{\star}$ on the sphere and correct the interpolation approximations discussed in Section~\ref{subsec:sec_3_subsec_2}. Moreover, this approach aligns with standard reconstruction theory, where using a forward operator and its adjoint is essential to achieve accurate and stable reconstructions.\\

\noindent
Then, at each iteration $i\in\{1,2,...,\mathrm{I}\}$, the size of S-R2D2's training dataset $\mathrm{L}^{i}$ can be reduced using the same pruning procedure applied in R2D2. Indeed, in simulation, we generate the RI visibilities and the corresponding dirty images $\mathbf{x}_{\mathrm{p},l}^{\mathrm{d}}$ through the measurement equation \eqref{eqn:vis_discrete_2D_sphere}, and thus $\mathbf{r}_{\mathrm{p},l}^{i}$ reduces to:
\begin{equation}
    \label{eqn:s-r2d2_residual}
    \mathbf{r}_{\mathrm{p},l}^{i} =\mathrm{\kappa} \textrm{Re}\left[\mathbf{\Phi}_{\mathrm{p}}^{\dagger}\mathbf{\Phi}_{\mathrm{p}} \mathbf{\Gamma}(\mathbf{x}_{\mathrm{s},l}^{\star}-\mathbf{x}_{\mathrm{s},l}^{i})+ \mathbf{\Phi}_{\mathrm{p}}^{\dagger}\mathbf{n}_l\right].
\end{equation}
Then, as in Section~\ref{subsec:sec_2_subsec_2}, if $\mathbf{r}_{\mathrm{p},l}^{i}$ satisfies the pruning criterion \eqref{eqn:pruning criterion}, then $(\mathbf{x}_{\mathrm{s},l}^{i},(\mathbf{x}_{\mathrm{p},l}^{i},\mathbf{r}_{\mathrm{p},l}^{i}))$ is removed from the training dataset.\\

\noindent
During the reconstruction stage, we aim to recover the unknown spherical signal $\mathbf{x}_{\mathrm{s}}^{\star}$ from the measurements $\mathbf{y}$. Given the $\mathrm{I}$ trained DNNs $(\mathcal{N}_{\mathrm{p}}^{\boldsymbol{\theta}^i})_{i=1}^{\mathrm{I}}$ and starting with the initial estimation $\mathbf{x}_{\mathrm{p}}^{0}=0$ and residual $\mathbf{r}_{\mathrm{p}}^{0}=\mathbf{x}_{\mathrm{p}}^{\mathrm{d}}=\mathrm{\kappa} \textrm{Re}(\mathbf{\Phi}_{\mathrm{p}}^{\dagger}\mathbf{y})$, we build our final estimation $\mathbf{\hat{x}}_{\mathrm{s}} = \mathbf{x}_{\mathrm{s}}^{\mathrm{I}}$ using Algorithm~\ref{algo:S_R2D2_reconstruction}.

\noindent
\begin{minipage}{0.48\textwidth}

\begin{algorithm}[H]
  \caption{: S-R2D2 Training} 
  \label{algo:S_R2D2_training}
  \begin{algorithmic}[1]
  
  \small
      \State \textbf{Input:} spherical ground-truth signals $(\mathbf{x}_{\mathrm{s},l}^{\star})_{l=1}^{\mathrm{L}^{1}}$, initial back-projected visibility-data residuals $(\mathbf{r}_{\mathrm{p},l}^{0}=\mathbf{x}_{\mathrm{p},l}^{\mathrm{d}})_{l=1}^{\mathrm{L}^{1}}$, initial estimations $(\mathbf{x}_{\mathrm{p},l}^{0}=0)_{l=1}^{\mathrm{L}^{1}}$, interpolator $\mathbf{\Gamma}$, standard RI operator $\mathbf{\Phi}_{\mathrm{p}}$, and $\mathrm{\kappa}$ > 0.
      \State \textbf{While} $i\leq \mathrm{I}$:
        \State \quad \textbf{Repeat}
          \Statex \quad  Take gradient descent step on:
          \Statex \small \quad \quad $\nabla_{\theta^{i}} \left( \frac{1}{\mathrm{L}^{i}} \sum\limits_{l=1}^{\mathrm{L}^i} \left\lVert \mathbf{x}_{\mathrm{s},l}^{\star} - \left[\mathbf{\Gamma}^{\dagger} \left( \mathbf{x}_{\mathrm{p},l}^{i-1} +\mathcal{N}_{\mathrm{p}}^{\boldsymbol{\theta}^i} \left( \mathbf{r}_{\mathrm{p},l}^{i-1}, \mathbf{x}_{\mathrm{p},l}^{i-1} \right) \right) \right]_+ \right\rVert_1 \right)
            $
        \Statex \quad \textbf{Until} $\mathcal{N}_{\mathrm{p}}^{\boldsymbol{\theta}^i}$ converged
      \State \quad $\mathbf{x}_{\mathrm{p},l}^{i}= \mathbf{x}_{\mathrm{p},l}^{i-1}+\mathcal{N}_{\mathrm{p}}^{\boldsymbol{\theta}^i}(\mathbf{r}_{\mathrm{p},l}^{i-1}, \mathbf{x}_{\mathrm{p},l}^{i-1})$
      \State \quad  $\mathbf{x}_{\mathrm{s},l}^{i} =\left[ \mathbf{\Gamma}^{\dagger}\mathbf{x}_{\mathrm{p},l}^{i}\right]_+$
      \State \quad $\mathbf{r}_{\mathrm{p},l}^{i} = \mathbf{x}_{\mathrm{p},l}^{\mathrm{d}} - \mathrm{\kappa} \textrm{Re}(\mathbf{\Phi}_{\mathrm{p}}^{\dagger}\mathbf{\Phi}_{\mathrm{p}}\mathbf{\Gamma}\mathbf{x}_{\mathrm{s},l}^{i})$
      \State \quad Pruning Step: $\mathrm{L}^{i+1}=\alpha^i \mathrm{L}^{i}, \ \alpha^i\in[0,1]$
    \State \textbf{End}
  \end{algorithmic}
\end{algorithm}
\end{minipage}

\noindent
\begin{minipage}{0.48\textwidth}
\begin{algorithm}[H]
  \caption{: S-R2D2 Reconstruction} 
  \label{algo:S_R2D2_reconstruction}
  \begin{algorithmic}[1]
  
  \small
      \State \textbf{Input:} trained DNNs $(\mathcal{N}_{\mathrm{p}}^{\boldsymbol{\theta}^i})_{i=1}^{\mathrm{I}}$, probed visibilities $\mathbf{y}$, initial back-projected visibility-data residual $\mathbf{r}_{\mathrm{p}}^{0}=\mathbf{x}_{\mathrm{p}}^{\mathrm{d}}=\mathrm{\kappa}\textrm{Re}(\mathbf{\Phi}_{\mathrm{p}}^{\dagger}\mathbf{y})$, initial estimation $\mathbf{x}_{\mathrm{p}}^{0} = 0$, interpolator $\mathbf{\Gamma}$, measurement operator $\mathbf{\Phi}_{\mathrm{p}}$, and $\mathrm{\kappa}$ > 0, .
      \State \textbf{While} $i\leq \mathrm{I}$:
      \State \quad $\mathbf{x}_{\mathrm{p}}^{i}=\mathbf{x}_{\mathrm{p}}^{i-1} +\mathcal{N}_{\mathrm{p}}^{\boldsymbol{\theta}^i}(\mathbf{r}_{\mathrm{p}}^{i-1}, \mathbf{x}_{\mathrm{p}}^{i-1})$
      \State \quad  $\mathbf{x}_{\mathrm{s}}^{i} =\left[ \mathbf{\Gamma}^{\dagger}\mathbf{x}_{\mathrm{p}}^{i}\right]_+$
      \State \quad $\mathbf{r}_{\mathrm{p}}^{i} = \mathbf{x}_{\mathrm{p}}^{\mathrm{d}} - \mathrm{\kappa} \textrm{Re}(\mathbf{\Phi}_{\mathrm{p}}^{\dagger}\mathbf{\Phi}_{\mathrm{p}} \mathbf{\Gamma}\mathbf{x}_{\mathrm{s}}^{i})$
    \State \textbf{End}
    \State \textbf{Return:} $\mathbf{\hat{x}}_{\mathrm{s}} = \mathbf{x}_{\mathrm{s}}^{\mathrm{I}}.$
  \end{algorithmic}
\end{algorithm}
\end{minipage}

\section{Datasets Configuration}\label{sec:section_4}
\noindent
This section describes the procedure to generate a realistic dataset composed of spherical ground truths and corresponding dirty signals in a wide-field RI framework.
\subsection{Ground Truth Dataset Generation}\label{subsec:sec_4_subsec_1}
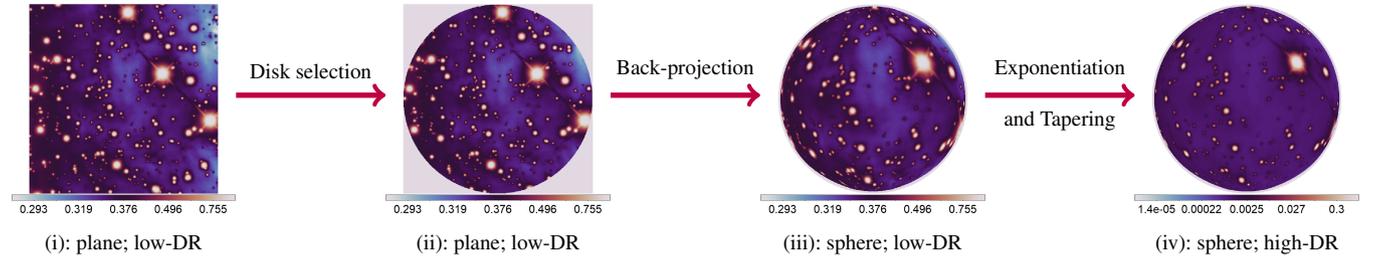
\begin{figure*}
\centering
\begin{tikzpicture}
    \node[anchor=south west, inner sep=0] (imageA) at (0,0) {
        \adjustbox{valign=t}{\includegraphics[width=0.165\textwidth]{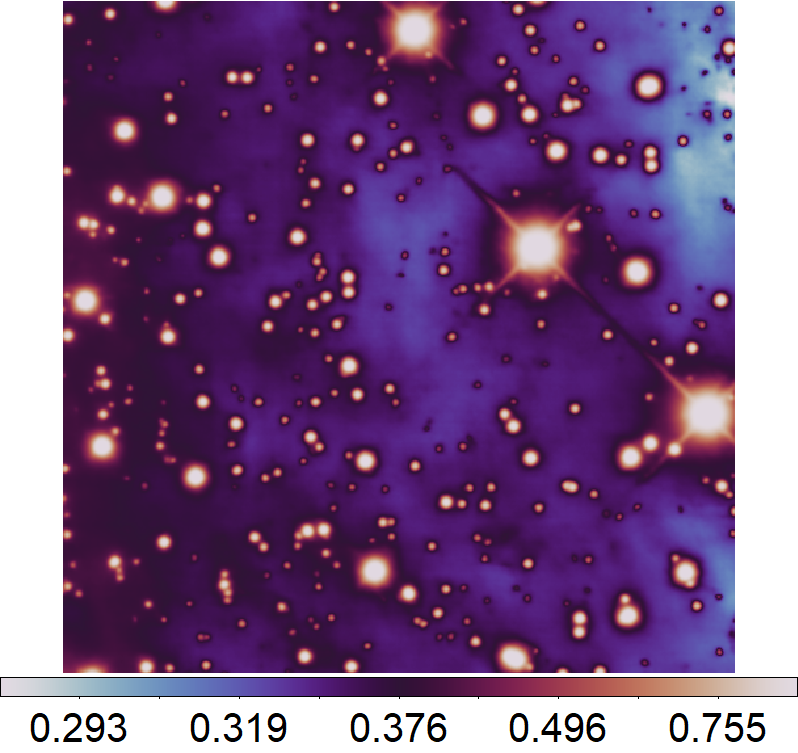}}
    };
    \node[below=0.5em of imageA.south, align=center] {(i): plane; low-$\textrm{DR}$};
    
    \node[anchor=south west, inner sep=0, right=7em of imageA] (imageB) {
        \adjustbox{valign=t}{\includegraphics[width=0.165\textwidth]{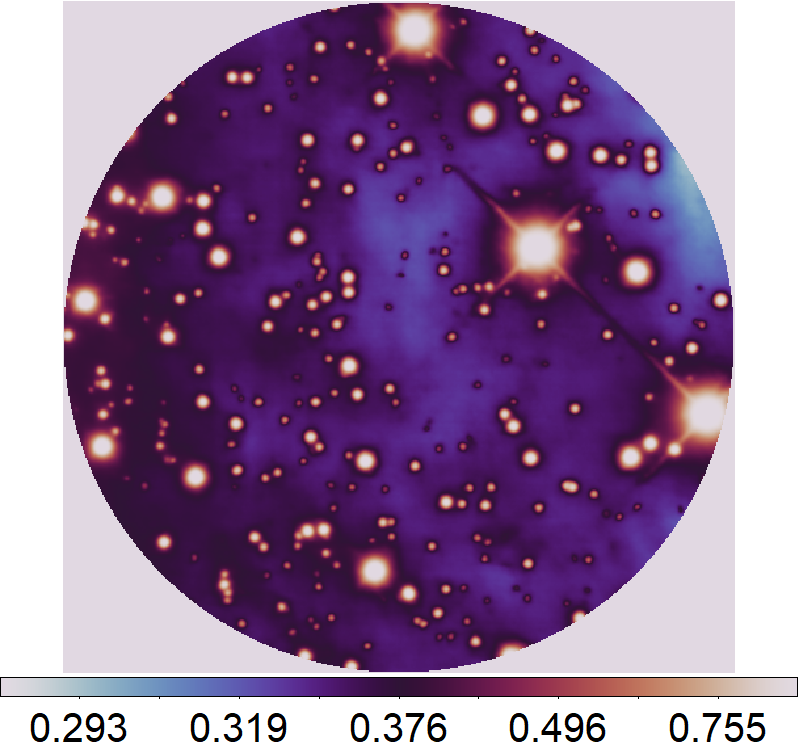}}
    };
    \node[below=0.5em of imageB.south, align=center] {(ii): plane; low-$\textrm{DR}$};

    \draw[->,  line width=0.7mm, purple] ([yshift=0.7em]imageA.east) -- ([yshift=0.7em]imageB.west)
        node[midway, above, sloped, black, yshift=0.3em] {Disk selection};

    \node[anchor=south west, inner sep=0, right=7em of imageB] (imageC) {
        \adjustbox{valign=t}{\includegraphics[width=0.165\textwidth]{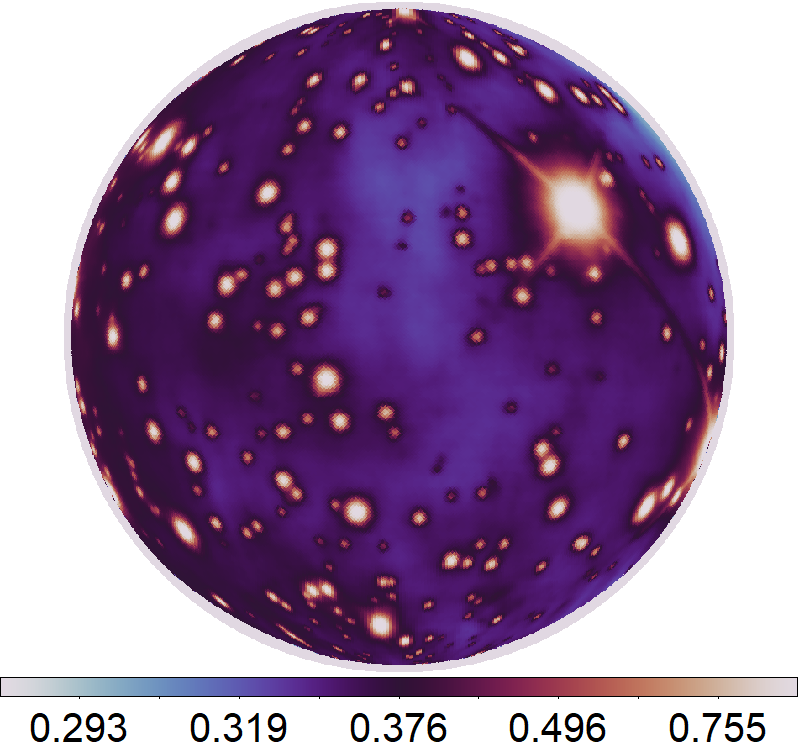}}
    };
    \node[below=0.5em of imageC.south, align=center] {(iii): sphere; low-$\textrm{DR}$};

    \draw[->,  line width=0.7mm, purple] ([yshift=0.7em]imageB.east) -- ([yshift=0.7em]imageC.west)
        node[midway, above, sloped, black, yshift=0.3em] {Back-projection};

    \node[anchor=south west, inner sep=0, right=7em of imageC] (imageD) {
        \adjustbox{valign=t}{\includegraphics[width=0.165\textwidth]{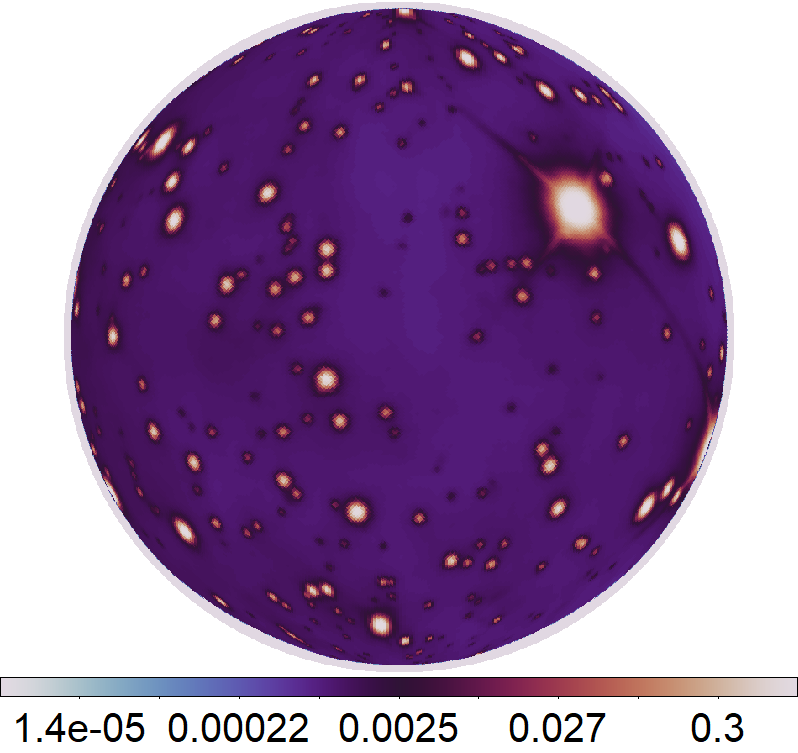}}
    };
    \node[below=0.5em of imageD.south, align=center] {(iv): sphere; high-$\textrm{DR}$};

    \draw[->,  line width=0.7mm, purple] ([yshift=0.7em]imageC.east) -- ([yshift=0.7em]imageD.west)
        node[midway, above, sloped, black, yshift=0.3em] {Exponentiation}
        node[midway, below, sloped, black, yshift=-0.3em] {and Tapering};

\end{tikzpicture}

\caption{Illustration of the generation of a high-$\textrm{DR}$ spherical ground truth (panel (iv), $\textrm{DR}=1.7\!\times\!10^5$), starting from the initial low-$\textrm{DR}$ planar image (panel (i), $\textrm{DR}=45$). The domain (plane or sphere) and the $\textrm{DR}$ are noted below each image. Firstly, we apply the disk selection step from (panel (i)) to (panel (ii)). Then, from (panel (ii)) to (panel (iii)), we back-project onto the sphere the disk with the ray-tracing operator which generates a low-$\textrm{DR}$ spherical signal. Finally, we apply the $\textrm{DR}$ exponentiation and tapering operations to produce the final high-$\textrm{DR}$ spherical signal (panel (iv)). All images are displayed in logarithmic scale, with the logarithmic exponent equals to their $\textrm{DR}$. For the spherical signals, we visualise the Northern hemisphere in the orthographic projection perspective.}
\label{fig:ray_tracing_visualization}
\end{figure*}

\noindent
The efficiency of supervised learning pipelines, such as R2D2 or S-R2D2, relies on a large and diverse dataset with different features and dynamic ranges ($\textrm{DR}$). To address the absence of a large realistic spherical RI dataset, we transformed an existing realistic low-$\textrm{DR}$ planar RI dataset into a high-$\textrm{DR}$ spherical one using the procedure illustrated in Figure~\ref{fig:ray_tracing_visualization}.\\ 

\begin{figure}
\centering
\adjustbox{valign=t, raise=-0.4em}{\begin{subfigure}[b]{0.193\textwidth}
\includegraphics[width=1.007\textwidth]{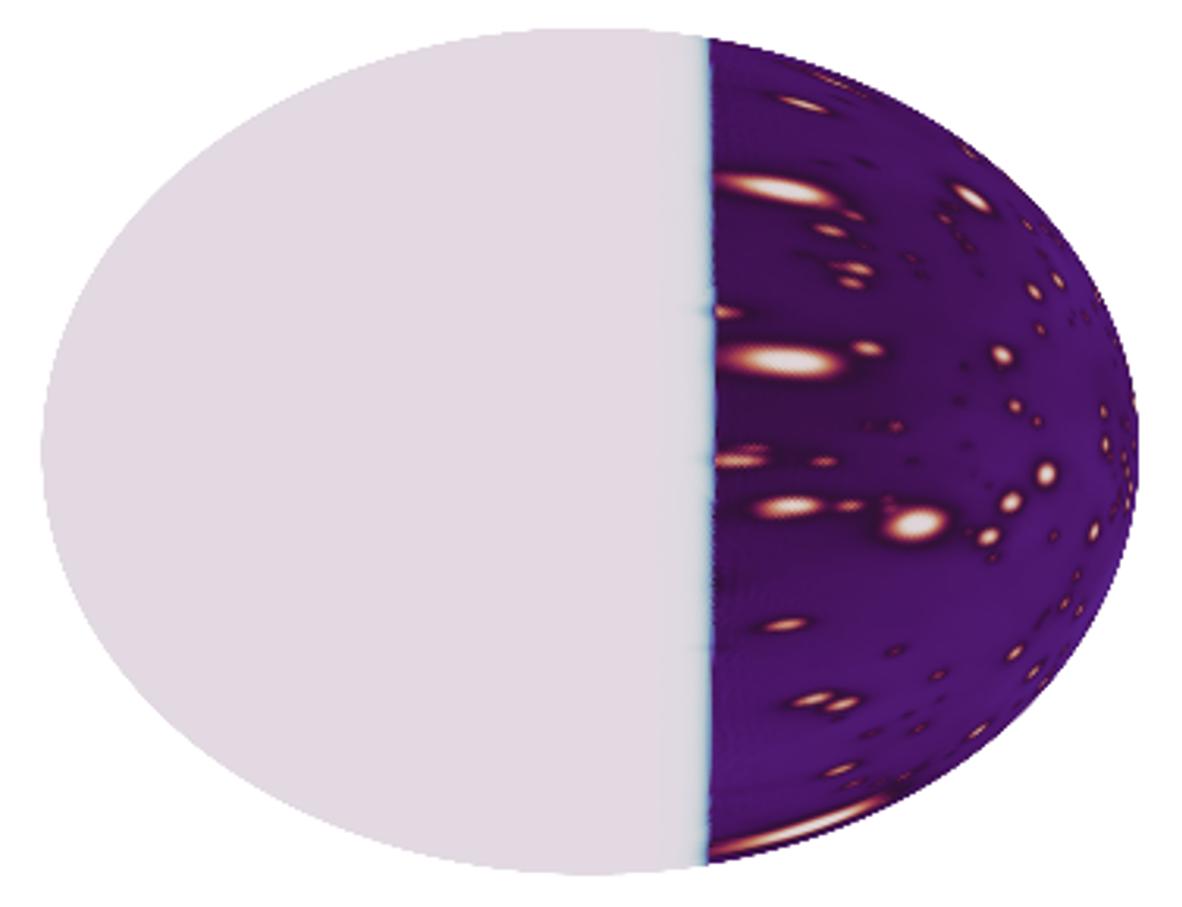}
\vspace{-1.92em}
\caption{}
\end{subfigure}}
\hspace{1.25em} 
\adjustbox{valign=t, raise=-0.4em}{\begin{subfigure}[b]{0.19\textwidth}
\includegraphics[width=1.007\textwidth]{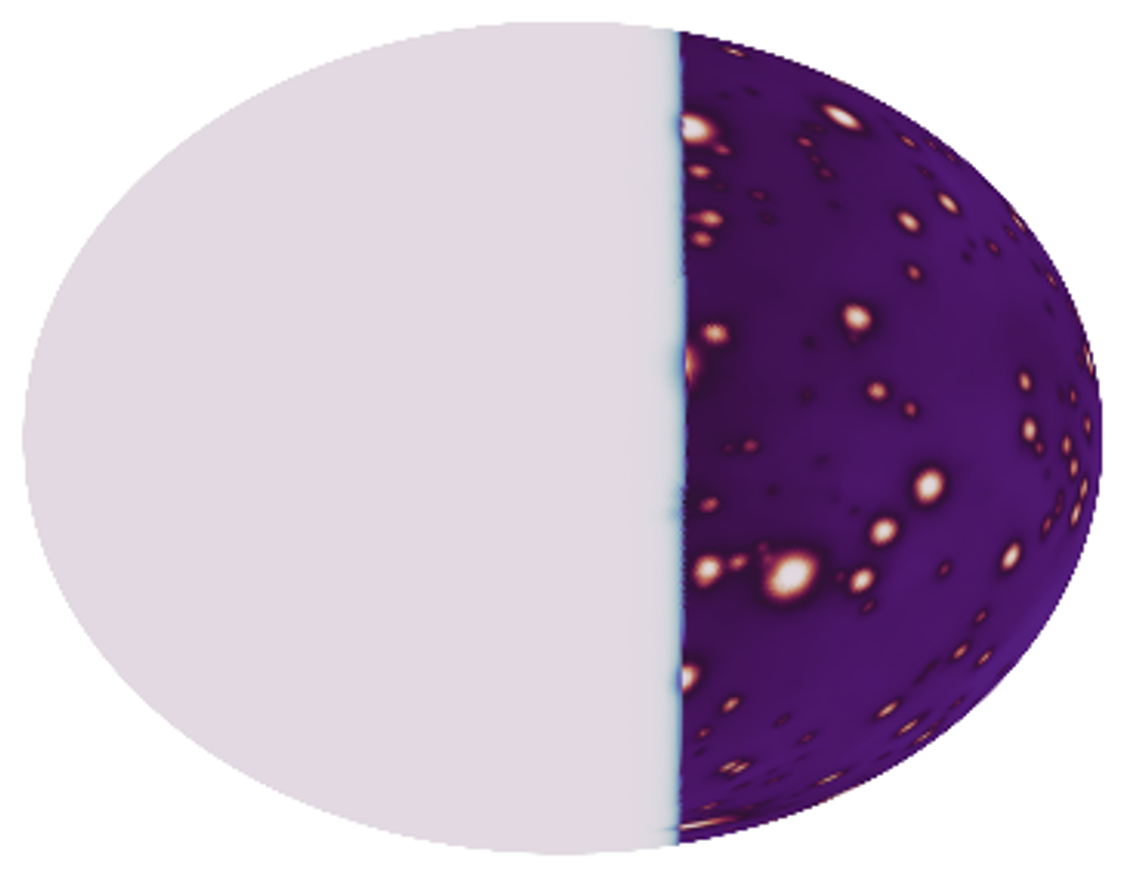}
\vspace{-1.92em}
\caption{}
\end{subfigure}}
\vspace{-0.3em}
\caption{3D lateral cut visualisation of a spherical signal, generated following the procedure illustrated in Figure~\ref{fig:ray_tracing_visualization} with the back-projection step performed either with the adjoint Fourier-based interpolator $\mathbf{\Gamma}^{\dagger}$ (panel (a)) or with the ray-tracing operator (panel (b)). The 2 spherical images originate from the same planar image. The ray-tracing operator produce realistic RI signals as it preserves the structures' roundness at the extents of the FOV, unlike $\mathbf{\Gamma}^{\dagger}$. We can observe the effect of the tapering operation \eqref{eqn:tapering_function} that creates smooth and faint signal at the very extents of the FOV.}
\label{fig:ray_tracing_vs_Fourier_3D}
\end{figure}

\noindent
Firstly, to promote a one to one correspondence between the equatorial plane and the sphere, we perform a disk selection on the low-$\textrm{DR}$ planar image. The selected disk is then back-projected onto the sphere within the FOV. Importantly, this back-projection must transfer the realistic structures of a planar RI image (\emph{e.g.}, round shapes) onto the sphere when mapping the disk onto it. For this purpose, the back-projection operator must intentionally distort the underlying continuous signal of the planar image, transforming it into a new, distinct continuous signal that we then represent on the HEALPix grid on the sphere. Indeed, as illustrated in Figure~\ref{fig:ray_tracing_vs_Fourier_3D}, using a plane-to-sphere interpolator that seeks to preserve the original continuous signal, such as $\mathbf{\Gamma}^{\dagger}$, creates unrealistic elliptical shapes at the extents of the FOV. Then, to perform such realistic plane-to-sphere back-projections, we rely on the ray-tracing algorithm developed by \citet{ray_tracing}. \\

\noindent
Finally, to simulate the high-$\textrm{DR}$ characteristic of RI signals, we apply to the generated low-$\textrm{DR}$ spherical image, the exponentiation procedure depicted in \citet{terris2024}. This pixel-wise transform enables to create a dataset with diverse random $\textrm{DR}$ between $10^3$ and $5\!\times\! 10^5$. More precisely, $\textrm{DR}$ values are uniformly distributed in the logarithmic scale such that $\log_{10}(\textrm{DR})\in [3,~5.69]$. Furthermore, we improve the physical realism of the high-$\textrm{DR}$ spherical image by radially tapering the intensity of the last $\theta_{t}^\circ$ of the FOV. This is achieved through a pixel-wise multiplication of the signal intensity by the arbitrary following function:
\begin{equation}
    \label{eqn:tapering_function}
    f(r)=1-\exp^\frac{(r - b)^3}{r-b_t},
\end{equation}
where $b=\cos\left(\mathrm{\theta}_{\textrm{FOV}}/2\right)$ and $b_t=\cos\left((\mathrm{\theta}_{\textrm{FOV}}-\theta_{t})/2\right)$. The effects of the tapering procedure can be observed in Figure~\ref{fig:ray_tracing_vs_Fourier_3D}. We note that the generated high-$\textrm{DR}$ spherical signals are not constrained to be band-limited on the sphere, which aligns with the observations made in Figure~\ref{fig:gamma_gamma_diaga}.

\subsection{Dirty Dataset Generation}\label{subsec:sec_4_subsec_2}
\noindent
To generate the Fourier visibilities and their corresponding dirty signals, we first apply the developed Fourier-based interpolator $\mathbf{\Gamma}$ to the spherical high-$\textrm{DR}$ ground truths following the measurement model \eqref{eqn:vis_discrete_2D_sphere}. The number of pixels $\mathrm{N}_{\mathrm{p}}$ of the planar interpolated image is fixed according to equation \eqref{eqn:SR_rule} by choosing some $\textrm{SR}$ factor and using the values of $\mathrm{N}_{\mathrm{p}}^{\min}$ and $\textrm{SR}_{\mathrm{0}}$ detailed in Section~\ref{subsec:sec_5_subsec_1}. Then, we apply to the planar interpolated image the standard RI operator $\mathbf{\Phi}_{\mathrm{p}}$ to generate visibilities $\mathbf{y}$ from this image. Furthermore, we use the MeqTrees software \citep{Noordam2010} to mimic VLA-type antenna configurations and generate realistic sampling patterns, as illustrated in Figure~\ref{fig:inverse problem}. The number of points in these sampling patterns vary within the range $[2 \!\times\!10^5,2 \!\times\! 10^6]$. Each pattern is unique, resulting in distinct RI inverse problems to be solved. Then, with the WSClean software \citep{offringa2014}, we integrate a Briggs weighting scheme to $\mathbf{\Phi}_{\mathrm{p}}$ and fix the Briggs weights to 0 to slightly favor resolution over sensitivity. Finally, we corrupted the simulated RI visibilities with an additive Gaussian random noise adapted to high-$\textrm{DR}$ signals, obtained based on the noise adaptation rule depicted in \citet{aghabiglou2024r2d2}. The dirty images $\mathbf{x}_{\mathrm{p}}^{\mathrm{d}}$ are then obtained by back-projection of the simulated visibilities $\mathbf{y}$ onto the plane as $\mathbf{x}_{\mathrm{p}}^{\mathrm{d}}=\mathrm{\kappa}\textrm{Re}(\mathbf{\Phi}_{\mathrm{p}}^{\dagger}\mathbf{y})$. The corresponding dirty spheres $\mathbf{x}_{\mathrm{s}}^{\mathrm{d}}$, illustrated in Figure~\ref{fig:inverse problem}, are obtained as $\mathbf{x}_{\mathrm{s}}^{\mathrm{d}}=\mathbf{\Gamma}^{\dagger}\mathbf{x}_{\mathrm{p}}^{\mathrm{d}}$.\\

\noindent
We implemented $\mathbf{\Gamma}$ using both pytorch\_finufft \citep{barnett2024finufft} and torch.FFT \citep{paszke2019pytorch} packages. Moreover, we rely on the implementation of $\mathbf{\Phi}_{\mathrm{p}}$ developed by \citet{aghabiglou2024r2d2,aghabiglou2025} that uses GPU modalities and the torchkbnufft package \citep{muckley_20} for the nonuniform FFT operation.

\subsection{Training, Validation and Test Datasets}\label{subsec:sec_4_subsec_3}
\noindent
For all datasets, we generate the high-$\textrm{DR}$ spherical\footnotemark \ ground truths, with varying $\textrm{DR}\in[10^3, 5\!\times\!10^5]$, from low-$\textrm{DR}$ planar images of $512\!\times\!512$ pixels, following Section~\ref{subsec:sec_4_subsec_1}.\\ 
\noindent
The training dataset contains $10^4$ high-$\textrm{DR}$ spherical images, generated from $10^4$ low-$\textrm{DR}$ planar optical astronomical images from the National Optical-Infrared Astronomy Research Laboratory and medical images from NYU fastMRI \citep{zbontar2018fastmri,knoll2020fastmri}. The inclusion of medical images is motivated by the works of \citet{terris2024} and \citet{aghabiglou2024r2d2}, who demonstrated their effectiveness in a RI deep learning context.\\
\noindent
The validation dataset contains 250 high-$\textrm{DR}$ spherical images, generated from 250 low-$\textrm{DR}$ planar radio astronomical images gathered from the National Radio Astronomy Observatory (NRAO) archives, and LOFAR surveys, namely the LOFAR HBA Virgo cluster survey \citep{edler2023} and the LoTSS-DR2 survey \citep{shimwell2022}. This difference in the nature of the images between the training and validation datasets enhances the generalizability of the DNNs.\\
\noindent
The test dataset contains 60 high-$\textrm{DR}$ spherical images, generated from 4 low-$\textrm{DR}$ real planar radio images: the giant radio galaxies 3C353 (sourced from the NRAO Archives) and Messier~106 \citep{shimwell2022}, the radio galaxy clusters Abell~2034 and PSZ2~G165.68+44.01 \citep{botteon2022}. More precisely, for each planar radio image, we generate 15 high-$\textrm{DR}$ spherical ground truths, with varying $\textrm{DR}$, following Section~\ref{subsec:sec_4_subsec_1}.\\

\noindent
Then, for each ground-truth dataset we generate the corresponding dirty dataset following Section~\ref{subsec:sec_4_subsec_2}. Therefore, each dataset is composed of pairs of high-$\textrm{DR}$ ground-truth spherical signals and associated dirty images. The $\textrm{DR}$ and the underlying sampling pattern are data-dependent, which results in creating diverse and different inverse problems to be solved. \footnotetext{The choice of the number of pixels, within the FOV, on the sphere $\mathrm{N}_{\mathrm{s}}$ is discussed in Section~\ref{subsec:sec_5_subsec_1}.}

\section{Experiments and Results} \label{sec:section_5} 
\noindent
This section presents a comparative evaluation of S-R2D2 and R2D2 in a wide-field context, detailing the experimental setup, performance metrics, and computational efficiency. Moreover, we analyse the precision-efficiency trade-off and assess the methods' ability to solve the wide-field RI inverse problem.
\subsection{Experiments}\label{subsec:sec_5_subsec_1}
\noindent
We aim to compare the performance of S-R2D2 with the baseline pipeline R2D2 to solve the wide-field RI inverse problem \eqref{eqn:vis_discrete_2D_sphere}, while analysing the precision-efficiency trade-off. Importantly, we must ensure consistency by firstly training and secondly testing both methods to solve the same underlying inverse problems \eqref{eqn:vis_discrete_2D_sphere}, linking the dirty images to their corresponding targets.\\

\noindent
Firstly, during the training stage, S-R2D2's networks learn to reconstruct the spherical targets $\mathbf{x}_{\mathrm{s}}^{\star}$ while R2D2's networks are trained to reconstruct their planar interpolated counterparts $\mathbf{x}_{\mathrm{p}}^{\star}=\mathbf{\Gamma}\mathbf{x}_{\mathrm{s}}^{\star}$. Importantly, both pipelines share identical dirty images $\mathbf{x}_{\mathrm{p}}^{\mathrm{d}}$ as:
\begin{equation}
    \label{eqn:dirty_consistency}
    \mathbf{x}_{\mathrm{p}}^{\mathrm{d}} = \mathrm{\kappa}\mathbf{\Phi}_{\mathrm{p}}^{\dagger}\left(\mathbf{\Phi}_{\mathrm{p}}\mathbf{\Gamma}\mathbf{x}_{\mathrm{s}}^{\star}+\mathbf{n}\right)=\mathrm{\kappa}\mathbf{\Phi}_{\mathrm{p}}^{\dagger}\left(\mathbf{\Phi}_{\mathrm{p}}\mathbf{x}_{\mathrm{p}}^{\star}+\mathbf{n}\right), \ \text{as} \ \mathbf{x}_{\mathrm{p}}^{\star}= \mathbf{\Gamma}\mathbf{x}_{\mathrm{s}}^{\star}.
\end{equation}
Therefore, the inverse problems linking dirty and ground-truth images (either spherical or interpolated) are the same. Moreover, although $\mathbf{x}_{\mathrm{s}}^{\star}$ is non-negative, the interpolated images $\mathbf{x}_{\mathrm{p}}^{\star}=\mathbf{\Gamma}\mathbf{x}_{\mathrm{s}}^{\star}$ can contain negative values. Thus, we removed the non-negativity constraints present in steps 3 and 5 of the R2D2 training Algorithm~\ref{algo:R2D2_training}. Furthermore, we note that in the S-R2D2 training Algorithm~\ref{algo:S_R2D2_training}, at each iteration $i\in\{1,2,...,\mathrm{I}\}$ the reconstruction is performed on the sphere (step 5) and the residual is computed with $\mathbf{\Gamma}$ (step 6). In contrast, in the R2D2 training Algorithm~\ref{algo:R2D2_training}, the reconstruction occurs on the plane (steps 4 and 5) and the residual is computed without interpolation (step 6). Therefore, the interpolated ground-truth dataset is inevitably the only connection between the spherical context and the R2D2 pipeline, limiting its iterative process to be significantly blind to this context.\\

\noindent
Secondly, in the reconstruction stage, we need to adapt the R2D2 reconstruction Algorithm~\ref{algo:R2D2_reconstruction} in order to ensure consistent comparisons in the spherical domain with the S-R2D2 reconstruction Algorithm~\ref{algo:S_R2D2_reconstruction}. Specifically, we first apply the R2D2 reconstruction Algorithm~\ref{algo:R2D2_reconstruction} until the final iteration $\mathrm{I}$, ensuring the removal of the non-negative constraint in step 4 for all iterations $i\in\{1,2,...,\mathrm{I}\}$. Then, the final planar reconstruction (iteration $\mathrm{I}$) undergoes a post-processing step. We first back-project it onto the sphere using the adjoint interpolator $\mathbf{\Gamma}^{\dagger}$, and then apply the non-negative constraint $\left[\cdot \right]_+$, as in S-R2D2. This corresponds to modifying the final step 7 of the R2D2 Algorithm~\ref{algo:R2D2_reconstruction} to return $\mathbf{\hat{x}} = \left[ \mathbf{\Gamma}^{\dagger}\mathbf{x}_{\mathrm{p}}^{\mathrm{I}}\right]_+$. We note that this post-processing procedure can be applied at any iteration $i\in\{1,2,...,\mathrm{I}\}$ to compare the reconstructions of S-R2D2 and R2D2 at a given iteration $i$.\\

\noindent
Finally, we also aim to study the precision-efficiency trade-off discussed in Section~\ref{subsec:sec_3_subsec_2}. Thus, we conducted consistent experiments for different values $\mathrm{N}_{\mathrm{p}}$, following equation \eqref{eqn:SR_rule}, for both S-R2D2 and R2D2. Firstly, we set the baseline $\textrm{SR}_{\mathrm{0}}$ factor to $1.5$, as in \citet{aghabiglou2024r2d2}, and the baseline resolution on the plane $\mathrm{N}_{\mathrm{p}}^{\min}$ to $400^{2}$, fixing the sampling pattern bandwidth to $266$ (corresponding to $\sqrt{400^2}/1.5$). More precisely, as $\mathrm{N}_{\mathrm{p}}^{\min}$ is the number of pixels that maintains the same pixel size on the sphere and the plane, we determined $\mathrm{N}_{\mathrm{p}}^{\min}$ by setting $\theta_{\textrm{FOV}} = 170^\circ$ and $\mathrm{N}_{\mathrm{s}}=89737$. Specifically, the number of pixels, within the FOV, $\mathrm{N}_{\mathrm{s}}$ is obtained using $\mathrm{N}_{\mathrm{s}}=6\mathrm{n}_\textrm{side}^2(1-\cos(\theta_{FOV}/2))$ and fixing the HEALPix parameter $\mathrm{n}_\textrm{side}=128$, which sets the total number of pixels on the Northern hemisphere to $6\mathrm{n}_\textrm{side}^2$. The value $\mathrm{\theta}_{\textrm{FOV}}=170^{\circ}$ ensures a safety margin, as the ray-tracing algorithm \citep{ray_tracing} maintains non-degenerative performance for FOV below $174^{\circ}$. Secondly, we consider the $\textrm{SR}$ enhancement power of our algorithms to be relevant up to $\textrm{SR}=3$, which is a standard limit in high-$\textrm{DR}$ frameworks for similar algorithms, such as CLEAN and uSARA \citep{Terris2022}. Therefore, we conducted experiments from $\mathrm{N}_{\mathrm{p}}=\mathrm{N}_{\mathrm{p}}^{\min}=400^2$ (corresponding to $\textrm{SR}=\textrm{SR}_{\mathrm{0}}=1.5$) to $\mathrm{N}_{\mathrm{p}}=800^2$ (corresponding to $\textrm{SR}=3$). We note that the number of pixels $\mathrm{N}_{\mathrm{p}}$ that satisfies the resolution rule \eqref{eqn:resolution_rule} is $2860^2$, corresponding to $\textrm{SR} \approx 7$, which is an impractical $\textrm{SR}$ enhancement value. Moreover, we arbitrarily set the tapering angle to $\theta_{t}=15^{\circ}$ in equation \eqref{eqn:tapering_function}.\\

\noindent
The DNNs were trained with a GPU-accelerated Python implementation using the PyTorch library \citep{paszke2019pytorch}. Each training was carried out on the Heriot-Watt high-performance computing facility (DMOG) cluster, where the utilised GPU nodes consisted of two 32-core AMD EPYC 7543 processors, two Nvidia A40 (48GB) GPUs, and 256GB of RAM. The DNNs share the same U-Net architecture and hyperparameters, and are implemented as in \citet{aghabiglou2024r2d2,aghabiglou2025}. We set the learning rate to $10^{-4}$. Moreover, during experimentation, we observed a minor decrease in performance for batch sizes larger than 1. Then, to ensure optimal performance, we set the batch size to 1 for all experiments, though this leaves room for improvement in computational efficiency.

\subsection{Metrics and Timings Evaluation}\label{subsec:sec_5_subsec_2} 
\noindent
As in \citet{aghabiglou2024r2d2,aghabiglou2025}, we quantitatively evaluate our reconstructions in the image domain, using the signal-to-noise ratio metric in linear (SNR) and logarithmic ($\textrm{logSNR}$) scales, and in the data domain using a residual-based metric $\textrm{RDR}$. All metrics are computed using quantities defined on the sphere. More precisely, considering a ground truth $\mathbf{x}^{\star}_{\mathrm{s}}$ and a signal estimate $\mathbf{\hat{x}}_{\mathrm{s}}$, the $\textrm{SNR}$ metric is formulated as:
\begin{equation}
\textrm{SNR}(\mathbf{x}^{\star}_{\mathrm{s}},\mathbf{\hat{x}}_{\mathrm{s}}) = 20\textrm{log}_{10}\left( \frac{\| \mathbf{x}^{\star}_{\mathrm{s}}\|_2 }{\|\mathbf{x}^{\star}_{\mathrm{s}} - \mathbf{\hat{x}}_{\mathrm{s}}\|_2}\right).
\end{equation}
The $\textrm{logSNR}$ metric quantifies the reconstruction quality with a particular emphasis on faint structures and is expressed as: 
\begin{equation}
\label{eqn:logSNR_metric}
 \textrm{logSNR}(\mathbf{x}^{\star}_{\mathrm{s}}, \mathbf{\hat{x}}_{\mathrm{s}}, \mathrm{a}) = \textrm{SNR}\left(\textrm{rlog}(\mathbf{x}^{\star}_{\mathrm{s}},\mathrm{a}), \textrm{rlog}(\mathbf{\hat{x}}_{\mathrm{s}},\mathrm{a})\right),
\end{equation}
where the logarithmic transform $\textrm{rlog}$, parameterised by $\mathrm{a}>0$, corresponds to:
\begin{equation}
\textrm{rlog}(\mathbf{x}_{\mathrm{s}},\mathrm{a}) = \mathrm{x}_{\mathrm{s}}^{\max}\textrm{log}_{\mathrm{a}}\left({\frac{\mathrm{a}}{\mathrm{x}_{\mathrm{s}}^{\max}}}~ \mathbf{x}_{\mathrm{s}}+\textbf{1}\right),
\end{equation}
where $\mathrm{x}_{\mathrm{s}}^{\max}$ is the maximum pixel value of the signal $\mathbf{x}_{\mathrm{s}}$ and $\textbf{1}$ is a vector with all elements equal to 1. Throughout the article, we have fixed $\mathrm{a}$ to the $\textrm{DR}$ of the target signal to evaluate the reconstruction quality down to its faintest features.\\
\noindent
The residual-to-dirty image ratio ($\textrm{RDR}$) metric evaluates the fidelity of the reconstruction to the visibility data in the image domain. For a dirty signal $\mathbf{x}^{\mathrm{d}}_{\mathrm{s}}$ and a residual $\mathbf{r}_{\mathrm{s}}$, the metric $\textrm{RDR}$ is formulated as:
\begin{equation} 
\textrm{RDR}(\mathbf{r}_{\mathrm{s}}, \mathbf{x}^{\mathrm{d}}_{\mathrm{s}})=\frac{\|\mathbf{r}_{\mathrm{s}}\|_2}{\|\mathbf{x}^{\mathrm{d}}_{\mathrm{s}}\|_{2}}.
\label{eq:data_fidelity}
\end{equation}
We recall that to compute the spherical dirty and spherical residual signals, we back-project onto the sphere the planar quantities, defined in Section~\ref{subsec:sec_2_subsec_2} and Section~\ref{subsec:sec_3_subsec_3}, using $\mathbf{\Gamma}^{\dagger}$ operator. A decrease in the value of $\textrm{RDR}$ over the course of the iterations of the R2D2 and S-R2D2 algorithms corresponds to an iterative decrease of the residual norm $\|\mathbf{r}_{\mathrm{s}}\|_2$. Thus, the corresponding performed reconstruction is increasingly satisfying the measurement equation \eqref{eqn:vis_discrete_2D_sphere}, which enhances its fidelity to the visibility data.\\

\noindent
We also evaluate the computational time across all experiments in GPU hours using a Nvidia A40 (48GB) GPU of the Heriot-Watt high-performance computing facility (DMOG) cluster. The total time, denoted as $\mathrm{t}_{\textrm{tot.}}$, is the time required to complete all $\mathrm{I}$ iterations leading up to the final spherical reconstruction. The total time is expressed as $\mathrm{t}_{\textrm{tot.}} = \mathrm{I}\!\times\!(\mathrm{t}_{\textrm{reg.}}+\mathrm{t}_{\textrm{dat.}})$, where $\mathrm{t}_{\textrm{reg.}}$ is the time required for a single regularisation step, 
and $\mathrm{t}_{\textrm{dat.}}$ is the time required for a single visibility-data fidelity step. The time $\mathrm{t}_{\textrm{reg.}}$ 
corresponds to steps 3 and 4 of both the S-R2D2 Algorithm~\ref{algo:S_R2D2_reconstruction} and the R2D2 Algorithm~\ref{algo:R2D2_reconstruction} (with the modification that we compute $\left[ \mathbf{\Gamma}^{\dagger}\mathbf{x}_{\mathrm{p}}^{i}\right]_+$ instead of $\mathbf{x}_{\mathrm{p}}^{i}$ only for the last iteration). The time $\mathrm{t}_{\textrm{dat.}}$ corresponds to step 5 of both algorithms.

\subsection{Results}\label{subsec:sec_5_subsec_3}
\begin{figure}
    \vspace{-0.4cm}
    \centering
    \includegraphics[width=0.8\linewidth]{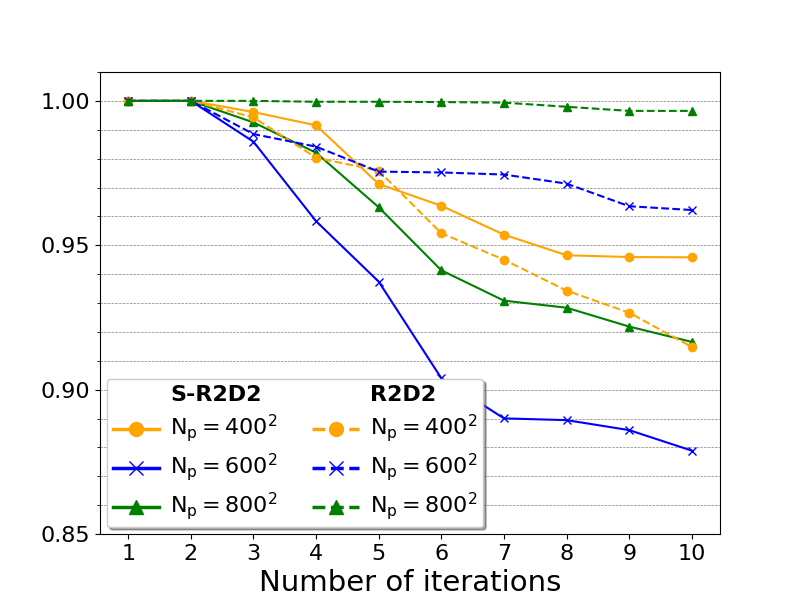}
    \caption{Illustration of the effect of the pruning procedure on the size of the training dataset throughout the iterations, depending on $\mathrm{N}_{\mathrm{p}}$ and the employed method (R2D2 or S-R2D2). The pruning procedure corresponds to step 7 in both the R2D2 Algorithm~\ref{algo:R2D2_training} and the S-R2D2 Algorithm~\ref{algo:S_R2D2_training}. At a given iteration $i$, the y-axis value represents the training dataset size as a fraction of its initial size, given by $\Pi_{k=1}^{i}\alpha_{k}$ where $(\alpha_{k})_{k=1}^{\mathrm{I}}$ is defined in Section~\ref{subsec:sec_2_subsec_2}.}
    \label{fig:pruning}
\end{figure}

\begin{table*}
\centering
\begin{tabular}{cccccccccc}
\toprule
\multirow{2}{*}{Method} & \multirow{2}{*}{Resolution $\mathrm{N}_{\mathrm{p}}$} & \multicolumn{3}{c}{Metrics} & \multicolumn{3}{c}{Computation Times} \\
\cmidrule(lr){3-5} \cmidrule(lr){6-8}
 & & $\textrm{SNR}$ (dB) & $\textrm{logSNR}$ (dB) & $ \textrm{RDR} \!\times\! 10^{-2}$ & $\mathrm{t}_{\textrm{tot.}}$  (s)& $\mathrm{t}_{\textrm{dat.}} \!\times\! 10^{-1}$ (s) & $\mathrm{t}_{\textrm{reg.}} \!\times\! 10^{-2}$ (s) \\
\midrule
\multirow{3}{*}{R2D2}   & $400^2$ & $6.6$ $\pm$ $1.5$ & $7.3$ $\pm$ $3.2$ & $0.8$ $\pm$ $0.5$ & $2.1$ $\pm$ $0.8$ & $1.5$ $\pm$ $0.8$ & $5.7$ $\pm$ $0.4$ \\
                         & $600^2$ & $2.8$ $\pm$ $1.5$ & $5.2$ $\pm$ $1.5$ & $2.7$ $\pm$ $1.1$ & $2.3$ $\pm$ $0.8$ & $1.7$ $\pm$ $0.8$ & $5.9$ $\pm$ $0.9$ \\
                         & $800^2$ & $1.9$ $\pm$ $1.5$ & $1.6$ $\pm$ $1.3$ & $14$ $\pm$ $7.9$ & $2.5$ $\pm$ $0.8$ & $1.9$ $\pm$ $0.8$ & $6.1$ $\pm$ $1.4$ \\
\midrule
\multirow{3}{*}{S-R2D2} & $400^2$ & $21.7$ $\pm$ $4.3$ & $17.5$ $\pm$ $3.2$ & $1.5$ $\pm$ $1.3$ & $3.0$ $\pm$ $0.8$ & $2.2$ $\pm$ $0.8$ & $7.3$ $\pm$ $0.3$ \\
                         & $600^2$ & $21.2$ $\pm$ $3.9$ & $18.7$ $\pm$ $2.9$ & $1.2$ $\pm$ $1.1$ & $3.1$ $\pm$ $0.8$ & $2.3$ $\pm$ $0.8$ & $8.5$ $\pm$ $0.4$ \\
                         & $800^2$ & $20.8$ $\pm$ $3.7$ & $15.9$ $\pm$ $3.5$ & $2.1$ $\pm$ $1.6$ & $3.3$ $\pm$ $0.8$ & $2.3$ $\pm$ $0.8$ & $10.3$ $\pm$ $0.3$ \\
\bottomrule
\end{tabular}
\caption{Quantitative comparison of R2D2 and S-R2D2 across three different resolution values ($\mathrm{N}_{\mathrm{p}} = 400^2$, $\mathrm{N}_{\mathrm{p}} = 600^2$, $\mathrm{N}_{\mathrm{p}} = 800^2$).  The reported values correspond to the means $\pm$ the standard deviations (std), computed over 60 different test inverse problems, with different target $\textrm{DR}$ randomly varying in $[10^3,5\!\times\!10^5]$, at the final iteration $\mathrm{I} = 10$. The values are computed for the three performance metrics ($\textrm{SNR}$, log$\textrm{SNR}$, and $\textrm{RDR}$) and the three computational time metrics ($\mathrm{t}_{\textrm{tot.}}$, $\mathrm{t}_{\textrm{dat.}}$, and $\mathrm{t}_{\textrm{reg.}}$) defined in Section~\ref{subsec:sec_5_subsec_2}.}
\label{table:results}
\end{table*}

\begin{figure*}
\vspace{-0.25cm}
 \setlength{\tabcolsep}{-5.5pt}
 \begin{tabular}{ccc}
 \includegraphics[width = 0.35\textwidth]{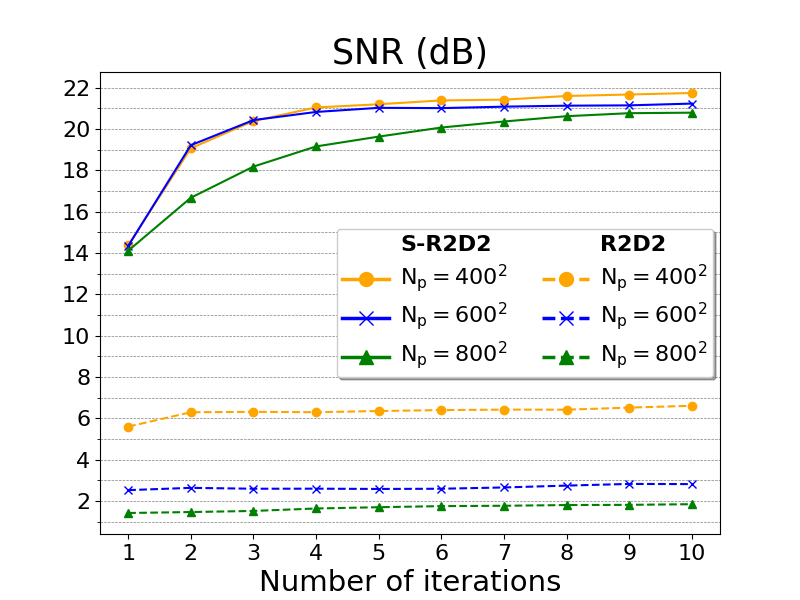}&
 \includegraphics[width = 0.35\textwidth]{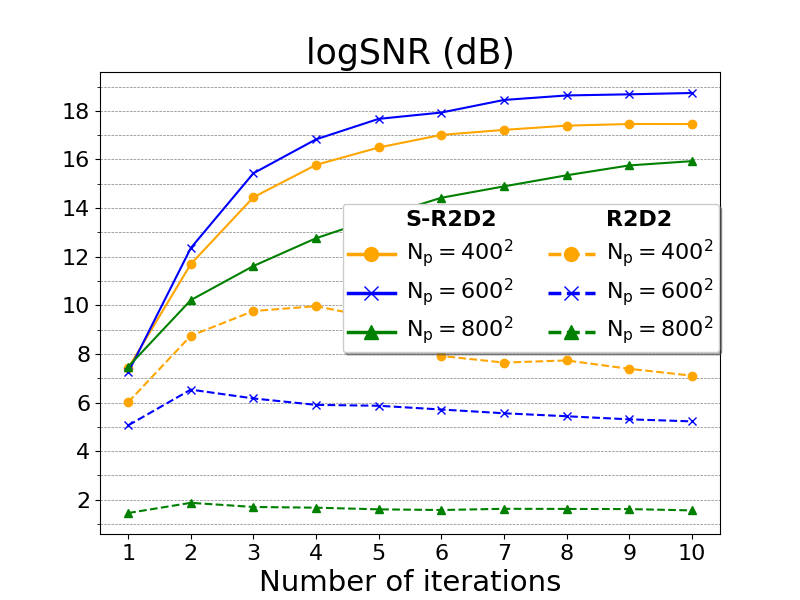}&
 \includegraphics[width = 0.35\textwidth]{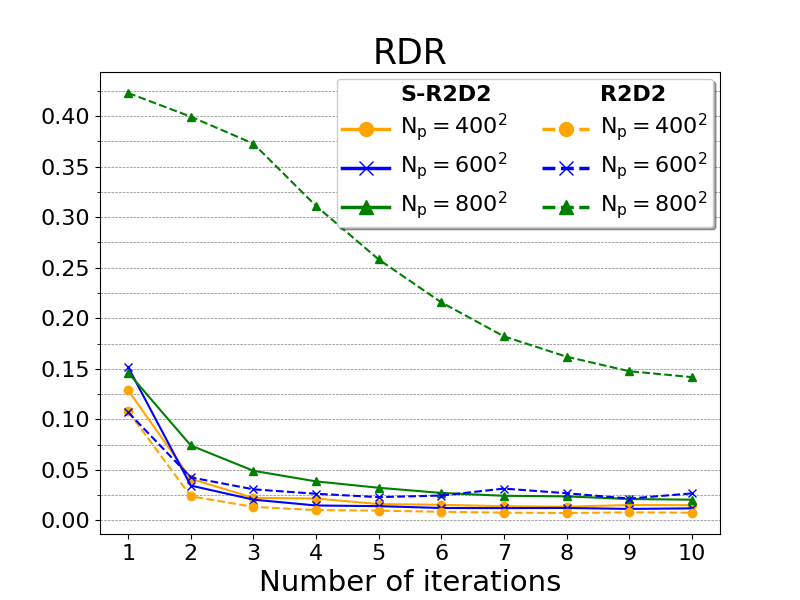}
 \end{tabular}
\caption{Progress of the reconstruction quality for the different metrics introduced in Section~\ref{subsec:sec_5_subsec_2} (left column: $\textrm{SNR}$, middle column: $\textrm{logSNR}$, right column: $\textrm{RDR}$) across iterations, depending on $\mathrm{N}_{\mathrm{p}}$ and the employed method (R2D2 or S-R2D2). For each metric, the reported values are averages over the test dataset composed of 60 different inverse problems, with different $\textrm{DR}$ values ranging from $10^3$ to $5\!\times\!10^5$.}
\label{fig:result_evolution}
\end{figure*}

\noindent
S-R2D2 and R2D2 are parameter-free pipelines, depending only on the number of iterations $\mathrm{I}$. We chose to fix\footnotemark\footnotetext{Recent R2D2 developments \citep{aghabiglou2025} use a convergence criterion for robust early stopping, which we have not implemented here.} $\mathrm{I}=10$ as none of the experiments showed a noticeable improvement in reconstruction quality, based on the $\textrm{logSNR}$ metric, beyond this point.\\

\noindent
Figure~\ref{fig:pruning} illustrates the evolution of training dataset sizes for each conducted experiment. A significant decrease in the size of the dataset over the course of iterations indicates an enhanced capability of the algorithm to solve the wide-field inverse problem $\eqref{eqn:vis_discrete_2D_sphere}$. Indeed, as discussed in Sections~\ref{subsec:sec_2_subsec_2} and~{\ref{subsec:sec_3_subsec_3}, a pair of data $(\mathbf{x}_{\mathrm{s},l}^{i},(\mathbf{x}_{\mathrm{p},l}^{i},\mathbf{r}_{\mathrm{p},l}^{i}))$ is pruned at iteration $i$ if the underlying inverse problem is considered solved. Moreover, the pruning procedure is less effective here than in \citet{aghabiglou2024r2d2,aghabiglou2025} due to the increased complexity of the wide-field inverse problem compared to the small-field one.\\

\noindent
In a small-field framework, \citet{aghabiglou2024r2d2, aghabiglou2025} demonstrated the overall superior speed of R2D2 compared to state-of-the-art methods, typically superseding uSARA, AIRI \citep{terris2024}, and CLEAN by more than an order of magnitude. Firstly, in a wide-field framework, Table~\ref{table:results} showcases that R2D2 preserves its computational performance. Secondly, it indicates that S-R2D2 is less than 1.5 times slower than R2D2, with the entire reconstruction process requiring approximately 3 seconds, which demonstrates S-R2D2's computational efficiency. This slight speed difference stems from the incorporation of the interpolator $\mathbf{\Gamma}^{\dagger}$ in the regularisation step 4 (reflected in $\mathrm{t}_{\textrm{reg.}}$) and of $\mathbf{\Gamma}$ in the data-fidelity step 5 (reflected in $\mathrm{t}_{\textrm{dat.}}$), in the S-R2D2 Algorithm~\ref{algo:S_R2D2_reconstruction} compared to the R2D2 Algorithm~\ref{algo:R2D2_reconstruction}. This additional computational cost is minimal due to the efficiency of the Fourier-based interpolators that rely on FFT and nonuniform FFT operations. Moreover, computational time naturally slightly increases with the resolution $\mathrm{N}_{\mathrm{p}}$ due to the larger data size and higher processing demands.\\ 

\noindent
Final numerical reconstruction results are summarized in Table~\ref{table:results}, while Figure~\ref{fig:result_evolution} illustrates the numerical progression of both the S-R2D2 and R2D2 methods through iterations. The mean values in Table~\ref{table:results} correspond to the values of the final iteration $\mathrm{I}=10$ in Figure~\ref{fig:result_evolution}. The $\textrm{SNR}$ and $\textrm{logSNR}$ values highlight the overall significant superior performance of S-R2D2 compared to R2D2. The first iteration of both algorithms corresponds to an end-to-end reconstruction, where the DNN was trained using their respective ground-truth datasets and loss functions. Compared to R2D2, S-R2D2 effectively improves the suboptimal end-to-end reconstruction, with $\textrm{SNR}$ and $\textrm{logSNR}$ values at the first iteration being at least 7 to 11~dB lower than at the final iteration. Moreover, as illustrated in Figure~\ref{fig:evolution_S-R2D2}, the significant increase of the $\textrm{logSNR}$ value through iterations demonstrates that S-R2D2's iterative structure progressively recovers the full dynamic range and frequency content, enhancing high imaging precision. This improvement stems from the integration of the interpolators at each iteration in the DNN loss function and in the computation of the residual $\mathbf{r}_{\mathrm{p}}^{i}$, effectively informing S-R2D2 of the spherical context. Consequently, S-R2D2's networks learn to simultaneously solve the wide-field inverse problem and correct the interpolation limitations, illustrated in Figure~\ref{fig:gamma_gamma_diaga}. In contrast, the R2D2 pipeline, which is inherently overly blind to the spherical context (Section~\ref{subsec:sec_5_subsec_1}), suffers from uncorrected interpolation errors, which ultimately limit its reconstruction quality. As a result, interpolation inaccuracies, whether corrected or uncorrected, play a critical role as they iteratively guide S-R2D2 towards improved performance, while hindering R2D2's effectiveness.\\
\begin{figure*}
 \centering
 \begin{tikzpicture}
   \node[anchor=south west, inner sep=0] (table) at (0,0) {
     \begin{tabular}{ccccc}
       \includegraphics[width=0.178\linewidth]{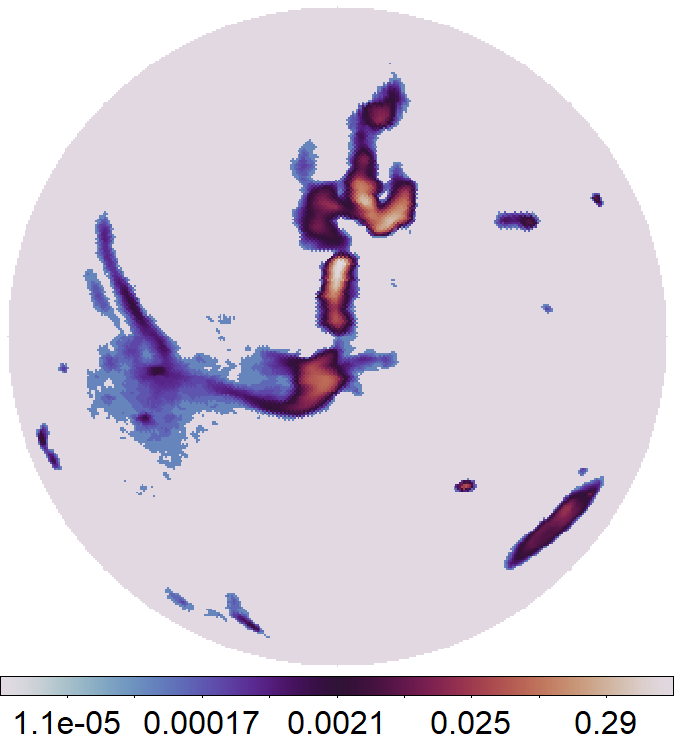} &
       \includegraphics[width=0.178\linewidth]{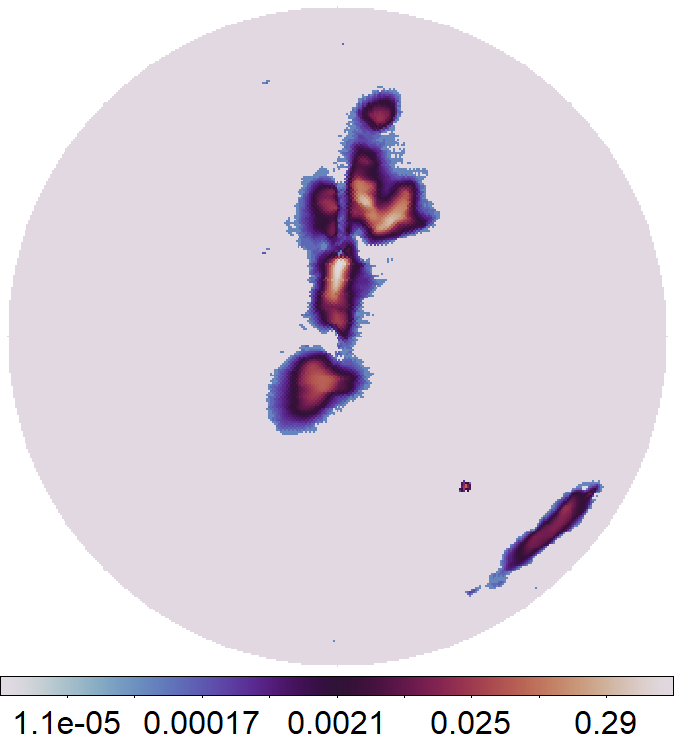} &
       \includegraphics[width=0.178\linewidth]{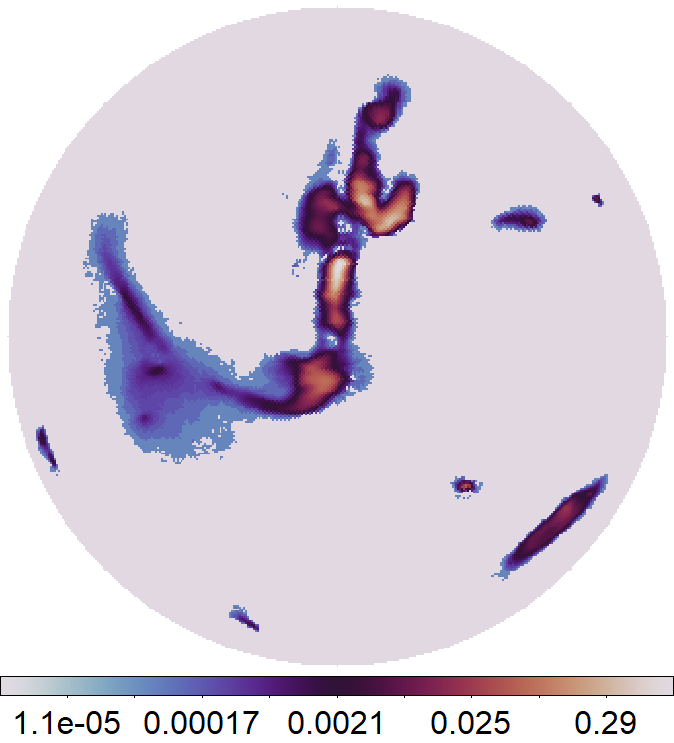} &
       \includegraphics[width=0.178\linewidth]{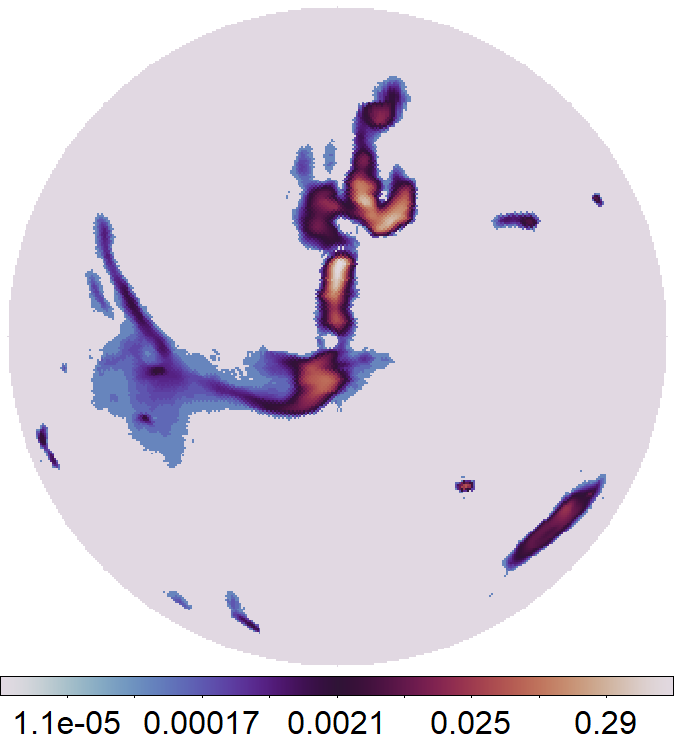} &
       \includegraphics[width=0.178\linewidth]{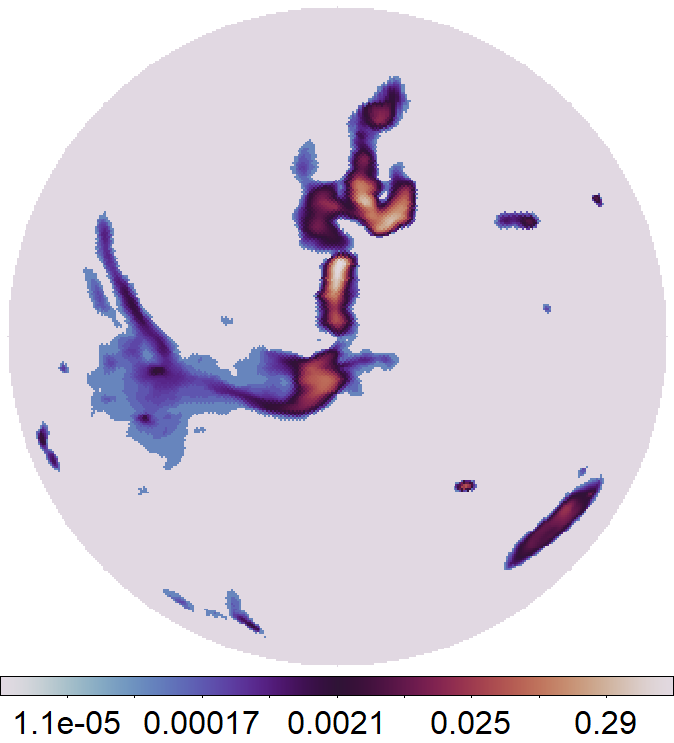}\\
       Ground Truth & $i=1$; (19.4, 4.3)~dB & $i=2$; (22.9, 9.4)~dB & $i=3$; (23.8, 13.6)~dB & $i=10$; (25.6, 18.1)~dB\\
     \end{tabular}
   };

   \draw[->, line width=1mm, purple] 
     ([xshift=0.214\textwidth, yshift=-15.25em]table.north west) -- 
     ([xshift=1\textwidth, yshift=-15.25em]table.north west);
 \end{tikzpicture}
 \caption{Visual progression of reconstructions performed by the S-R2D2 Algorithm~\ref{algo:S_R2D2_reconstruction} to recover the target signal displayed on the left. The purple arrow highlights the progression through iterations, leading to the final reconstruction on the right. For each iteration $i$, the values of ($\textrm{SNR}$, $\textrm{logSNR}$) are displayed below the corresponding signal. The spherical ground truth was generated with a $\textrm{DR}= 2.3\!\times\!10^5$ from the planar galaxy cluster Abell~2034 \citep{botteon2022} image following the procedure depicted in Section~\ref{subsec:sec_4_subsec_1}. The results are obtained with $\textrm{SR}=2.25$ (corresponding to $\mathrm{N}_{\mathrm{p}}=600^2$). For all images, we visualise the Northern hemisphere in the orthographic projection perspective and in the logarithmic scale, with the logarithmic exponent equals to $\textrm{DR}$.} 
 \label{fig:evolution_S-R2D2}
\end{figure*}

\noindent
In Figures~\ref{fig:P138}-\ref{fig:PSZ}, we provide a visual comparison of the results for 4 different selected signals from the test dataset, with different $\textrm{DR}$. S-R2D2 shifts the precision-efficiency trade-off towards higher values of $\textrm{SR}$ by taking advantage of more accurate interpolators as $\textrm{SR}$ increases. However, this shift is limited by its inherent capability of super-resolving, as discussed in Section~\ref{subsec:sec_3_subsec_2}. The results obtained with $\textrm{SR}=2.25$, corresponding to $\mathrm{N}_{\mathrm{p}}=600^2$, represents an optimal balance and are considered the best overall. While this value does not yield the top performance across all metrics, it achieves the best results for the $\textrm{logSNR}$ metric and comparable performance for the other metrics. Consistent with its $\textrm{logSNR}$ performance, it visually delivers the best reconstruction especially for the faintest features. Unlike S-R2D2, R2D2's performance weakens as the resolution $\mathrm{N}_{\mathrm{p}}$ increases. This performance decline occurs since R2D2 does not integrate $\mathbf{\Gamma}$ and $\mathbf{\Gamma}^{\dagger}$ in its pipeline and thus does not benefit from the enhanced interpolation capabilities available at higher resolutions. Consequently, R2D2 grapples with the challenge of super-resolving.\\

\noindent
Figures~\ref{fig:P138}-\ref{fig:PSZ} also illustrate the final visibility-data residuals back-projected onto the sphere using $\mathbf{\Gamma}^{\dagger}$. Compared to the dirty image, which is the initial residual, the final residuals have significantly lower intensity values, for both R2D2 and S-R2D2, which is quantified by the $\textrm{RDR}$ metric (Table~\ref{table:results}, Figure~\ref{fig:result_evolution}). Bright regions in the residuals capture areas that need further improvements in the image domain, particularly well quantified by the $\textrm{logSNR}$ metric. Importantly, a low $\textrm{RDR}$ value indicates that the corresponding reconstruction is faithful to the visibility data, as discussed in Section~\ref{subsec:sec_5_subsec_2}. Notably, S-R2D2 consistently delivers reconstructions with enhanced data fidelity for all resolutions $\mathrm{N}_{\mathrm{p}}$ and achieves its best performance at $\mathrm{N}_{\mathrm{p}} = 600^2$. In contrast, R2D2 showcases poor data-fidelity performance at higher resolutions, $\mathrm{N}_{\mathrm{p}}=600^2$ and $\mathrm{N}_{\mathrm{p}}=800^2$, due to the increased challenge of super-resolving. However, interestingly, R2D2 achieves the best overall performance at resolution $\mathrm{N}_{\mathrm{p}} = 400^2$. Nevertheless, due to the ill-posed nature of the inverse problem \eqref{eqn:vis_discrete_2D_sphere}, stronger data fidelity does not automatically translate to better image-domain reconstructions. Therefore, while R2D2 can ensure enhanced data fidelity, S-R2D2’s scheme, which effectively integrates $\mathbf{\Gamma}$ and $\mathbf{\Gamma}^{\dagger}$, goes further by learning more effective regularisers, leading to superior image-domain reconstructions.\\

\noindent
Furthermore, although better overall performance is achieved at $\mathrm{N}_{\mathrm{p}} = 600^2$, the results indicate that S-R2D2 still performs satisfactory high-quality reconstructions at $\mathrm{N}_{\mathrm{p}} = \mathrm{N}_{\mathrm{p}}^{\min} = 400^2$, which is the resolution that maintains the same pixel size on the sphere and the plane. Therefore, despite the reduced accuracy of the interpolators at $\mathrm{N}_{\mathrm{p}}^{\min} $, working at this resolution on the plane grants S-R2D2 greater flexibility in managing computational constraints, ultimately facilitating higher-resolution reconstructions on the sphere.

\begin{figure*}
 \centering
 \setlength\tabcolsep{4pt}
 \begin{tabular}{c ccc}

 \multirow{2}{*}{\begin{minipage}{0.21\linewidth}
     \vspace{-2cm} 
     \hspace{-0.4cm}
     \centering
     \includegraphics[width=\linewidth]{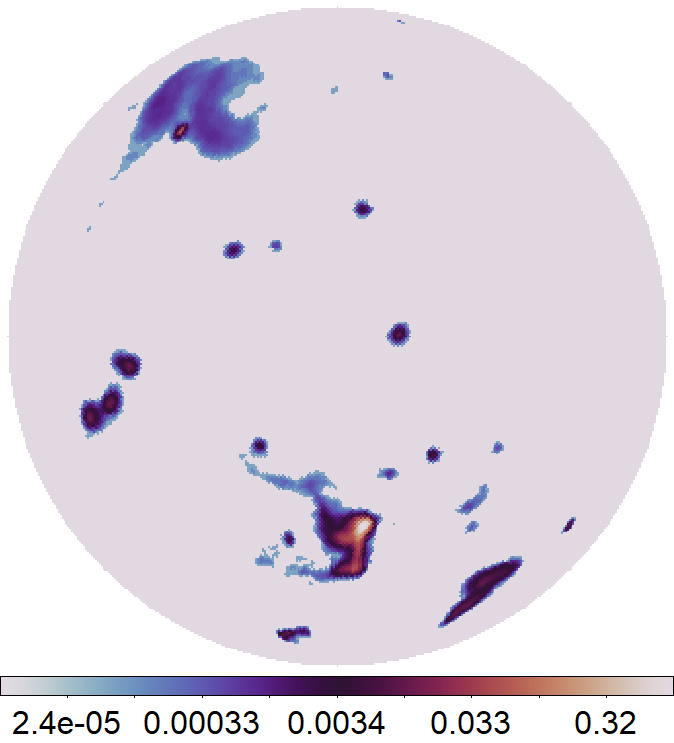} \\
     Ground Truth; $\mathbf{x}_{\mathrm{s}}^{\star}$
 \end{minipage}} &
 \hspace{0.8cm}
 \includegraphics[width=0.21\linewidth]{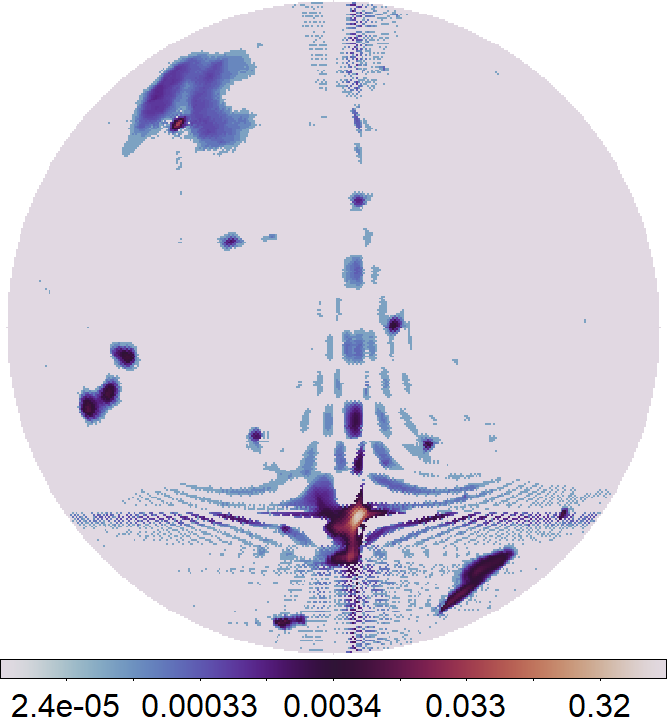} &
 \includegraphics[width=0.21\linewidth]{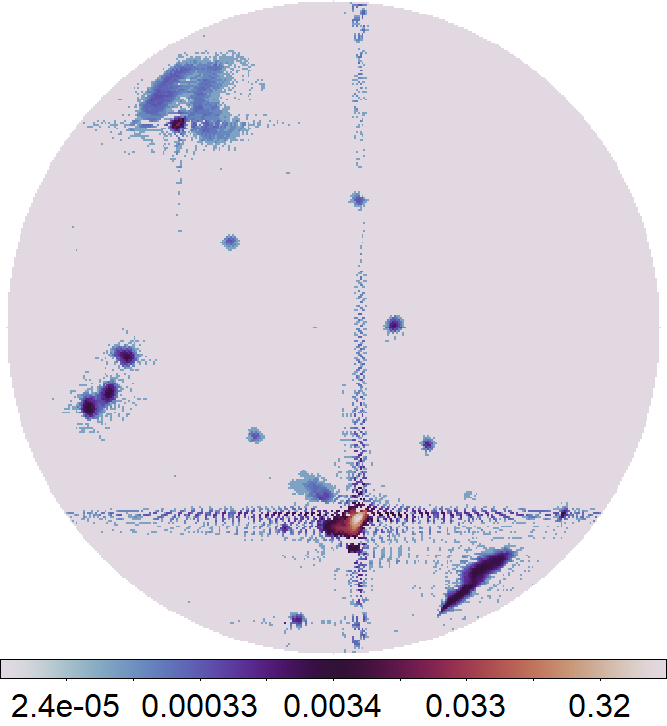} &
 \includegraphics[width=0.21\linewidth]{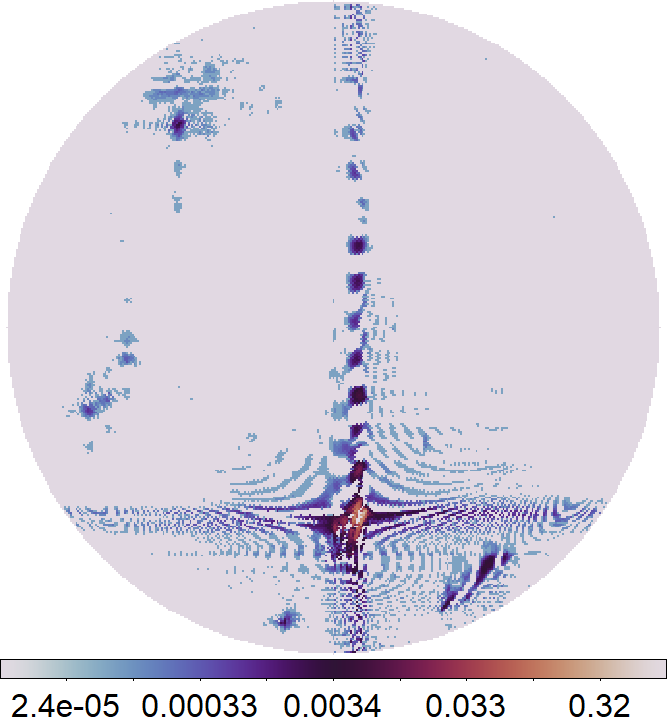} \\
 
  & \hspace{0.8cm} R2D2; $\mathrm{N}_{\mathrm{p}}=400^2$; (6.3, 2.3)~dB & R2D2; $\mathrm{N}_{\mathrm{p}}=600^2$; (4.7, 3.2)~dB & R2D2; $\mathrm{N}_{\mathrm{p}}=800^2$; (1.0, -0.2)~dB \\
\addlinespace[4pt]
 & \hspace{0.8cm}
 \includegraphics[width=0.21\linewidth]{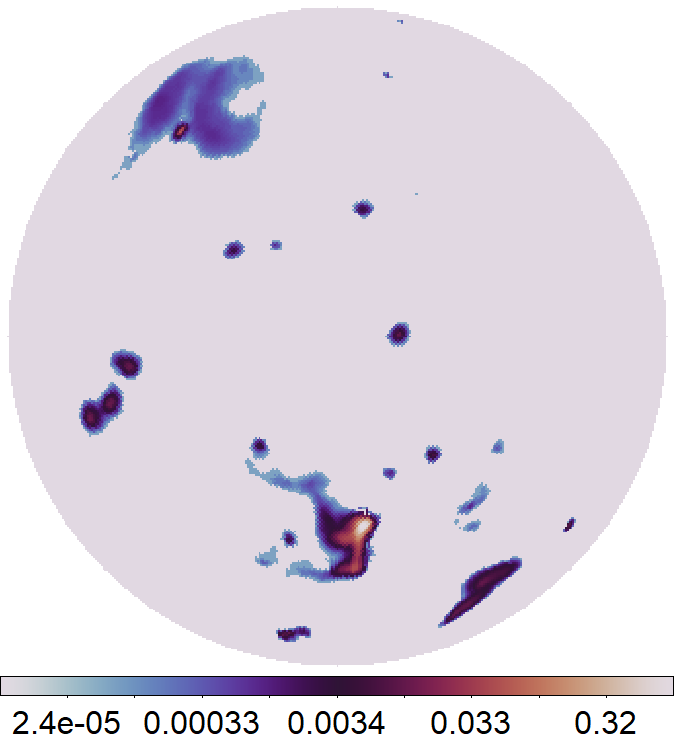} &
 \includegraphics[width=0.21\linewidth]{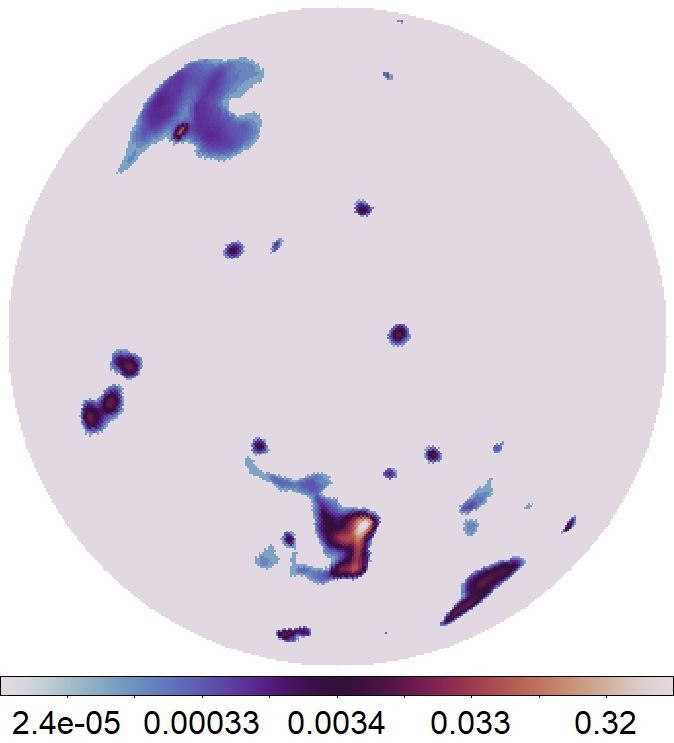} &
 \includegraphics[width=0.21\linewidth]{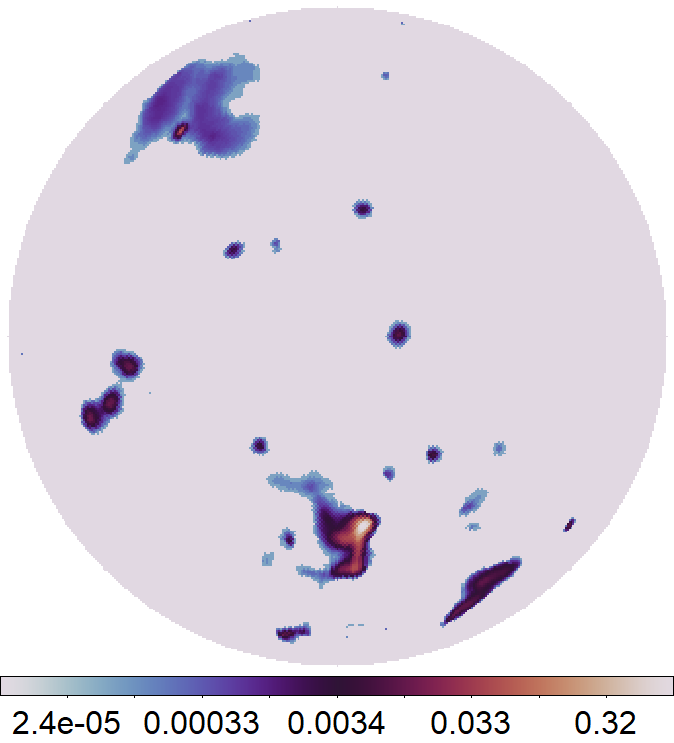} \\
  & \hspace{0.8cm} S-R2D2; $\mathrm{N}_{\mathrm{p}}=400^2$; (24.3, 17.1)~dB & S-R2D2; $\mathrm{N}_{\mathrm{p}}=600^2$; (21.9, 16.9)~dB & S-R2D2; $\mathrm{N}_{\mathrm{p}}=800^2$; (22.5, 15.2)~dB\\
\addlinespace[16pt]

 \multirow{2}{*}{\begin{minipage}{0.21\linewidth}
     \vspace{-2cm}
     \hspace{-0.4cm}
     \centering
     \includegraphics[width=\linewidth]{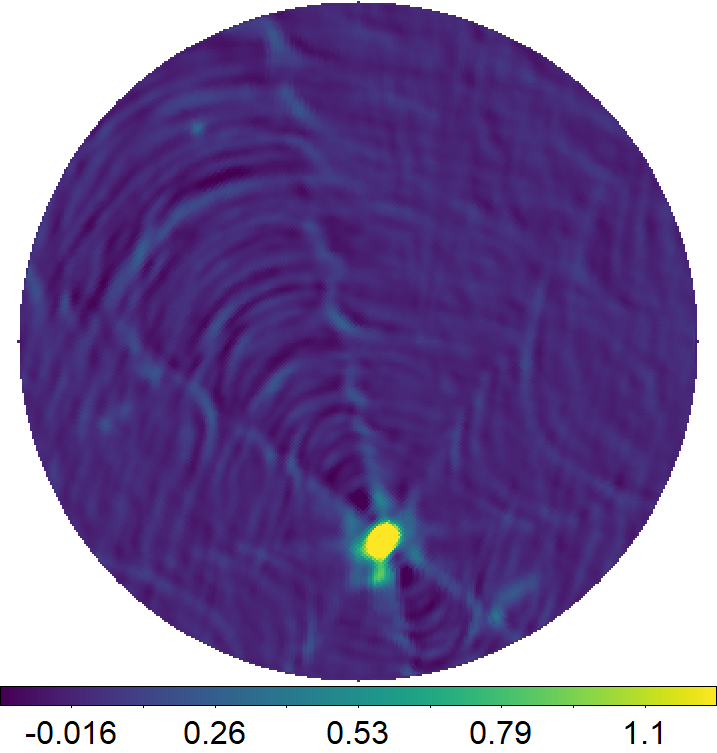} \\
     Spherical Dirty; $\mathbf{x}_s^{\mathrm{d}}$
 \end{minipage}} &
\hspace{0.8cm}
 \includegraphics[width=0.21\linewidth]{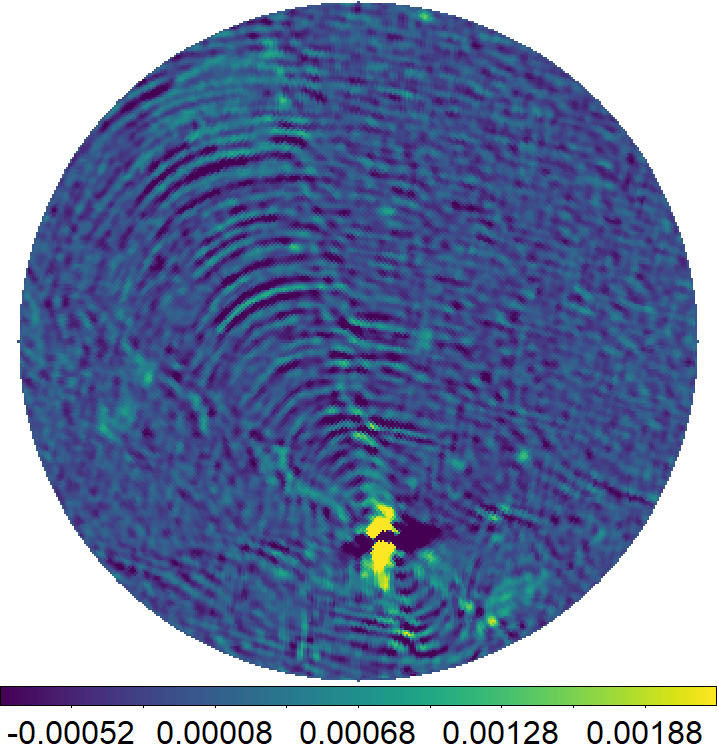} &
 \includegraphics[width=0.21\linewidth]{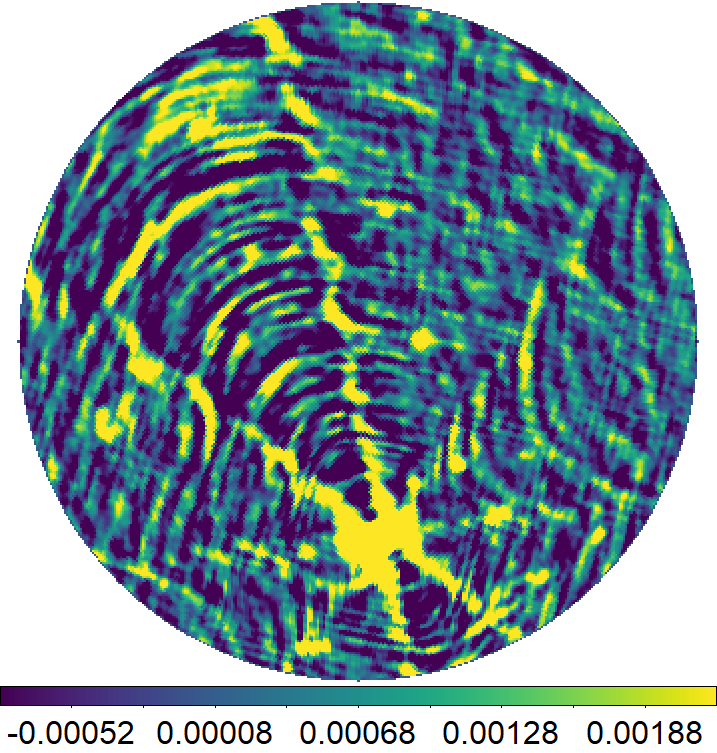} &
 \includegraphics[width=0.21\linewidth]{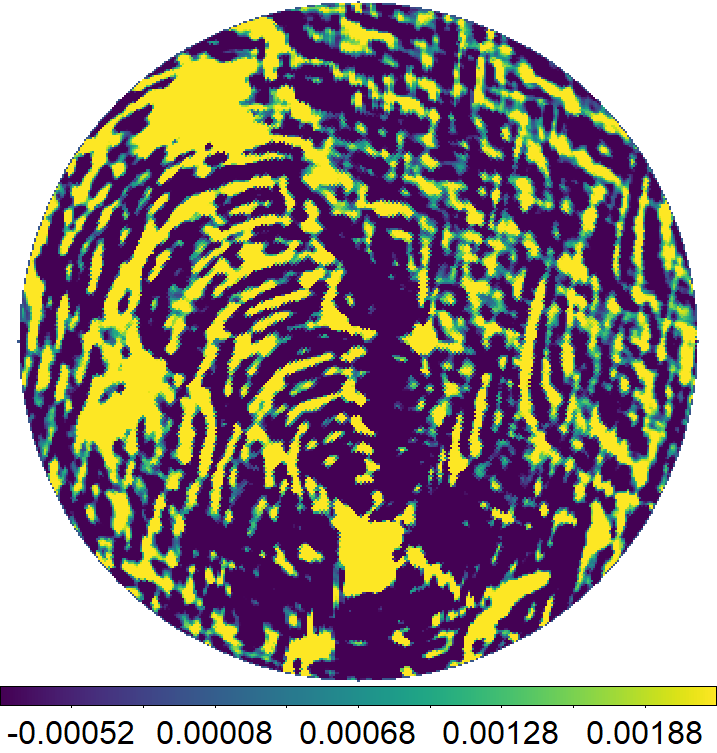} \\
 & \hspace{0.8cm} R2D2; $\mathrm{N}_{\mathrm{p}}=400^2$; $3.0\!\times\!10^{-3}$ & R2D2; $\mathrm{N}_{\mathrm{p}}=600^2$; $2.3\!\times\!10^{-2}$ & R2D2; $\mathrm{N}_{\mathrm{p}}=800^2$; $7.0\!\times\!10^{-2}$\\
\addlinespace[4pt]
 & \hspace{0.8cm}
 \includegraphics[width=0.21\linewidth]{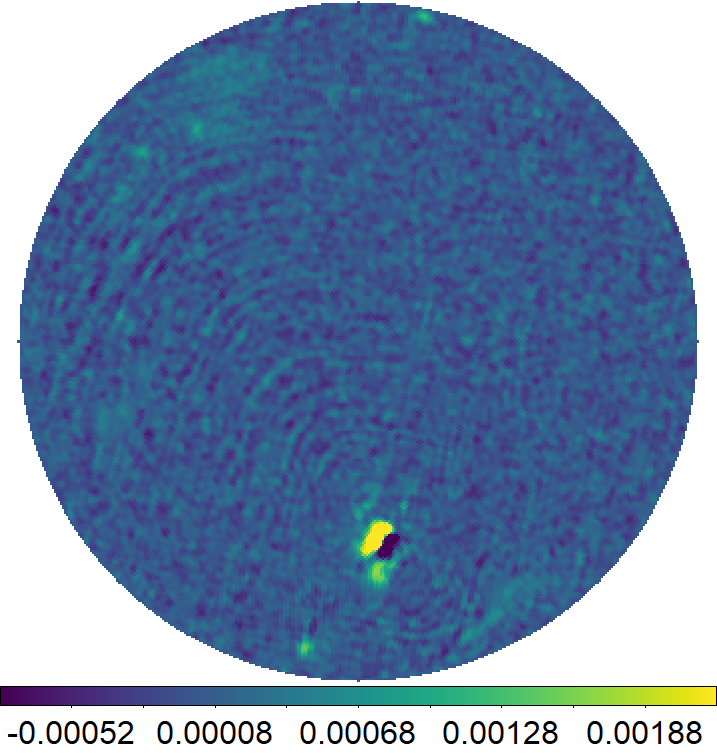} &
 \includegraphics[width=0.21\linewidth]{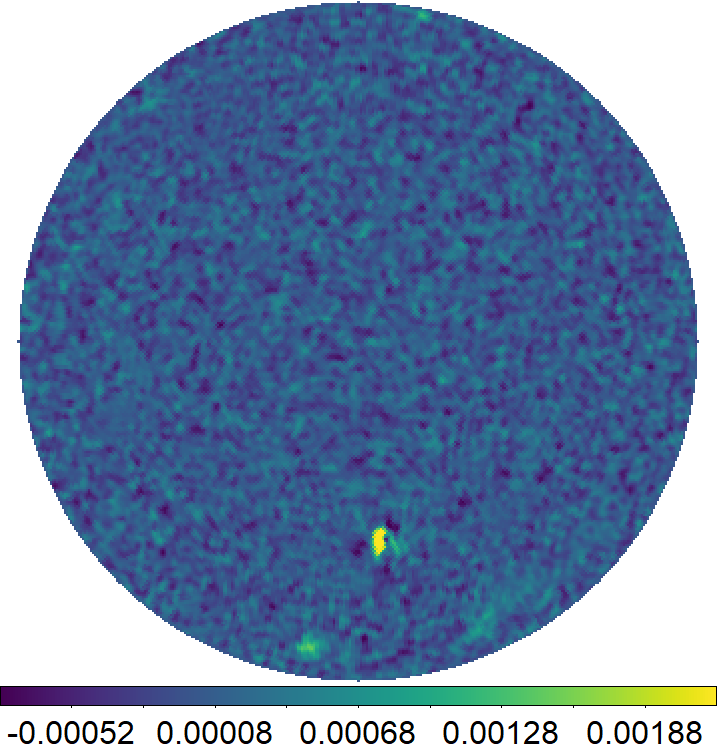} &
 \includegraphics[width=0.21\linewidth]{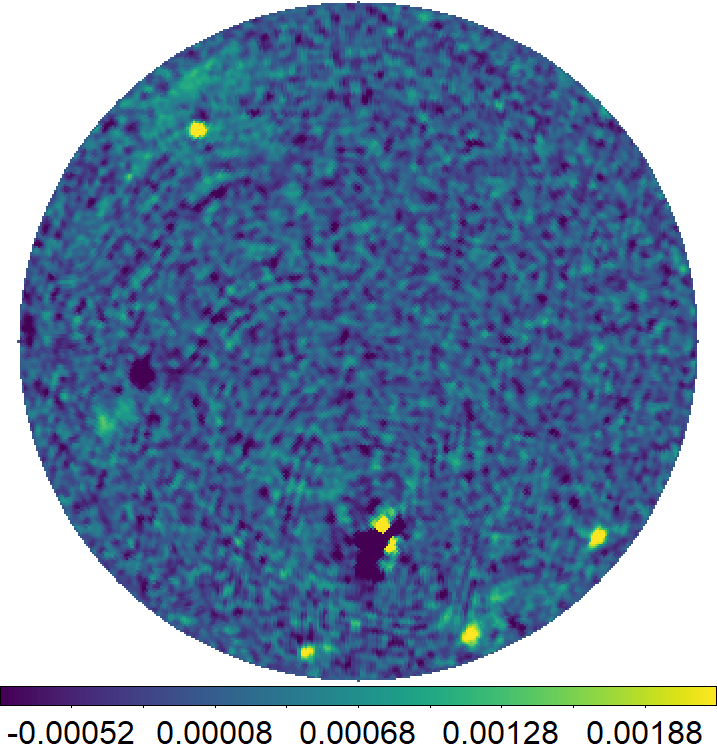} \\
 & \hspace{0.8cm} S-R2D2; $\mathrm{N}_{\mathrm{p}}=400^2$; $1.4\!\times\!10^{-3}$ & S-R2D2; $\mathrm{N}_{\mathrm{p}}=600^2$; $1.3\!\times\!10^{-3}$ & S-R2D2; $\mathrm{N}_{\mathrm{p}}=800^2$; $2.5\!\times\!10^{-3}$ \\
 \end{tabular}
 \caption{Visual final reconstructions of the spherical ground truth $\mathbf{x}_{\mathrm{s}}^{\star}$, displayed in the top-left corner, obtained with the S-R2D2 Algorithm~\ref{algo:S_R2D2_reconstruction} or the R2D2 Algorithm~\ref{algo:R2D2_reconstruction} (adapted following the procedure depicted in Section~\ref{subsec:sec_5_subsec_1}) for different resolutions on the plane $\mathrm{N}_{\mathrm{p}}$. The spherical ground truth was generated with a $\textrm{DR}=8.9\!\times\!10^4$ from the planar galaxy cluster PSZ2~G165.68+44.01 \citep{botteon2022} image following the procedure depicted in Section~\ref{subsec:sec_4_subsec_1}. From this spherical ground truth, we then generate its corresponding spherical dirty signal (Section~\ref{subsec:sec_4_subsec_2}). The two top row show the estimated signals displayed in logarithmic scale, with the logarithmic exponent equals to $\textrm{DR}$. Values of ($\textrm{SNR}$, $\textrm{logSNR}$) metrics are reported below each estimated image. The two bottom rows show the dirty signal and the residuals displayed on the sphere using $\mathbf{\Gamma}^{\dagger}$ and in linear scale. The value of the metric $\textrm{RDR}$ is indicated below each residuals. For all spherical signals, we visualise the Northern hemisphere in the orthographic projection perspective.} 
 \label{fig:P138}
\end{figure*}

\begin{figure*}
 \centering
 \setlength\tabcolsep{4pt}
 \begin{tabular}{c ccc}

 \multirow{2}{*}{\begin{minipage}{0.21\linewidth}
     \vspace{-2cm} 
     \hspace{-0.4cm}
     \centering
     \includegraphics[width=\linewidth]{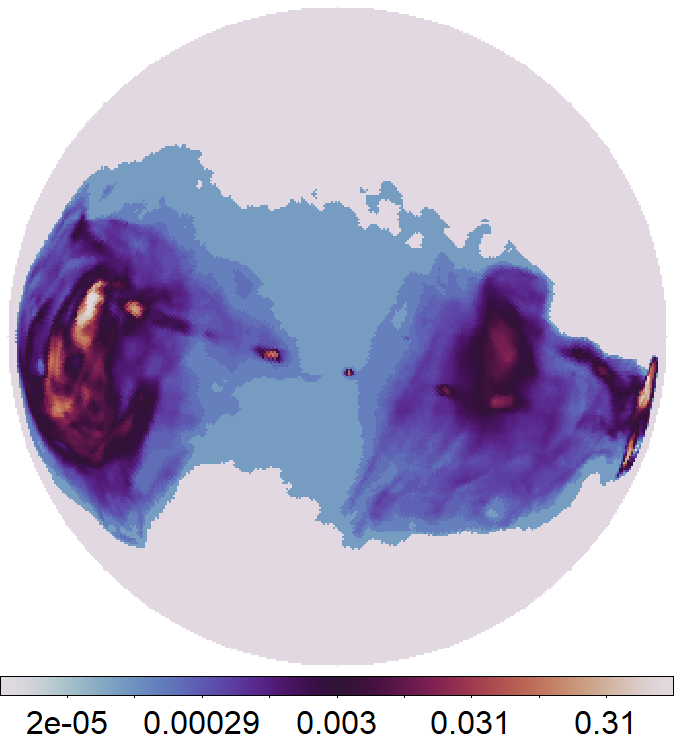} \\
     Ground Truth; $\mathbf{x}_{\mathrm{s}}^{\star}$
 \end{minipage}} &
 \hspace{0.8cm}
 \includegraphics[width=0.21\linewidth]{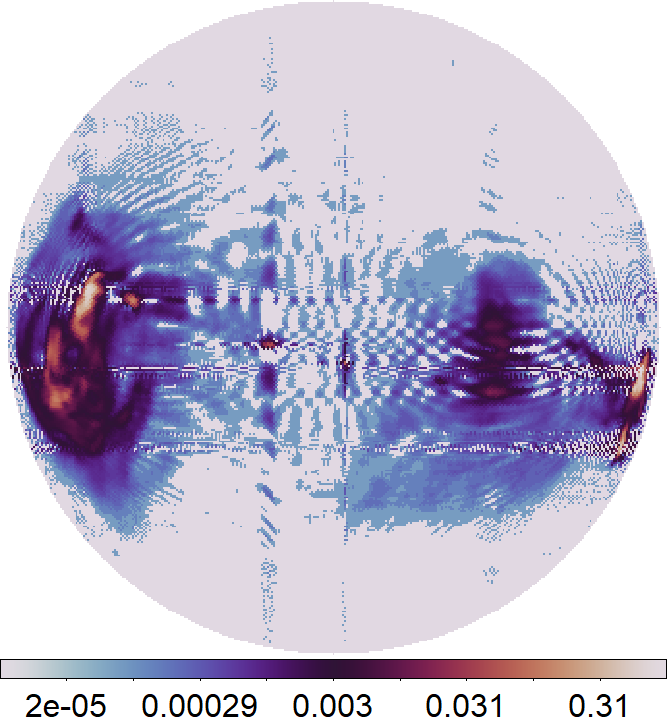} &
 \includegraphics[width=0.21\linewidth]{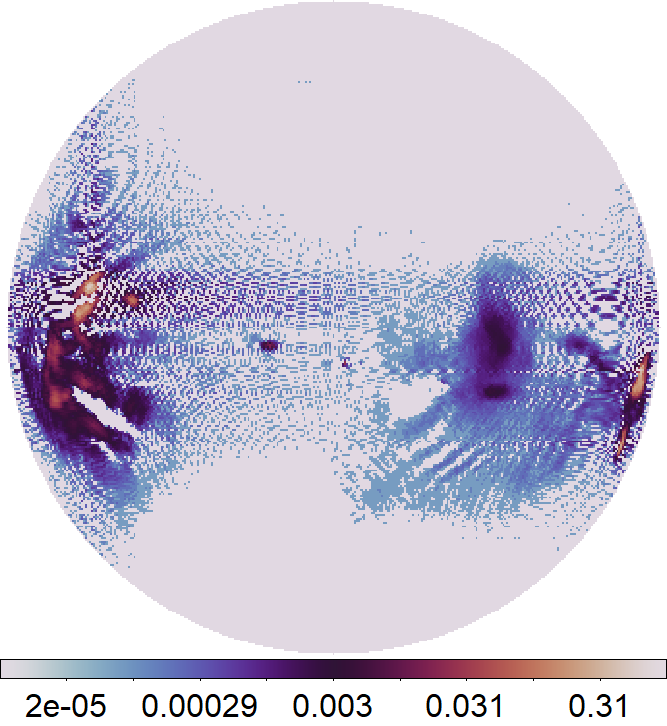} &
 \includegraphics[width=0.21\linewidth]{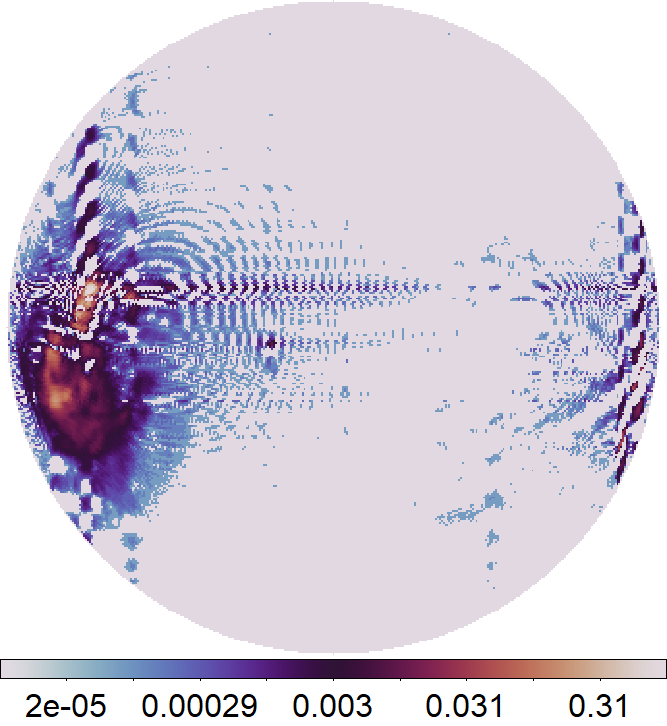} \\
 & \hspace{0.8cm} R2D2; $\mathrm{N}_{\mathrm{p}}=400^2$; (8.4, 7.9)~dB & R2D2; $\mathrm{N}_{\mathrm{p}}=600^2$; (2.9, 5.0)~dB & R2D2; $\mathrm{N}_{\mathrm{p}}=800^2$; (2.5, 2.5)~dB \\
\addlinespace[4pt]
 & \hspace{0.8cm}
 \includegraphics[width=0.21\linewidth]{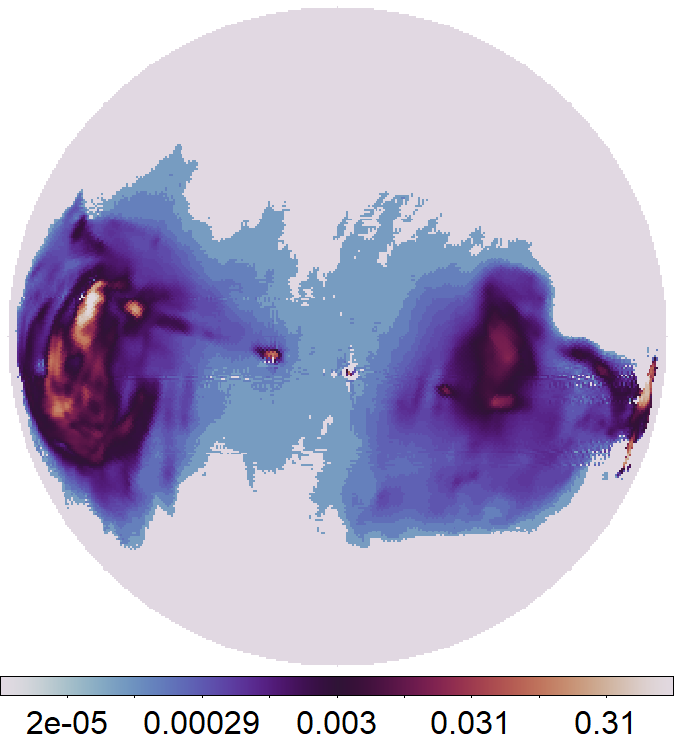} &
 \includegraphics[width=0.21\linewidth]{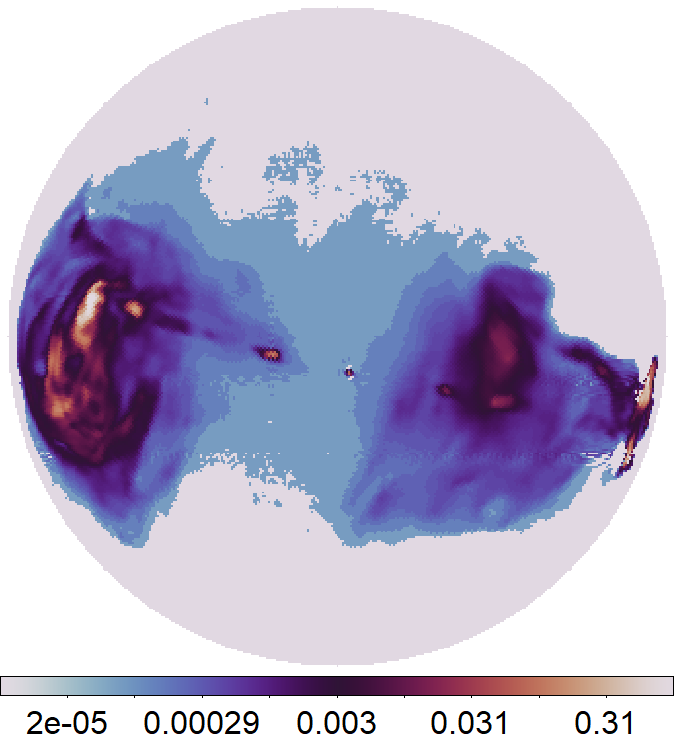} &
 \includegraphics[width=0.21\linewidth]{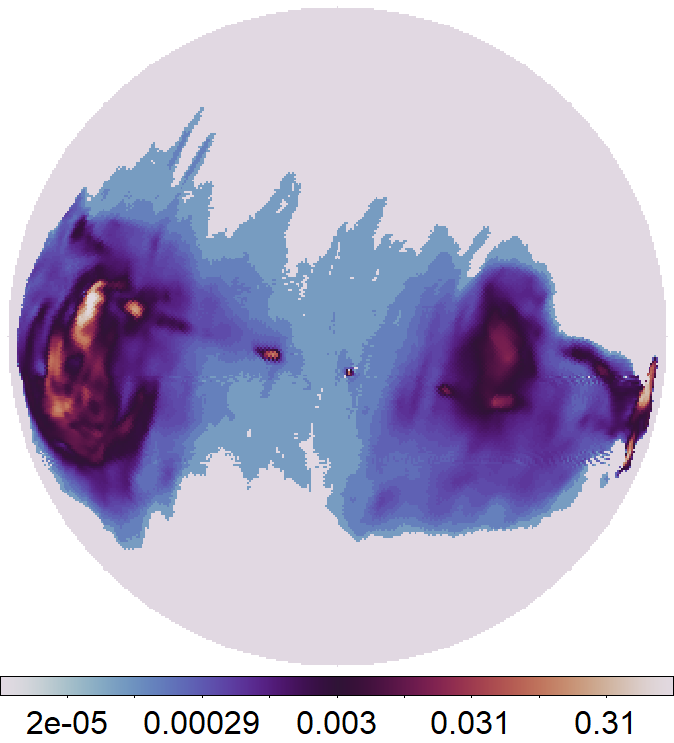} \\
 & \hspace{0.8cm} S-R2D2; $\mathrm{N}_{\mathrm{p}}=400^2$; (18.6, 14.2)~dB & S-R2D2; $\mathrm{N}_{\mathrm{p}}=600^2$; (17.6, 18.5)~dB & S-R2D2; $\mathrm{N}_{\mathrm{p}}=800^2$; (19.2, 15.1)~dB\\
\addlinespace[16pt]
 \multirow{2}{*}{\begin{minipage}{0.21\linewidth}
     \vspace{-2cm} 
     \hspace{-0.4cm}
     \centering
     \includegraphics[width=\linewidth]{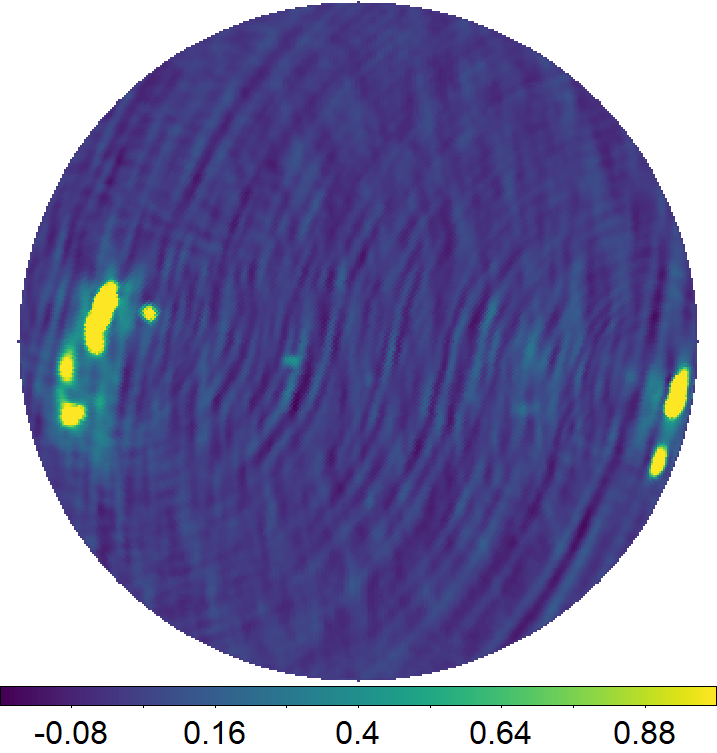} \\
     Spherical Dirty; $\mathbf{x}_s^{\mathrm{d}}$
 \end{minipage}} &
 \hspace{0.8cm}
 \includegraphics[width=0.21\linewidth]{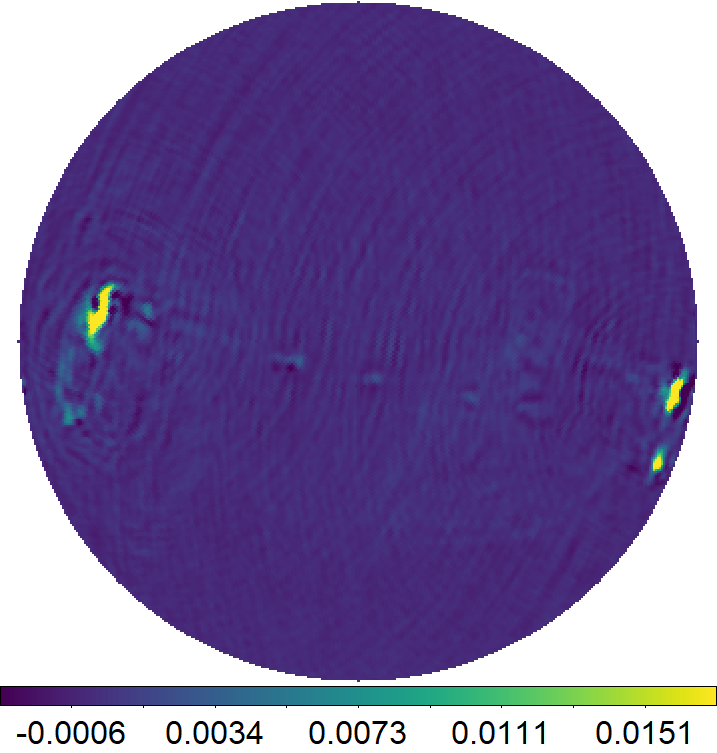} &
 \includegraphics[width=0.21\linewidth]{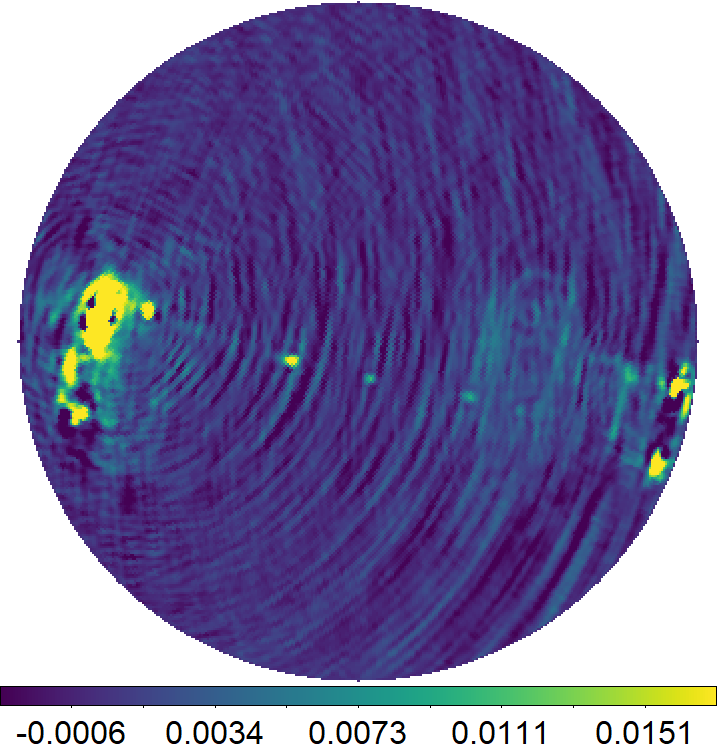} &
 \includegraphics[width=0.21\linewidth]{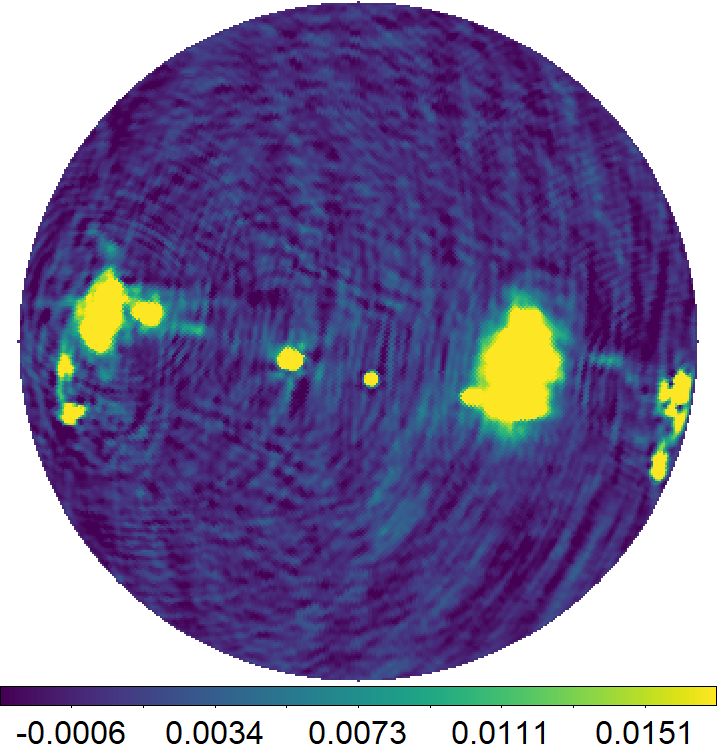} \\
 & \hspace{0.8cm} R2D2; $\mathrm{N}_{\mathrm{p}}=400^2$; $2.0\!\times\!10^{-2}$ & R2D2; $\mathrm{N}_{\mathrm{p}}=600^2$; $1.9\!\times\!10^{-2}$ & R2D2; $\mathrm{N}_{\mathrm{p}}=800^2$; $2.7\!\times\!10^{-2}$\\
\addlinespace[4pt]
 & \hspace{0.8cm}
 \includegraphics[width=0.21\linewidth]{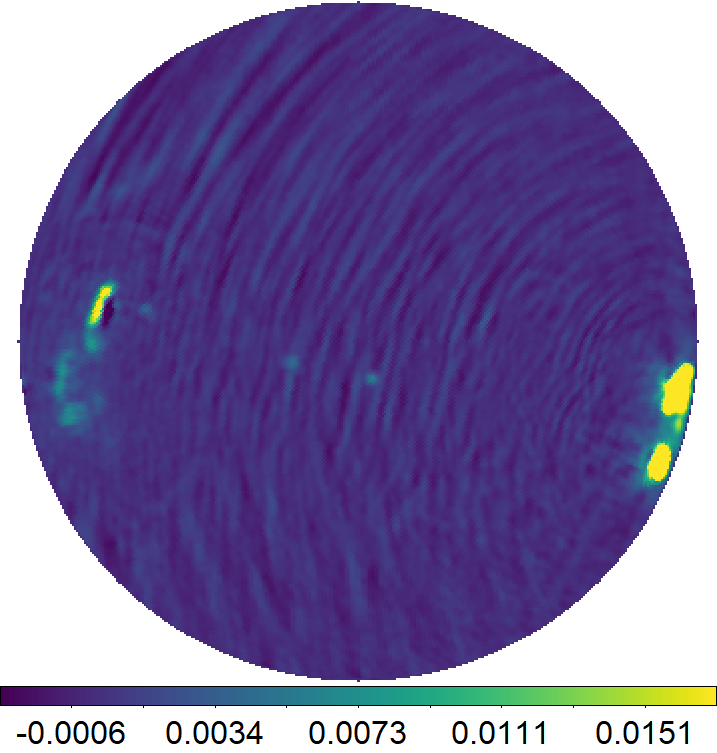} &
 \includegraphics[width=0.21\linewidth]{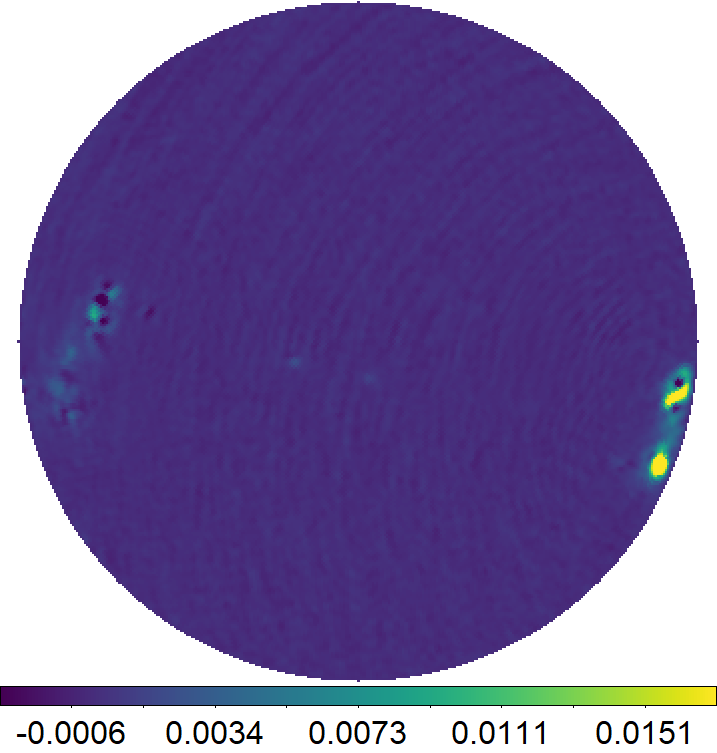} &
 \includegraphics[width=0.21\linewidth]{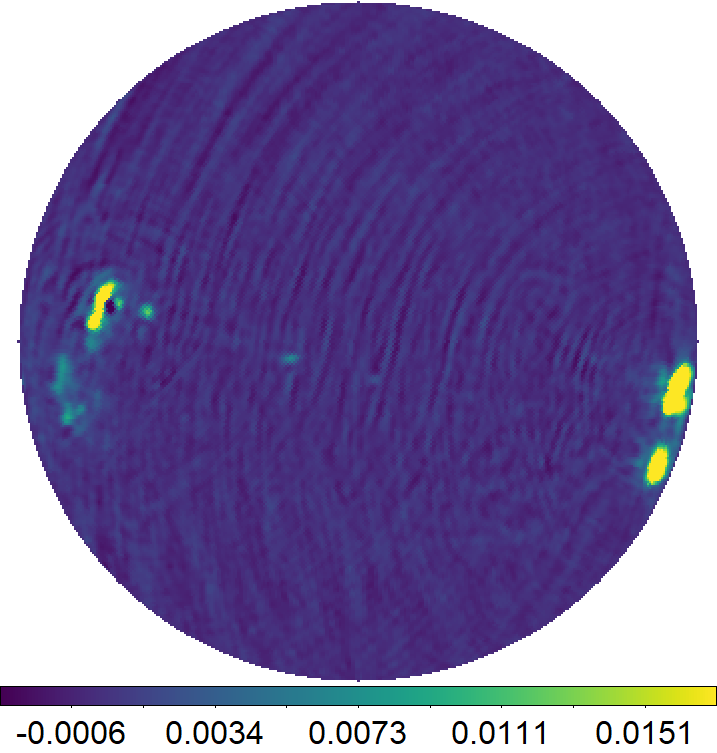} \\
 & \hspace{0.8cm} S-R2D2; $\mathrm{N}_{\mathrm{p}}=400^2$; $1.6\!\times\!10^{-2}$ & S-R2D2; $\mathrm{N}_{\mathrm{p}}=600^2$; $4.0\!\times\!10^{-3}$ & S-R2D2; $\mathrm{N}_{\mathrm{p}}=800^2$; $1.1\!\times\!10^{-2}$ \\
 \end{tabular}
 \caption{Visual final reconstructions of the spherical ground truth $\mathbf{x}_{\mathrm{s}}^{\star}$, displayed in the top-left corner, obtained with the S-R2D2 Algorithm~\ref{algo:S_R2D2_reconstruction} or the R2D2 Algorithm~\ref{algo:R2D2_reconstruction} (adapted following the procedure depicted in Section~\ref{subsec:sec_5_subsec_1}) for different resolutions on the plane $\mathrm{N}_{\mathrm{p}}$. The spherical ground truth was generated with a $\textrm{DR}=1.1\!\times\!10^5$ from the radio galaxy 3c354 (NRAO Archives) image following the procedure depicted in Section~\ref{subsec:sec_4_subsec_1}. From this spherical ground truth, we then generate its corresponding spherical dirty signal (Section~\ref{subsec:sec_4_subsec_2}). The two top row show the estimated signals displayed in logarithmic scale, with the logarithmic exponent equals to $\textrm{DR}$. Values of ($\textrm{SNR}$, $\textrm{logSNR}$) metrics are reported below each estimated image. The two bottom rows show the dirty signal and the residuals displayed on the sphere using $\mathbf{\Gamma}^{\dagger}$ and in linear scale. The value of the metric $\textrm{RDR}$ is indicated below each residuals. For all spherical signals, we visualise the Northern hemisphere in the orthographic projection perspective.}  
 \label{fig:3c354}
\end{figure*}

\begin{figure*}
 \centering
 \setlength\tabcolsep{4pt}
 \begin{tabular}{c ccc}

 \multirow{2}{*}{\begin{minipage}{0.21\linewidth}
     \vspace{-2cm}
     \hspace{-0.4cm}
     \centering
     \includegraphics[width=\linewidth]{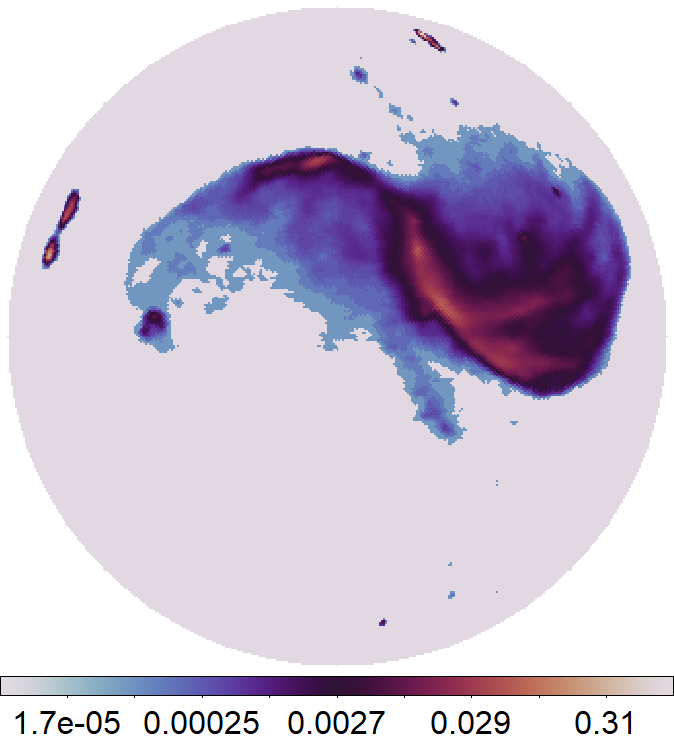} \\
     Ground Truth; $\mathbf{x}_{\mathrm{s}}^{\star}$
 \end{minipage}} & 
 \hspace{0.8cm}
 \includegraphics[width=0.21\linewidth]{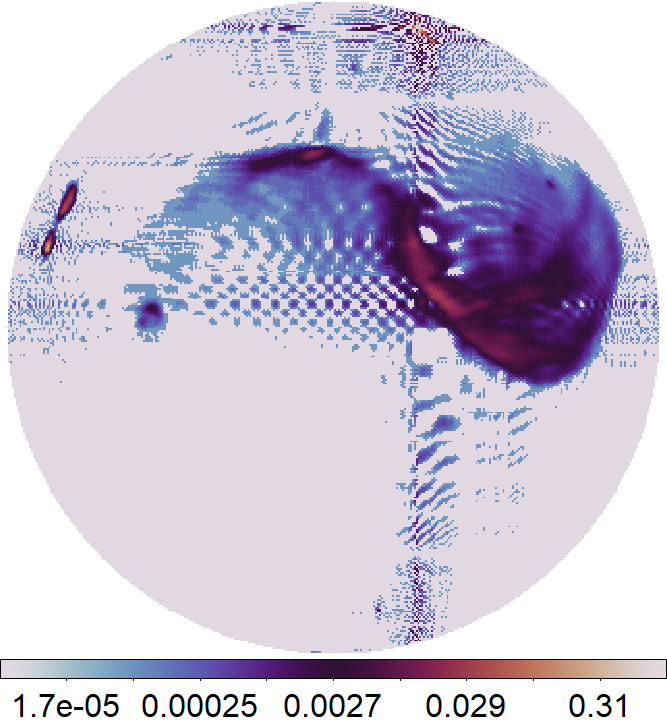} &
 \includegraphics[width=0.21\linewidth]{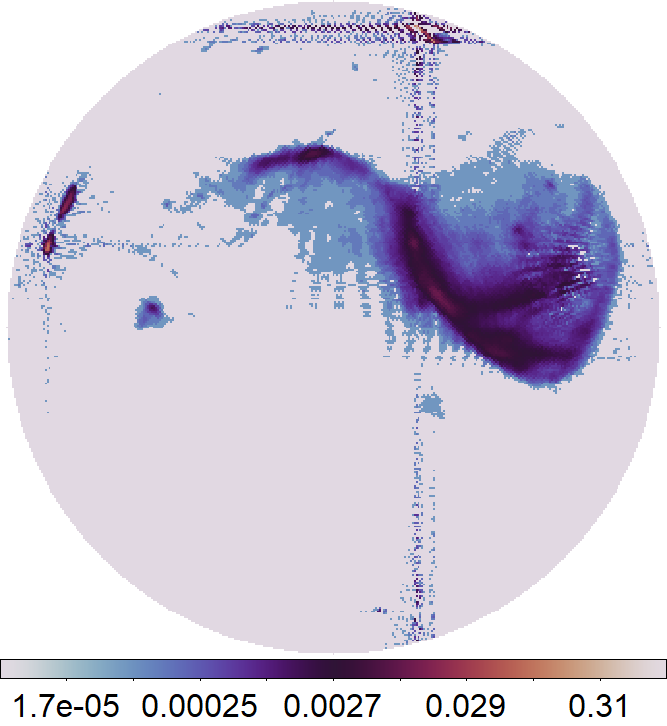} &
 \includegraphics[width=0.21\linewidth]{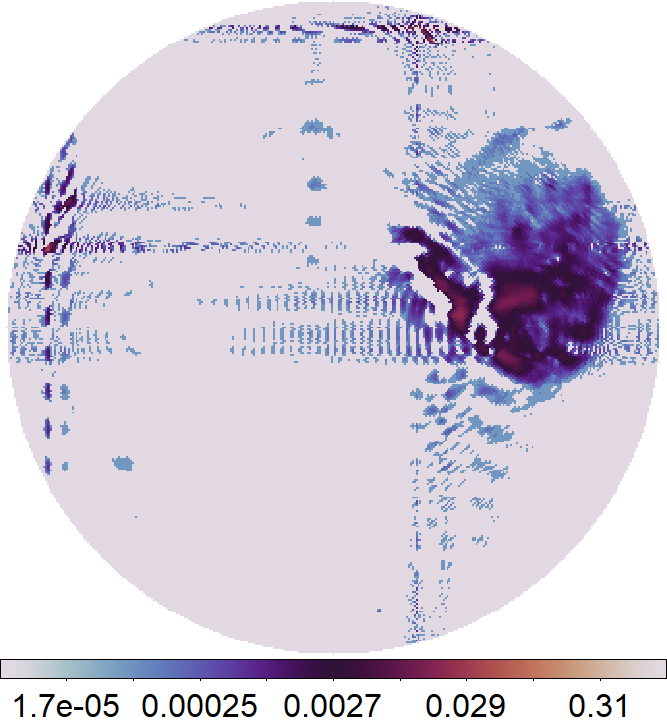} \\

 & \hspace{0.8cm} R2D2; $\mathrm{N}_{\mathrm{p}}=400^2$; (6.0, 5.2)~dB & R2D2; $\mathrm{N}_{\mathrm{p}}=600^2$; (2.9, 6.0)~dB & R2D2; $\mathrm{N}_{\mathrm{p}}=800^2$; (0.3, 2.4)~dB \\
\addlinespace[4pt]

 & \hspace{0.8cm}
 \includegraphics[width=0.21\linewidth]{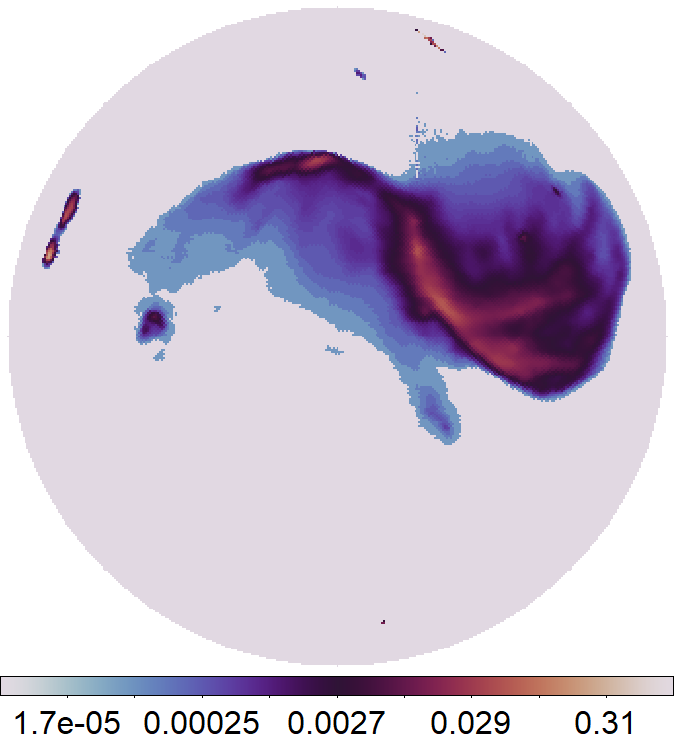} &
 \includegraphics[width=0.21\linewidth]{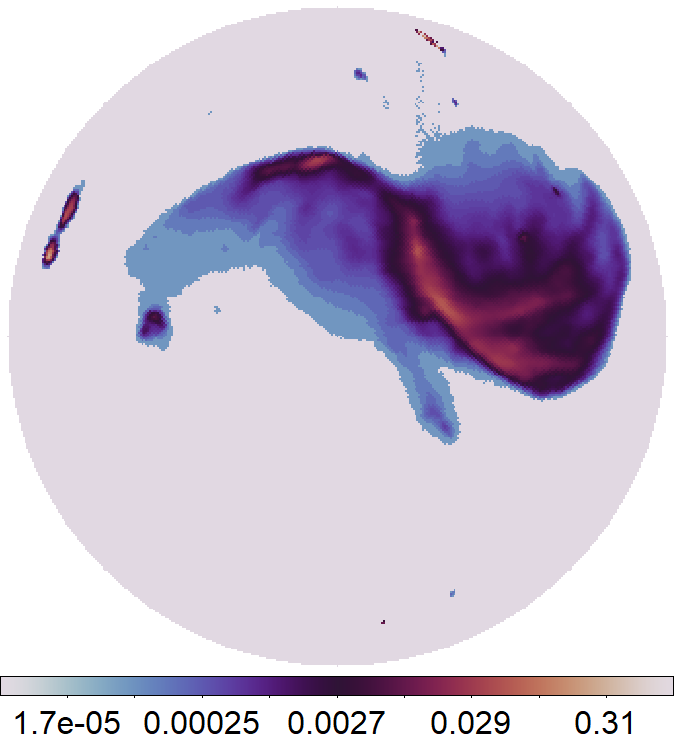} &
 \includegraphics[width=0.21\linewidth]{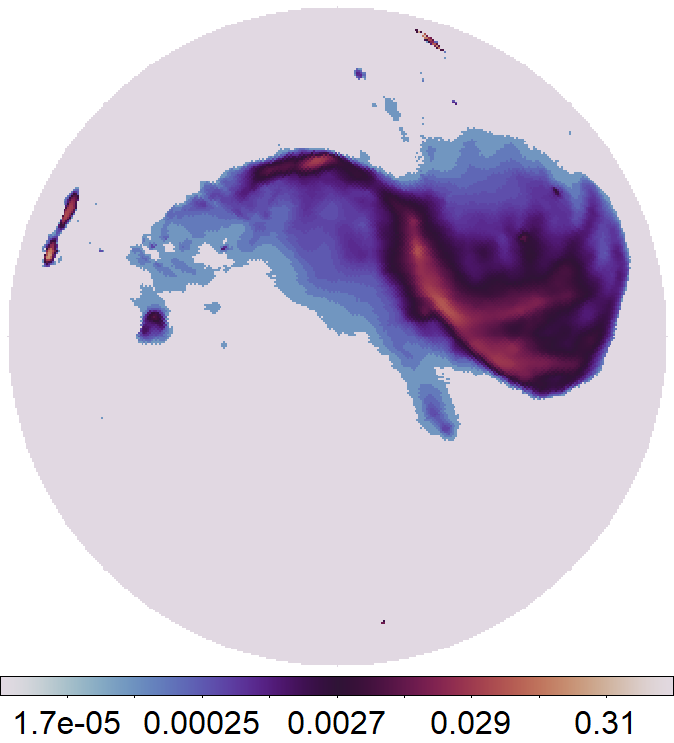} \\
 
 & \hspace{0.8cm} S-R2D2; $\mathrm{N}_{\mathrm{p}}=400^2$; (17.3, 16.7)~dB & S-R2D2; $\mathrm{N}_{\mathrm{p}}=600^2$; (14.8, 18.9)~dB & S-R2D2; $\mathrm{N}_{\mathrm{p}}=800^2$; (16.2,17.3)~dB\\
\addlinespace[16pt]

 \multirow{2}{*}{\begin{minipage}{0.21\linewidth}
     \vspace{-2cm}
     \hspace{-0.4cm}
     \centering
     \includegraphics[width=\linewidth]{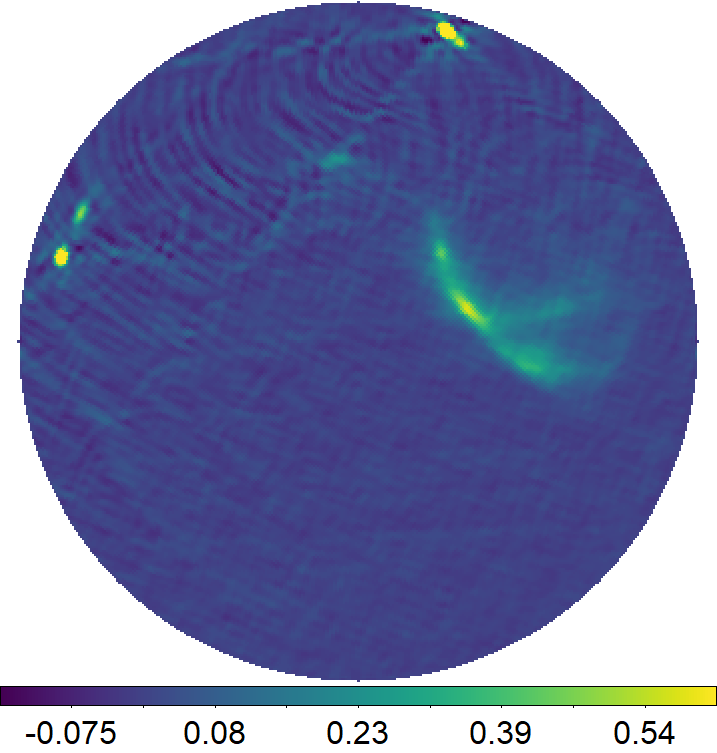} \\
     Spherical Dirty; $\mathbf{x}_s^{\mathrm{d}}$
 \end{minipage}} & \hspace{0.8cm}
 \includegraphics[width=0.21\linewidth]{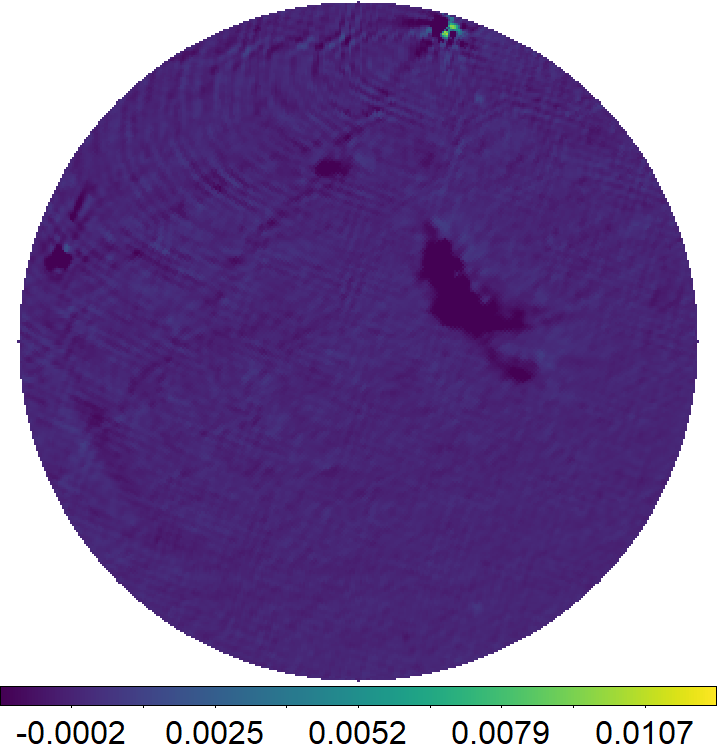} &
 \includegraphics[width=0.21\linewidth]{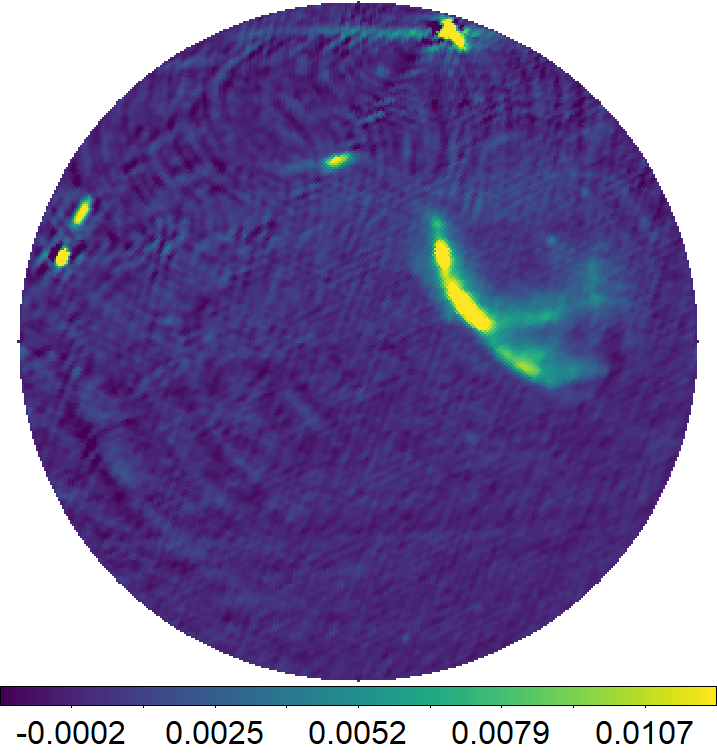} &
 \includegraphics[width=0.21\linewidth]{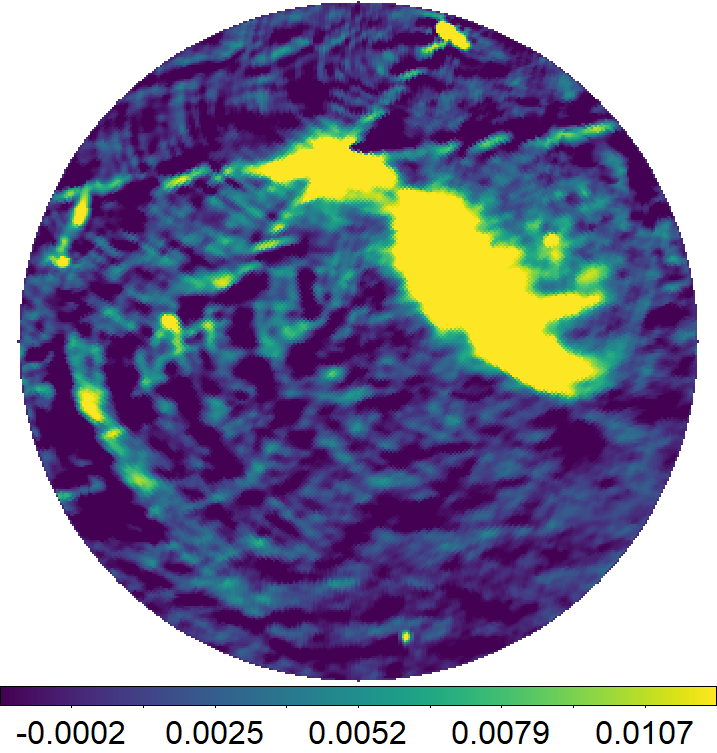} \\

 & \hspace{0.8cm} R2D2; $\mathrm{N}_{\mathrm{p}}=400^2$; $7.9\!\times\!10^{-3}$ & R2D2; $\mathrm{N}_{\mathrm{p}}=600^2$; $2.5\!\times\!10^{-2}$ & R2D2; $\mathrm{N}_{\mathrm{p}}=800^2$; $1.6\!\times\!10^{-1}$\\
\addlinespace[4pt]

 & \hspace{0.8cm}
 \includegraphics[width=0.21\linewidth]{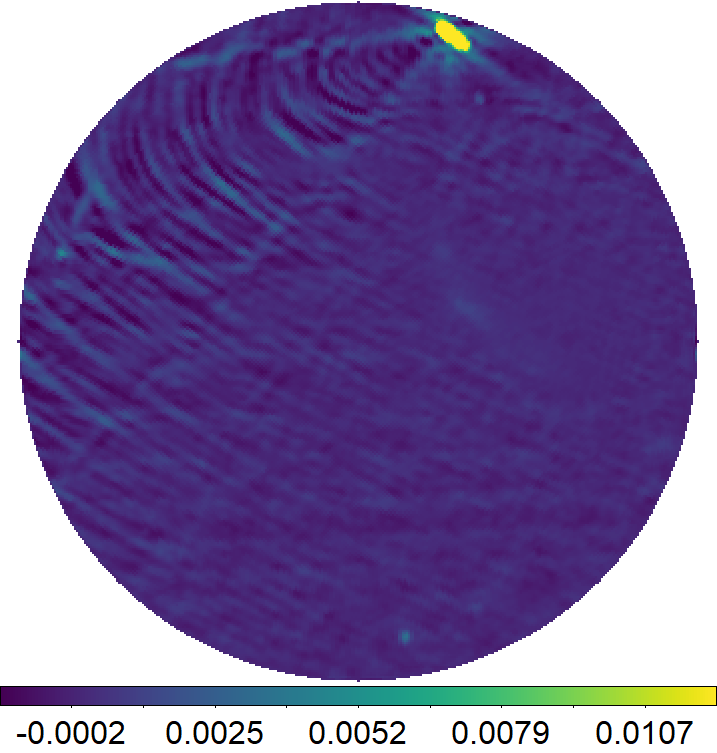} &
 \includegraphics[width=0.21\linewidth]{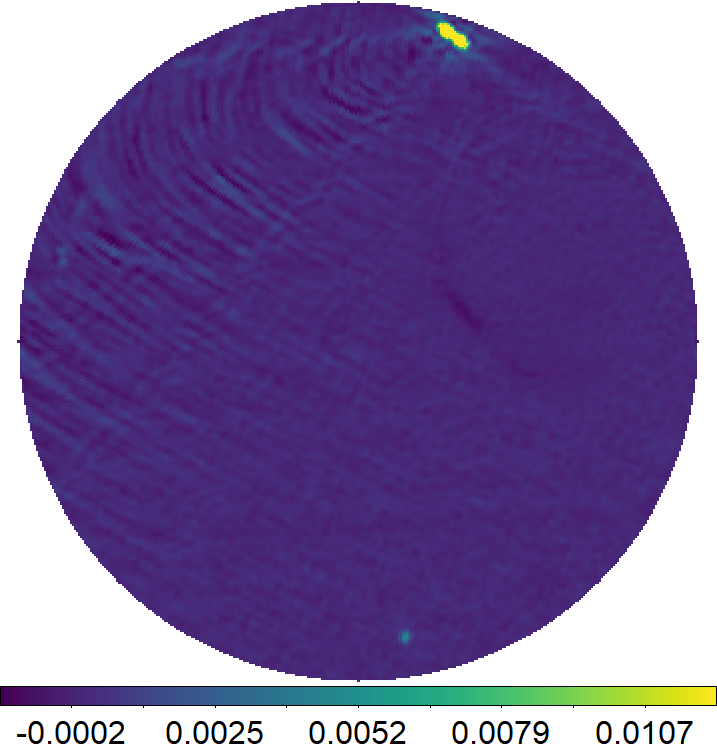} &
 \includegraphics[width=0.21\linewidth]{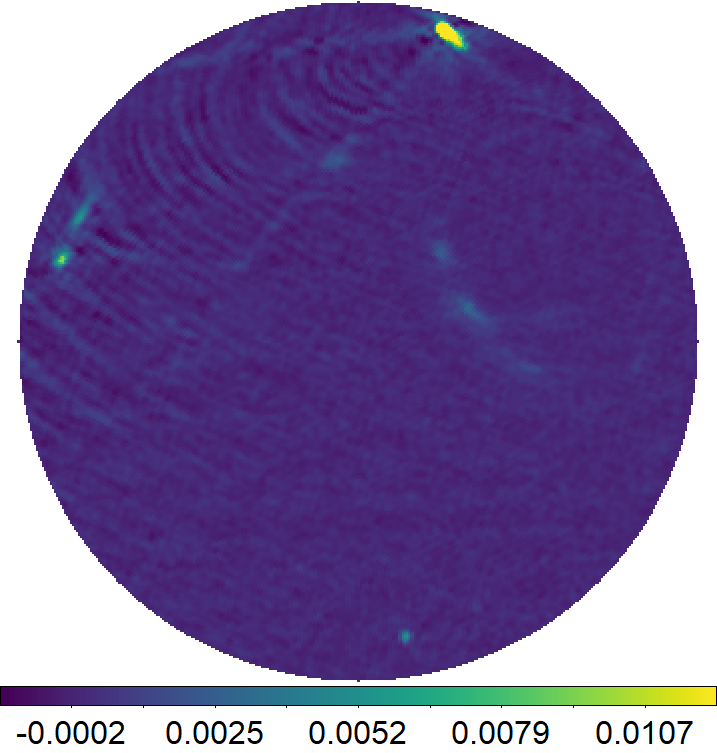} \\

 & \hspace{0.8cm} S-R2D2; $\mathrm{N}_{\mathrm{p}}=400^2$; $2.8\!\times\!10^{-2}$ & S-R2D2; $\mathrm{N}_{\mathrm{p}}=600^2$; $1.2\!\times\!10^{-2}$ & S-R2D2; $\mathrm{N}_{\mathrm{p}}=800^2$; $1.4\!\times\!10^{-2}$ \\
 \end{tabular}
 \caption{Visual final reconstructions of the spherical ground truth $\mathbf{x}_{\mathrm{s}}^{\star}$, displayed in the top-left corner, obtained with the S-R2D2 Algorithm~\ref{algo:S_R2D2_reconstruction} or the R2D2 Algorithm~\ref{algo:R2D2_reconstruction} (adapted following the procedure depicted in Section~\ref{subsec:sec_5_subsec_1}) for different resolutions on the plane $\mathrm{N}_{\mathrm{p}}$. The spherical ground truth was generated with a $\textrm{DR}=1.4\!\times\!10^5$ from the radio galaxy Messier~106 \citep{shimwell2022} following the procedure depicted in Section~\ref{subsec:sec_4_subsec_1}. From this spherical ground truth, we then generate its corresponding spherical dirty signal (Section~\ref{subsec:sec_4_subsec_2}). The two top row show the estimated signals displayed in logarithmic scale, with the logarithmic exponent equals to $\textrm{DR}$. Values of ($\textrm{SNR}$, $\textrm{logSNR}$) metrics are reported below each estimated image. The two bottom rows show the dirty signal and the residuals displayed on the sphere using $\mathbf{\Gamma}^{\dagger}$ and in linear scale. The value of the metric $\textrm{RDR}$ is indicated below each residuals. For all spherical signals, we visualise the Northern hemisphere in the orthographic projection perspective.} 
 \label{fig:P23}
\end{figure*}

\begin{figure*}
 \centering
 \setlength\tabcolsep{4pt}
 \begin{tabular}{c ccc}

 \multirow{2}{*}{\begin{minipage}{0.21\linewidth}
     \vspace{-2cm} 
     \hspace{-0.4cm}
     \centering
     \includegraphics[width=\linewidth]{images/experiments/images/Reconstruction/Fourier/GDTH/PSZ.png} \\
     Ground Truth; $\mathbf{x}_{\mathrm{s}}^{\star}$
 \end{minipage}} & \hspace{0.8cm}
 \includegraphics[width=0.21\linewidth]{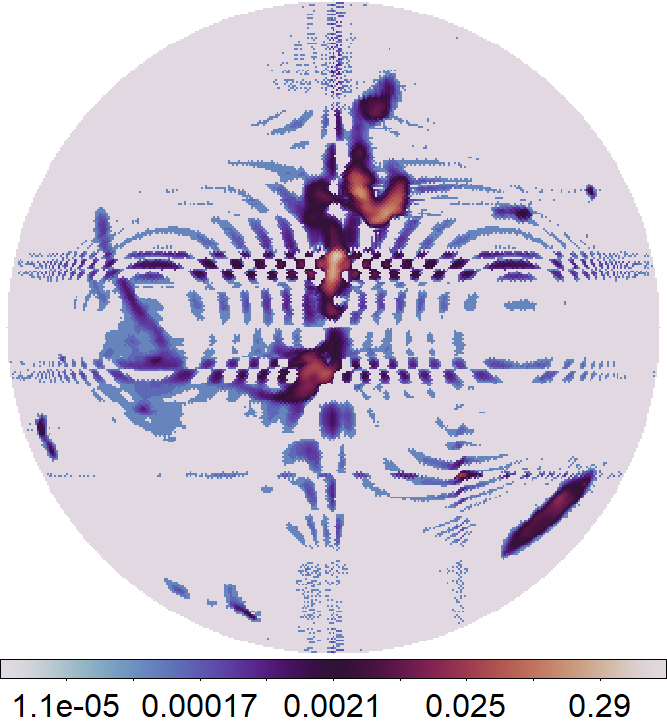} &
 \includegraphics[width=0.21\linewidth]{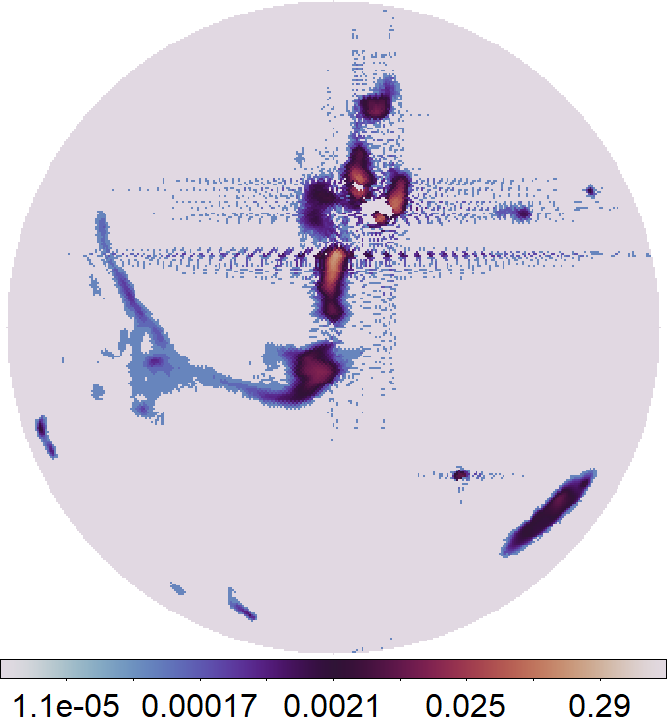} &
 \includegraphics[width=0.21\linewidth]{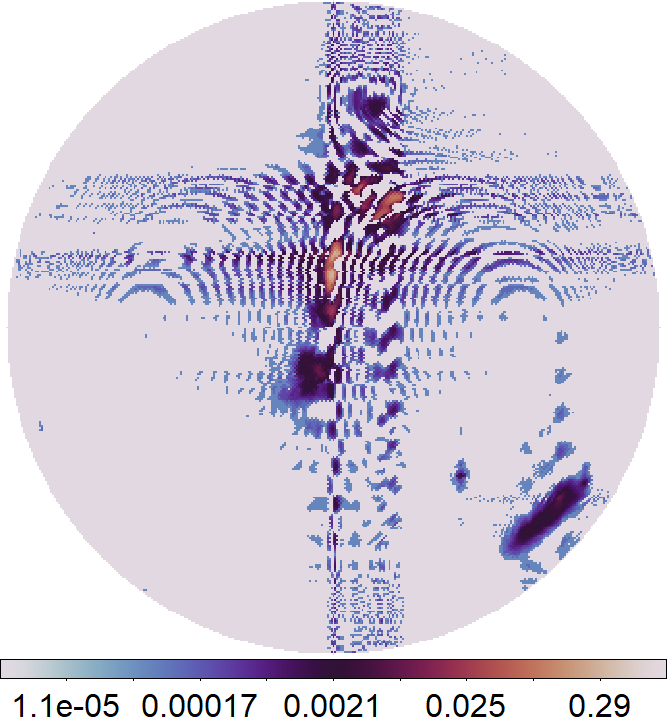} \\

 & \hspace{0.8cm} R2D2; $\mathrm{N}_{\mathrm{p}}=400^2$; (4.3, 1.9)~dB & R2D2; $\mathrm{N}_{\mathrm{p}}=600^2$; (1.4, 5.5)~dB & R2D2; $\mathrm{N}_{\mathrm{p}}=800^2$; (1.5, -0.2)~dB \\
\addlinespace[4pt]

 & \hspace{0.8cm}
 \includegraphics[width=0.21\linewidth]{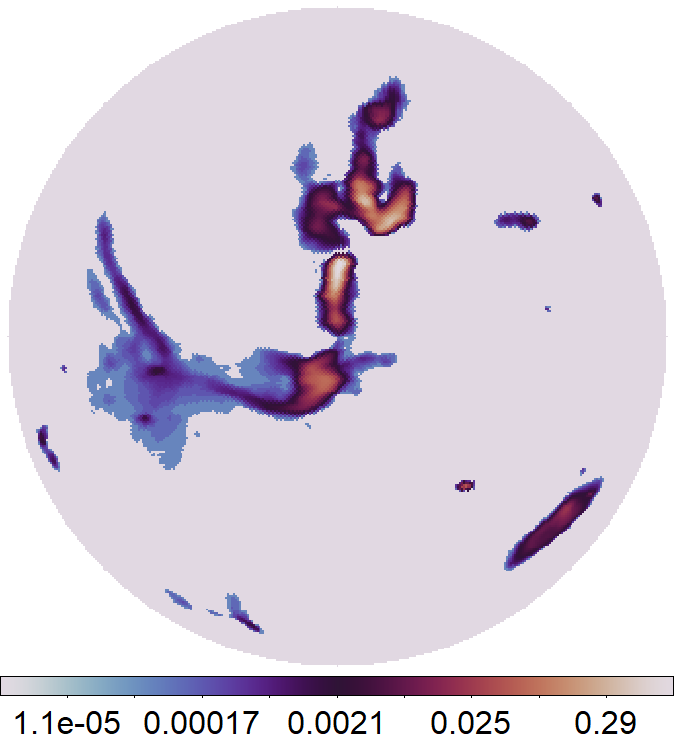} &
 \includegraphics[width=0.21\linewidth]{images/experiments/images/Reconstruction/Fourier/S-R2D2/600_new/PSZ.png} &
 \includegraphics[width=0.21\linewidth]{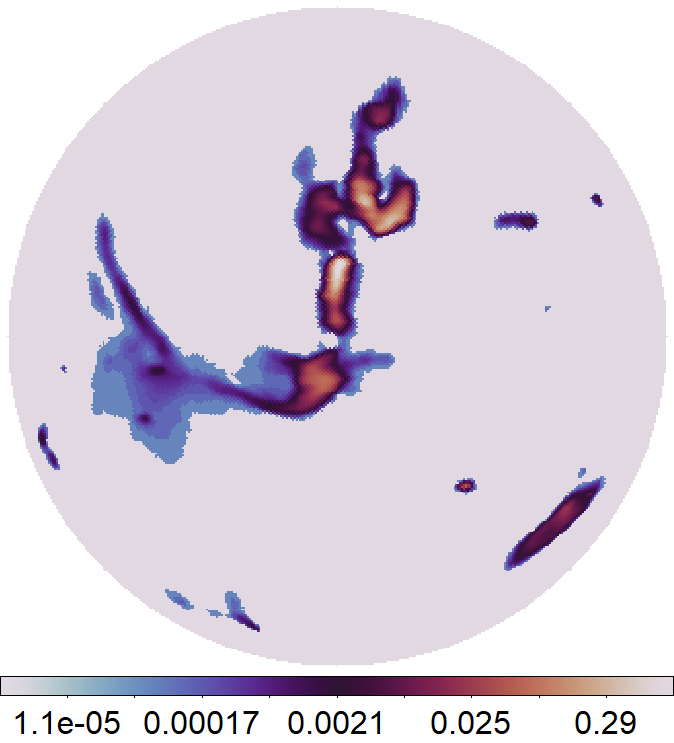} \\

 & \hspace{0.8cm} S-R2D2; $\mathrm{N}_{\mathrm{p}}=400^2$; (24.4, 15.3)~dB & S-R2D2; $\mathrm{N}_{\mathrm{p}}=600^2$; (25.6, 18.1)~dB & S-R2D2; $\mathrm{N}_{\mathrm{p}}=800^2$; (24.5, 15.1)~dB\\
\addlinespace[16pt]

 \multirow{2}{*}{\begin{minipage}{0.21\linewidth}
     \vspace{-2cm} 
     \hspace{-0.4cm}
     \centering
     \includegraphics[width=\linewidth]{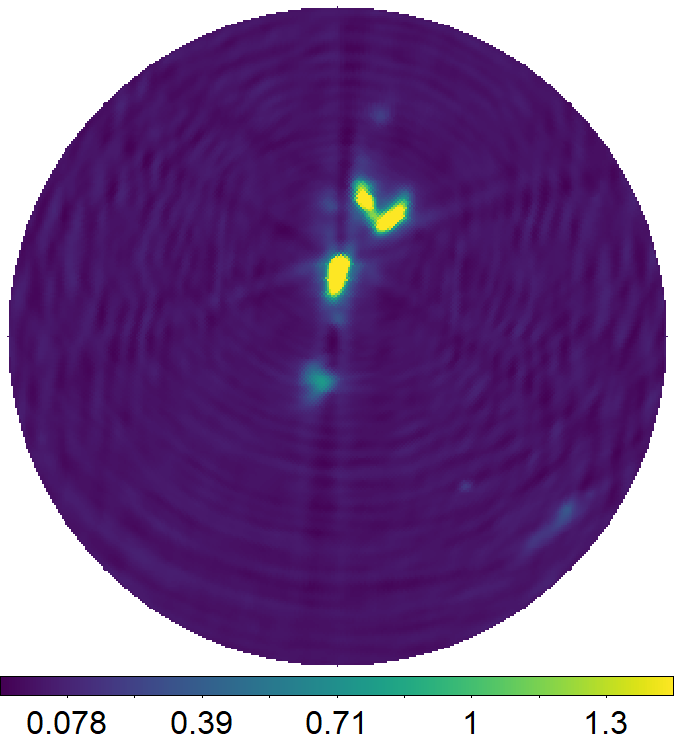} \\
     Spherical Dirty; $\mathbf{x}_s^{\mathrm{d}}$
 \end{minipage}} & \hspace{0.8cm}
 \includegraphics[width=0.21\linewidth]{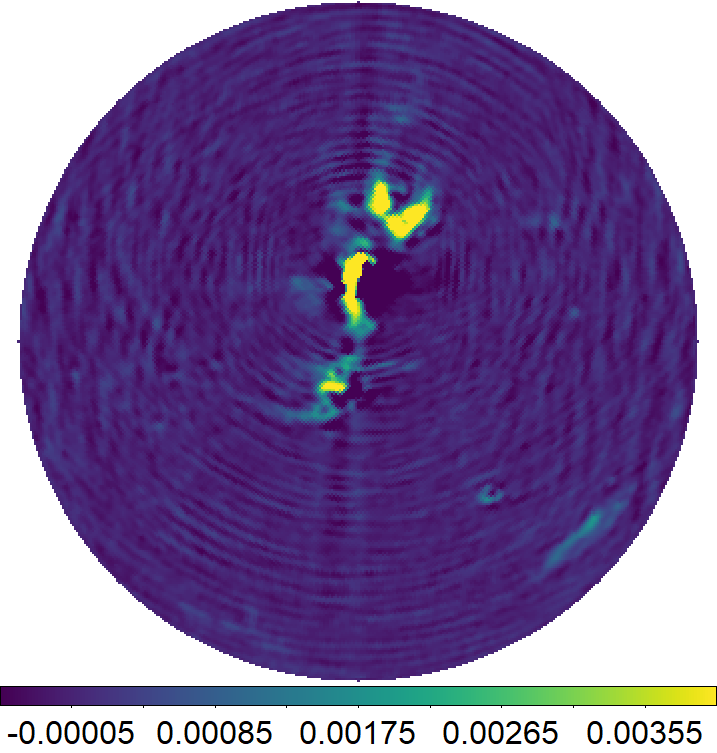} &
 \includegraphics[width=0.21\linewidth]{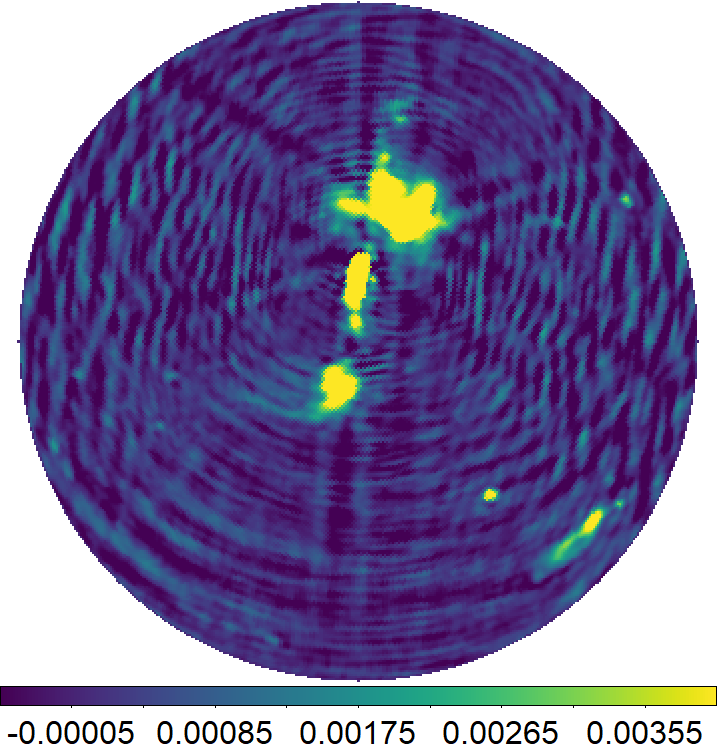} &
 \includegraphics[width=0.21\linewidth]{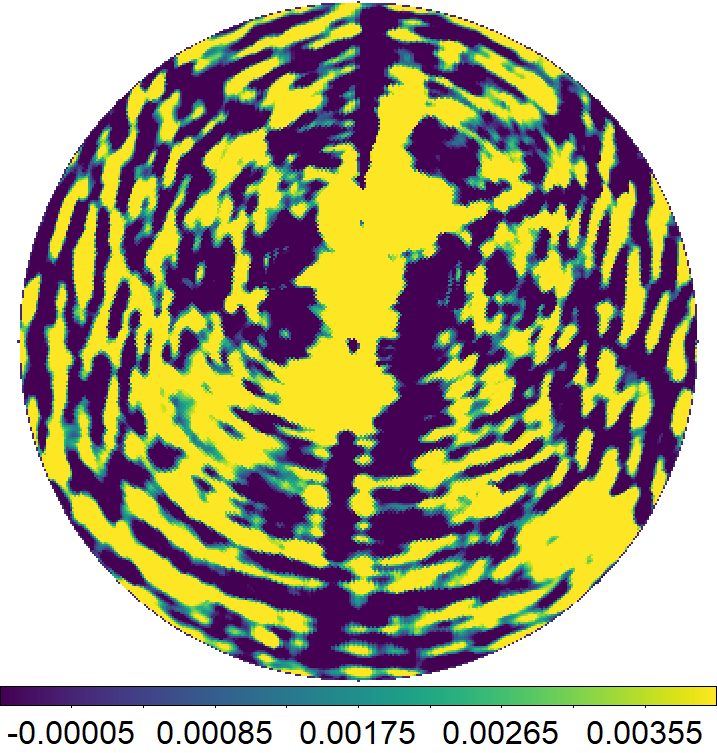} \\

 & \hspace{0.8cm} R2D2; $\mathrm{N}_{\mathrm{p}}=400^2$; $3.0\!\times\!10^{-3}$ & R2D2; $\mathrm{N}_{\mathrm{p}}=600^2$; $2.0\!\times\!10^{-2}$ & R2D2; $\mathrm{N}_{\mathrm{p}}=800^2$; $2.5\!\times\!10^{-1}$\\
\addlinespace[4pt]

 & \hspace{0.8cm}
 \includegraphics[width=0.21\linewidth]{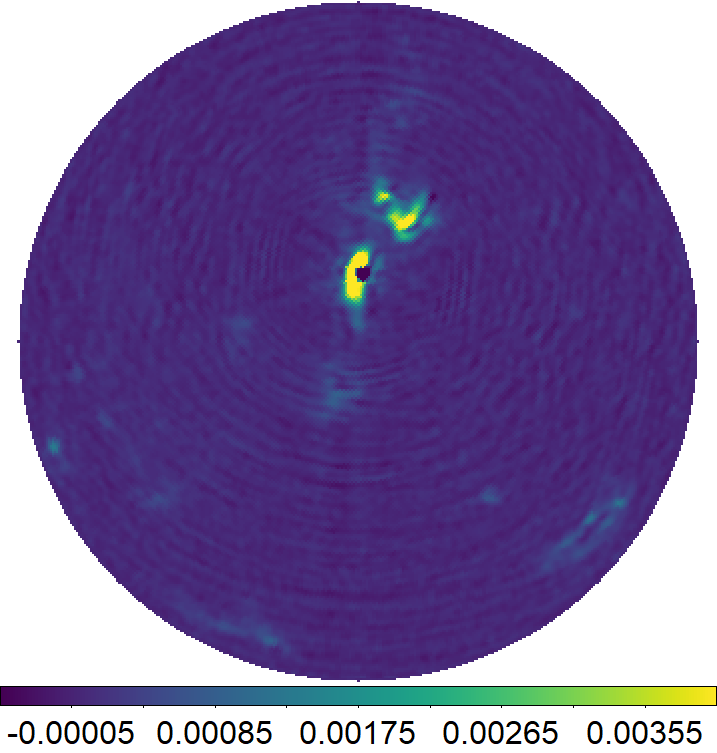} &
 \includegraphics[width=0.21\linewidth]{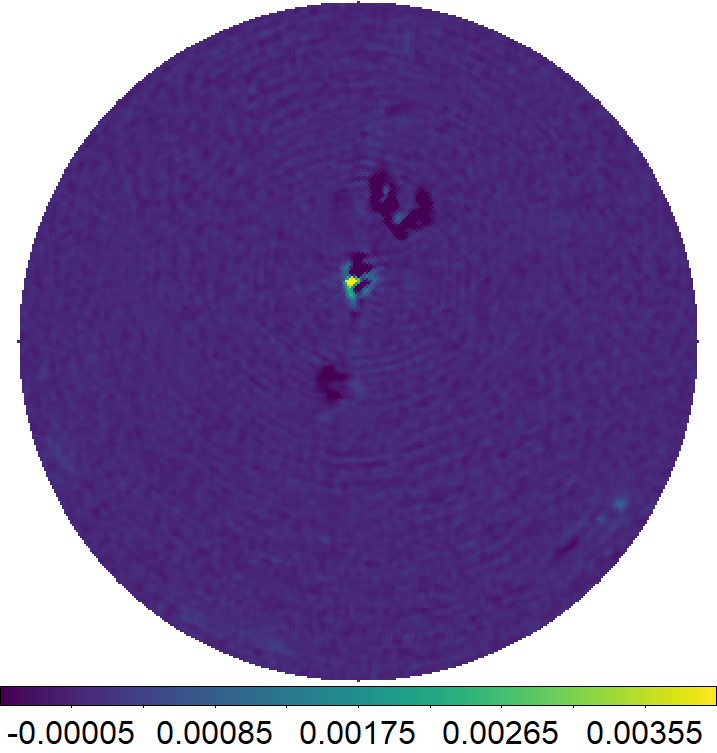} &
 \includegraphics[width=0.21\linewidth]{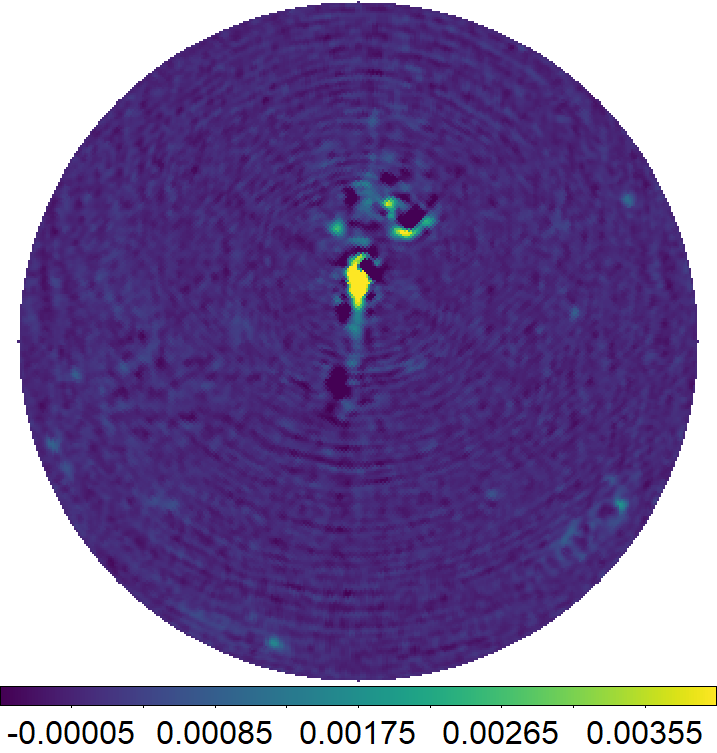} \\

 & \hspace{0.8cm} S-R2D2; $\mathrm{N}_{\mathrm{p}}=400^2$; $1.7\!\times\!10^{-3}$ & S-R2D2; $\mathrm{N}_{\mathrm{p}}=600^2$; $7.9\!\times\!10^{-4}$ & S-R2D2; $\mathrm{N}_{\mathrm{p}}=800^2$; $2.0\!\times\!10^{-3}$\\
 \end{tabular}
 \caption{
 Visual final reconstructions of the spherical ground truth $\mathbf{x}_{\mathrm{s}}^{\star}$, displayed in the top-left corner, obtained with the S-R2D2 Algorithm~\ref{algo:S_R2D2_reconstruction} or the R2D2 Algorithm~\ref{algo:R2D2_reconstruction} (adapted following the procedure depicted in Section~\ref{subsec:sec_5_subsec_1}) for different resolutions on the plane $\mathrm{N}_{\mathrm{p}}$. The spherical ground truth was generated with a $\textrm{DR}=2.3\!\times\!10^5$ from the galaxy cluster Abell~2034 \citep{botteon2022} following the procedure depicted in Section~\ref{subsec:sec_4_subsec_1}. From this spherical ground truth, we then generate its corresponding spherical dirty signal (Section~\ref{subsec:sec_4_subsec_2}). The two top row show the estimated signals displayed in logarithmic scale, with the logarithmic exponent equals to $\textrm{DR}$. Values of ($\textrm{SNR}$, $\textrm{logSNR}$) metrics are reported below each estimated image. The two bottom rows show the dirty signal and the residuals displayed on the sphere using $\mathbf{\Gamma}^{\dagger}$ and in linear scale. The value of the metric $\textrm{RDR}$ is indicated below each residuals. For all spherical signals, we visualise the Northern hemisphere in the orthographic projection perspective.} 
 \label{fig:PSZ}
\end{figure*}

\section{Conclusion}\label{sec:conclusion}
\noindent
We have introduced S-R2D2, an extension of the supervised learning pipeline R2D2, designed to solve the wide-field RI inverse problem on the sphere. Its efficiency stems from two key components. Firstly, it integrates an efficient implementation of the wide-field measurement model which incorporates a Fourier-based sphere-to-plane interpolator, $\mathbf{\Gamma}$. Secondly, at each iteration, a 2D-Euclidean U-Net DNN is efficiently trained to capture residual image structures on the sphere, with a loss that enforces direct consistency between the spherical ground truth and the DNN output back-projected onto the sphere using the plane-to-sphere adjoint $\mathbf{\Gamma}^{\dagger}$. Importantly, the fast FFT-based implementations of $\mathbf{\Gamma}$ and $\mathbf{\Gamma}^{\dagger}$, which are also compatible with automatic differentiation, enable efficient training and reconstruction stages. \\

\noindent  
Through simulations, we demonstrated that S-R2D2 preserves R2D2's cutting-edge computational efficiency while achieving significantly higher imaging precision in a wide-field regime. Although $\mathbf{\Gamma}$ and $\mathbf{\Gamma}^{\dagger}$ inevitably bypass the resolution rule and operate at a lower-than-optimal resolution on the plane, their effective integration into the scheme results in the DNNs correcting interpolation approximations and ultimately learning effective regularisers in the spherical domain. Specifically, S-R2D2's reconstruction quality is maintained at the lowest intermediate resolution on the plane, corresponding to maintaining the same pixel size on the plane and the sphere. This enables a significant reduction in computational complexity and ultimately facilitates reconstructions at higher resolutions on the sphere, thus enhancing S-R2D2's scalability.\\

\noindent
Future work will focus on expanding S-R2D2's applicability to simulations in realistic wide-field settings, including the integration of the $w$-component in the measurement model. Additionally, validating S-R2D2 on real data from telescopes such as LOFAR and MWA will be a crucial step towards its practical deployment.

\section*{Data Availability}
\noindent
R2D2 codes are available in the BASPLib\footnotemark \ code library on GitHub\footnotetext{\href{https://basp-group.github.io/BASPLib/}{https://basp-group.github.io/BASPLib/}}. S-R2D2 codes will be made available as part of a future release. BASPLib is developed and maintained by the Biomedical and Astronomical Signal Processing Laboratory (BASP\footnotemark)\footnotetext{\href{https://basp.site.hw.ac.uk/}{https://basp.site.hw.ac.uk/}}.


\section*{Acknowledgments}
\noindent
The authors thank Chung San Chu for his assistance with the training of the DNNs and Arwa Dabbech for insightful discussions on the implementation of the measurement model. The research of AT was supported by the EPFL Center for Imaging. The research of AA and YW was supported by the UK Research and Innovation under the EPSRC grant EP/T028270/1 and the STFC grant ST/W000970/1. The authors acknowledge the use of the Heriot-Watt high-performance computing facility (DMOG) and associated support services.


\bibliographystyle{mnras}
\bibliography{references} 

\begin{thebibliography}{}
\makeatletter
\relax
\def\mn@urlcharsother{\let\do\@makeother \do\$\do\&\do\#\do\^\do\_\do\%\do\~}
\def\mn@doi{\begingroup\mn@urlcharsother \@ifnextchar [ {\mn@doi@} {\mn@doi@[]}}
\def\mn@doi@[#1]#2{\def\@tempa{#1}\ifx\@tempa\@empty \href {http://dx.doi.org/#2} {doi:#2}\else \href {http://dx.doi.org/#2} {#1}\fi \endgroup}
\def\mn@eprint#1#2{\mn@eprint@#1:#2::\@nil}
\def\mn@eprint@arXiv#1{\href {http://arxiv.org/abs/#1} {{\tt arXiv:#1}}}
\def\mn@eprint@dblp#1{\href {http://dblp.uni-trier.de/rec/bibtex/#1.xml} {dblp:#1}}
\def\mn@eprint@#1:#2:#3:#4\@nil{\def\@tempa {#1}\def\@tempb {#2}\def\@tempc {#3}\ifx \@tempc \@empty \let \@tempc \@tempb \let \@tempb \@tempa \fi \ifx \@tempb \@empty \def\@tempb {arXiv}\fi \@ifundefined {mn@eprint@\@tempb}{\@tempb:\@tempc}{\expandafter \expandafter \csname mn@eprint@\@tempb\endcsname \expandafter{\@tempc}}}

\bibitem[\protect\citeauthoryear{Aghabiglou, San~Chu, Dabbech  \& Wiaux}{Aghabiglou et~al.}{2024}]{aghabiglou2024r2d2}
Aghabiglou A.,  San~Chu C.,  Dabbech A.,   Wiaux Y.,  2024, The Astrophysical Journal Supplement Series, 273, 3

\bibitem[\protect\citeauthoryear{Aghabiglou, San~Chu, Chao, Dabbech  \& Wiaux}{Aghabiglou et~al.}{2025}]{aghabiglou2025}
Aghabiglou A.,  San~Chu C.,  Chao T.,  Dabbech A.,   Wiaux Y.,  2025, preprint researchportal.hw.ac.uk:145493649

\bibitem[\protect\citeauthoryear{Barnett, Magland  \& Klinteberg}{Barnett et~al.}{2024}]{barnett2024finufft}
Barnett A.~H.,  Magland J.~F.,   Klinteberg L.~A.,  2024, Flatiron Institute Nonuniform Fast Fourier Transform Libraries (FINUFFT), \url{http://github.com/flatironinstitute/finufft}

\bibitem[\protect\citeauthoryear{Botteon et~al.,}{Botteon et~al.}{2022}]{botteon2022}
Botteon A.,  et~al., 2022, \aap, 660, A78

\bibitem[\protect\citeauthoryear{{Braun}, {Bourke}, {Green}, {Keane}  \& {Wagg}}{{Braun} et~al.}{2015}]{SKA_Low_1}
{Braun} R.,  {Bourke} T.,  {Green} J.~A.,  {Keane} E.,   {Wagg} J.,  2015, in Advancing Astrophysics with the Square Kilometre Array (AASKA14). p.~174, \mn@doi{10.22323/1.215.0174}

\bibitem[\protect\citeauthoryear{{Clark}}{{Clark}}{1980}]{Clark1980}
{Clark} B.~G.,  1980, \aap, \href {https://ui.adsabs.harvard.edu/abs/1980A&A....89..377C} {89, 377}

\bibitem[\protect\citeauthoryear{Connor, Bouman, Ravi  \& Hallinan}{Connor et~al.}{2022}]{Connor_2022}
Connor L.,  Bouman K.~L.,  Ravi V.,   Hallinan G.,  2022, \mn@doi [Monthly Notices of the Royal Astronomical Society] {10.1093/mnras/stac1329}, 514, 2614

\bibitem[\protect\citeauthoryear{{Cornwell} \& {Perley}}{{Cornwell} \& {Perley}}{1992}]{Cornwell_Perley_1992}
{Cornwell} T.~J.,  {Perley} R.~A.,  1992, \aap, \href {https://ui.adsabs.harvard.edu/abs/1992A&A...261..353C} {261, 353}

\bibitem[\protect\citeauthoryear{Dabbech, Wolz, Pratley, McEwen  \& Wiaux}{Dabbech et~al.}{2017}]{dabbech_2017}
Dabbech A.,  Wolz L.,  Pratley L.,  McEwen J.~D.,   Wiaux Y.,  2017, \mn@doi [Monthly Notices of the Royal Astronomical Society] {10.1093/mnras/stx1775}, 471, 4300

\bibitem[\protect\citeauthoryear{Dabbech, Terris, Jackson, Ramatsoku, Smirnov  \& Wiaux}{Dabbech et~al.}{2022}]{Dabbech_2022}
Dabbech A.,  Terris M.,  Jackson A.,  Ramatsoku M.,  Smirnov O.~M.,   Wiaux Y.,  2022, \mn@doi [The Astrophysical Journal Letters] {10.3847/2041-8213/ac98af}, 939, L4

\bibitem[\protect\citeauthoryear{Dabbech, Aghabiglou, Chu  \& Wiaux}{Dabbech et~al.}{2024}]{Dabbech_2024}
Dabbech A.,  Aghabiglou A.,  Chu C.~S.,   Wiaux Y.,  2024, \mn@doi [The Astrophysical Journal Letters] {10.3847/2041-8213/ad41df}, 966, L34

\bibitem[\protect\citeauthoryear{Eastwood et~al.,}{Eastwood et~al.}{2018}]{Eastwood_2018}
Eastwood M.~W.,  et~al., 2018, \mn@doi [The Astronomical Journal] {10.3847/1538-3881/aac721}, 156, 32

\bibitem[\protect\citeauthoryear{Edler et~al.,}{Edler et~al.}{2023}]{edler2023}
Edler H.,  et~al., 2023, \aap, 676, A24

\bibitem[\protect\citeauthoryear{{Garsden, H.} et~al.,}{{Garsden, H.} et~al.}{2015}]{Gardsen}
{Garsden, H.} et~al., 2015, \mn@doi [A&A] {10.1051/0004-6361/201424504}, 575, A90

\bibitem[\protect\citeauthoryear{Gorski, Hivon, Banday, Wandelt, Hansen, Reinecke  \& Bartelmann}{Gorski et~al.}{2005}]{Gorski_2005}
Gorski K.~M.,  Hivon E.,  Banday A.~J.,  Wandelt B.~D.,  Hansen F.~K.,  Reinecke M.,   Bartelmann M.,  2005, \mn@doi [The Astrophysical Journal] {10.1086/427976}, 622, 759–771

\bibitem[\protect\citeauthoryear{Ha \& Lyu}{Ha \& Lyu}{2022}]{SPHARM_Net}
Ha S.,  Lyu I.,  2022, \mn@doi [IEEE Transactions on Medical Imaging] {10.1109/TMI.2022.3168670}, 41, 2739

\bibitem[\protect\citeauthoryear{Hivon, Hansen, Wandelt, Górski, Banday  \& Reinecke}{Hivon et~al.}{2010}]{hivon_2010}
Hivon E.,  Hansen F.,  Wandelt B.,  Górski K.,  Banday A.,   Reinecke M.,  2010, HEALPix Fortran Facility User Guidelines, V2.15a

\bibitem[\protect\citeauthoryear{H{\"o}gbom}{H{\"o}gbom}{1974}]{hogbom1974aperture}
H{\"o}gbom J.,  1974, Astronomy and Astrophysics Supplement, Vol. 15, p. 417, 15, 417

\bibitem[\protect\citeauthoryear{{Junklewitz H.,}, {Bell M. R.,}, {Selig M.,}  \& {Enßlin T. A.}}{{Junklewitz H.,} et~al.}{2016}]{Junklewitz}
{Junklewitz H.,} {Bell M. R.,} {Selig M.,}  {Enßlin T. A.} 2016, \mn@doi [A&A] {10.1051/0004-6361/201323094}, 586, A76

\bibitem[\protect\citeauthoryear{Kashani, Queralt, Jarret  \& Simeoni}{Kashani et~al.}{2023}]{kashani2023}
Kashani S.,  Queralt J.~R.,  Jarret A.,   Simeoni M.,  2023, HVOX: Scalable Interferometric Synthesis and Analysis of Spherical Sky Maps (\mn@eprint {arXiv} {2306.06007}), \url {https://arxiv.org/abs/2306.06007}

\bibitem[\protect\citeauthoryear{Knoll et~al.,}{Knoll et~al.}{2020}]{knoll2020fastmri}
Knoll F.,  et~al., 2020, Radiol. Artif. Intell., 2, e190007

\bibitem[\protect\citeauthoryear{{Krachmalnicoff, N.} \& {Tomasi, M.}}{{Krachmalnicoff, N.} \& {Tomasi, M.}}{2019}]{ray_tracing}
{Krachmalnicoff, N.} {Tomasi, M.} 2019, \mn@doi [A\&A] {10.1051/0004-6361/201935211}, 628, A129

\bibitem[\protect\citeauthoryear{Kriele, Wayth, Bentum, Juswardy  \& Trott}{Kriele et~al.}{2022}]{Kriele2022}
Kriele M.~A.,  Wayth R.~B.,  Bentum M.~J.,  Juswardy B.,   Trott C.~M.,  2022, \mn@doi [Publications of the Astronomical Society of Australia] {10.1017/pasa.2022.2}, 39, e017

\bibitem[\protect\citeauthoryear{Lonsdale et~al.,}{Lonsdale et~al.}{2009}]{MWA1}
Lonsdale C.~J.,  et~al., 2009, \mn@doi [Proceedings of the IEEE] {10.1109/JPROC.2009.2017564}, 97, 1497

\bibitem[\protect\citeauthoryear{McEwen \& Wiaux}{McEwen \& Wiaux}{2011}]{McEwen_Wiaux_WFOV}
McEwen J.~D.,  Wiaux Y.,  2011, \mn@doi [Monthly Notices of the Royal Astronomical Society] {10.1111/j.1365-2966.2011.18217.x}, 413, 1318–1332

\bibitem[\protect\citeauthoryear{Monnier, Guibert, Tasse, Gac, Orieux, Raffin, Smirnov  \& Hugo}{Monnier et~al.}{2022}]{Monnier}
Monnier N.,  Guibert D.,  Tasse C.,  Gac N.,  Orieux F.,  Raffin E.,  Smirnov O.~M.,   Hugo B.~V.,  2022, in 2022 IEEE Workshop on Signal Processing Systems (SiPS). pp~1--6, \mn@doi{10.1109/SiPS55645.2022.9919239}

\bibitem[\protect\citeauthoryear{Morabito et~al.,}{Morabito et~al.}{2022}]{LOFAR2}
Morabito L.~K.,  et~al., 2022, \mn@doi [Monthly Notices of the Royal Astronomical Society] {10.1093/mnras/stac2129}, 515, 5758

\bibitem[\protect\citeauthoryear{Muckley, Stern, Murrell  \& Knoll}{Muckley et~al.}{2020}]{muckley_20}
Muckley M.~J.,  Stern R.,  Murrell T.,   Knoll F.,  2020, in ISMRM Workshop on Data Sampling \& Image Reconstruction.

\bibitem[\protect\citeauthoryear{{Noordam} \& {Smirnov}}{{Noordam} \& {Smirnov}}{2010}]{Noordam2010}
{Noordam} J.~E.,  {Smirnov} O.~M.,  2010, \aap, 524, A61

\bibitem[\protect\citeauthoryear{Ocampo, Price  \& McEwen}{Ocampo et~al.}{2023}]{ocampo2023}
Ocampo J.,  Price M.~A.,   McEwen J.~D.,  2023, Scalable and Equivariant Spherical CNNs by Discrete-Continuous (DISCO) Convolutions (\mn@eprint {arXiv} {2209.13603}), \url {https://arxiv.org/abs/2209.13603}

\bibitem[\protect\citeauthoryear{Offringa et~al.,}{Offringa et~al.}{2014}]{offringa2014}
Offringa A.,  et~al., 2014, \mnras, 444, 606

\bibitem[\protect\citeauthoryear{Paszke et~al.,}{Paszke et~al.}{2019}]{paszke2019pytorch}
Paszke A.,  et~al., 2019, in Wallach H.,  Larochelle H.,  Beygelzimer A.,  d\textquotesingle Alch\'{e}-Buc F.,  Fox E.,   Garnett R.,  eds,  NeurIPS Vol. 32, Advances in Neural Information Processing Systems. Curran Associates, Inc., \url {https://proceedings.neurips.cc/paper_files/paper/2019/file/bdbca288fee7f92f2bfa9f7012727740-Paper.pdf}

\bibitem[\protect\citeauthoryear{Perraudin, Defferrard, Kacprzak  \& Sgier}{Perraudin et~al.}{2019}]{Perraudin_2019}
Perraudin N.,  Defferrard M.,  Kacprzak T.,   Sgier R.,  2019, \mn@doi [Astronomy and Computing] {10.1016/j.ascom.2019.03.004}, 27, 130–146

\bibitem[\protect\citeauthoryear{Shaw, Sigurdson, Pen, Stebbins  \& Sitwell}{Shaw et~al.}{2014}]{Shaw_2014}
Shaw J.~R.,  Sigurdson K.,  Pen U.-L.,  Stebbins A.,   Sitwell M.,  2014, \mn@doi [The Astrophysical Journal] {10.1088/0004-637X/781/2/57}, 781, 57

\bibitem[\protect\citeauthoryear{Shimwell et~al.,}{Shimwell et~al.}{2022}]{shimwell2022}
Shimwell T.,  et~al., 2022, \aap, 659, A1

\bibitem[\protect\citeauthoryear{Strohmer}{Strohmer}{2000}]{STROHMER2000297}
Strohmer T.,  2000, \mn@doi [Journal of Computational and Applied Mathematics] {https://doi.org/10.1016/S0377-0427(00)00361-7}, 122, 297

\bibitem[\protect\citeauthoryear{Tasse et~al.,}{Tasse et~al.}{2018}]{Tasse_2018}
Tasse C.,  et~al., 2018, \mn@doi [Astronomy &amp; Astrophysics] {10.1051/0004-6361/201731474}, 611, A87

\bibitem[\protect\citeauthoryear{Terris, Dabbech, Tang  \& Wiaux}{Terris et~al.}{2022}]{Terris2022}
Terris M.,  Dabbech A.,  Tang C.,   Wiaux Y.,  2022, \mn@doi [Monthly Notices of the Royal Astronomical Society] {10.1093/mnras/stac2672}, 518, 604

\bibitem[\protect\citeauthoryear{Terris, Tang, Jackson  \& Wiaux}{Terris et~al.}{2025}]{terris2024}
Terris M.,  Tang C.,  Jackson A.,   Wiaux Y.,  2025, \mn@doi [Monthly Notices of the Royal Astronomical Society] {10.1093/mnras/staf022}, 537, 1608

\bibitem[\protect\citeauthoryear{{Thompson}, {Moran}  \& {Swenson}}{{Thompson} et~al.}{2017}]{Thompson2017}
{Thompson} A.~R.,  {Moran} J.~M.,   {Swenson} Jr. G.~W.,  2017, {Interferometry and Synthesis in Radio Astronomy, 3rd Edition}.
Springer, \mn@doi{10.1007/978-3-319-44431-4}

\bibitem[\protect\citeauthoryear{Tingay et~al.,}{Tingay et~al.}{2013}]{MWA2}
Tingay S.~J.,  et~al., 2013, \mn@doi [Publications of the Astronomical Society of Australia] {10.1017/pasa.2012.007}, 30, e007

\bibitem[\protect\citeauthoryear{Tolley et~al.,}{Tolley et~al.}{2025}]{TOLLEY2025100920}
Tolley E.,  et~al., 2025, \mn@doi [Astronomy and Computing] {https://doi.org/10.1016/j.ascom.2024.100920}, 51, 100920

\bibitem[\protect\citeauthoryear{Wayth et~al.,}{Wayth et~al.}{2018}]{MWA3}
Wayth R.,  et~al., 2018, \mn@doi [Publications of the Astronomical Society of Australia] {10.1017/pasa.2018.37}, 35

\bibitem[\protect\citeauthoryear{Wayth et~al.,}{Wayth et~al.}{2021}]{SKA_Low_2}
Wayth R.,  et~al., 2021, \mn@doi [Journal of Astronomical Telescopes, Instruments, and Systems] {10.1117/1.JATIS.8.1.011010}, 8, 011010

\bibitem[\protect\citeauthoryear{Wiaux, Jacques, Puy, Scaife  \& Vandergheynst}{Wiaux et~al.}{2009}]{wiaux2009compressed}
Wiaux Y.,  Jacques L.,  Puy G.,  Scaife A. M.~M.,   Vandergheynst P.,  2009, \mn@doi [Monthly Notices of the Royal Astronomical Society] {10.1111/j.1365-2966.2009.14665.x}, 395, 1733

\bibitem[\protect\citeauthoryear{Zbontar et~al.,}{Zbontar et~al.}{2018}]{zbontar2018fastmri}
Zbontar J.,  et~al., 2018, CoRR, abs/1811.08839

\bibitem[\protect\citeauthoryear{van Haarlem et~al.,}{van Haarlem et~al.}{2013}]{LOFAR1}
van Haarlem M.~P.,  et~al., 2013, \mn@doi [{Astronomy and Astrophysics - A\&A}] {10.1051/0004-6361/201220873}, A2, 53 pages

\makeatother
\end{thebibliography}



\bsp	
\label{lastpage}

\end{document}